\documentclass[aps,prb,groupedaddress,twocolumn,showpacs,floatfix]{revtex4-1}
\usepackage{graphicx}
\usepackage{gensymb}
\usepackage{amsmath}
\usepackage{amstext}
\usepackage{amssymb}
\usepackage{amsfonts}
\usepackage{amsbsy} 
\usepackage{dcolumn}
\usepackage{bm}
\usepackage[all]{xy}
\usepackage{braket}
\usepackage{color}
\usepackage[colorlinks,citecolor=blue]{hyperref}

\usepackage[bbgreekl]{mathbbol}
\usepackage{amscd}

\makeatletter
\newlength{\negph@wd}
\DeclareRobustCommand{\negphantom}[1]{%
  \ifmmode
    \mathpalette\negph@math{#1}%
  \else
    \negph@do{#1}%
  \fi
}
\newcommand{\negph@math}[2]{\negph@do{$\m@th#1#2$}}
\newcommand{\negph@do}[1]{%
  \settowidth{\negph@wd}{#1}%
  \hspace*{-\negph@wd}%
}
\makeatother

\begin{document}

\title{Twisted gauge theories in 3D Walker-Wang models}

\author{Zitao Wang}
\affiliation{Department of Physics and Institute for Quantum Information and Matter, California Institute of Technology, Pasadena, CA 91125, USA}

\author{Xie Chen}
\affiliation{Department of Physics and Institute for Quantum Information and Matter, California Institute of Technology, Pasadena, CA 91125, USA}

\begin{abstract}
Three dimensional gauge theories with a discrete gauge group can emerge from spin models as a gapped topological phase with fractional point excitations (gauge charge) and loop excitations (gauge flux). It is known that 3D gauge theories can be ``twisted'', in the sense that the gauge flux loops can have nontrivial braiding statistics among themselves and such twisted gauge theories are realized in models discovered by Dijkgraaf and Witten. A different framework to systematically construct three dimensional topological phases was proposed by Walker and Wang and a series of examples have been studied. Can the Walker Wang construction be used to realize the topological order in twisted gauge theories? This is not immediately clear because the Walker-Wang construction is based on a loop condensation picture while the Dijkgraaf-Witten theory is based on a membrane condensation picture. In this paper, we show that the answer to this question is Yes, by presenting an explicit construction of the Walker Wang models which realize both the twisted and untwisted gauge theories with gauge group $\mathbb{Z}_2 \times \mathbb{Z}_2$. We identify the topological order of the models by performing modular transformations on the ground state wave functions and show that the modular matrices exactly match those for the $\mathbb{Z}_2 \times \mathbb{Z}_2$ gauge theories. By relating the Walker-Wang construction to the Dijkgraaf-Witten construction, our result opens up a way to study twisted gauge theories with fermonic charges, and correspondingly strongly interacting fermionic symmetry protected topological phases and their surface states, through exactly solvable models.  
\end{abstract}

\maketitle

\section{Introduction}

It is an important problem in condensed matter physics to understand gapped quantum phases of matter. Two gapped systems are said to be equivalent if their Hamiltonians can be deformed into each other without closing the energy gap, or equivalently, if their ground states are related by a local unitary (LU) evolution.\cite{Chen2010} We define a topological phase as an equivalence class of gapped systems under such deformation of the Hamiltonian or evolution of the ground state. Note that all systems whose ground state can be transformed into a product state through an LU evolution lie in the same phase called the short-range entangled (SRE) phase. Systems that are not in the SRE phase are said to be in the long-range entangled (LRE) phase.

%

Substantial progress has been made in the study of 2D topological phases. Topological phases in 2D are characterized by, for example, robust ground state degeneracy on spaces with nontrivial topology,\cite{Haldane1985,Wen1989,WenNiu1990} gapless edge excitations,\cite{Wen1991,Wen1992} quasiparticle excitations with anyonic statistics,\cite{Wilczek1982Jan,Wilczek1982Oct,Arovas1984} and nonabelian Berry phases induced by modular transformations in the degenerate ground space on a torus (the $S$ and $T$ matrices),\cite{Wilczek1984,Wen1990,Vakkuri1993} which are directly related to the quasiparticle statistics. It was conjectured that the $S$ and $T$ matrices give complete description of a topological phase,\cite{Wen1990} and therefore serve as ``non-local order parameters'' of the phase.\cite{Bais2012} Another approach to study topological phases in 2D is from an effective field theory point of view. Assuming that the macroscopic properties of the system are described by a topological quantum field theory (TQFT),\cite{Atiyah1988,Witten1988,Witten1989,Turaev1994} which in two spatial dimensions is described by the mathematical construction of a modular tensor category (MTC),\cite{Turaev1994} one can have an algebraic description of anyons in the system in terms of MTC. A subclass of the systems --- those with vanishing thermal Hall conductivity (vanishing MTC central charge) and gapped boundary, admits a simple, exactly solvable Hamiltonian description in terms of the string-net models proposed by Levin and Wen.\cite{LevinWen2004}

%

\begin{figure}
\centering
\includegraphics[width=0.45\linewidth]{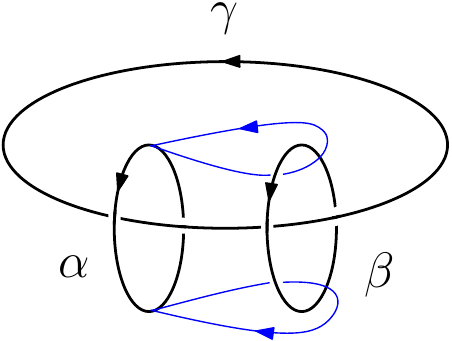}\caption{A three-loop braiding process. The process involves a flux loop $\alpha$ sweeping out a torus enclosing a flux loop $\beta$ while both linked with a ``base'' flux loop $\gamma$. The blue curves indicate the trajectory of two points on loop $\alpha$. If $\alpha$ and $\beta$ are identical, we can similarly define the process where $\alpha$ and $\beta$ are exchanged while both linked with $\gamma$.
}
\label{loop}
\end{figure}

What about 3D topological phases? A systematic understanding of topological phases in 3D systems is still lacking. An interesting family of 3D topological phases is discrete gauge theories and their twisted versions, which can be described by Dijkgraaf-Witten models.\cite{Dijkgraaf1990,Wan2014} The theory contains both point excitations and loop excitations, which are the gauge charges and flux loops, respectively. It was proposed in Ref. \onlinecite{Jiang2014} that for twisted gauge theories with abelian gauge groups, 3D modular transformations applied to the degenerate ground states of the system on a three-torus is related to certain three-loop braiding processes illustrated in Fig. \ref{loop} (such braiding process has also been discussed in Ref. \onlinecite{WangLevin2014}), and can be used to distinguish different 3D twisted gauge theories. Thus, the three-loop braiding statistics (or the 3D $S$ and $T$ matrices) can serve as ``non-local order parameters'' of 3D twisted gauge theories. Dijkgraaf-Witten models provide a systematic way to study 3D twisted gauge theories. However, they fail to describe theories with (at least one) fermionic gauge charges, so it would be nice to have some other exactly solvable models, which not only give us new perspectives on 3D twisted gauge theories, but also have the potential to describe theories involving fermionic gauge charges.

%

Walker-Wang models\cite{Walker2011,Keyserlingk2012} are viable candidates to describe 3D twisted gauge theories. Given the input of a set of anyons, they provide a way to write down exactly solvable models with 3D topological order. There are two types of Walker-Wang models: Those with a trivial (short-range entangled) bulk and those with a nontrivial (long-range entangled) bulk. Quasiparticle excitations in these models are well understood. First, there are anyons that appear only on the surface of both types of models. Secondly, there are deconfined quasiparticle excitations in the bulk of the second type of models, which can only be bosons or fermions. Besides quasiparticle excitations, Walker-Wang models also support loop excitations, but they are much less well understood. In this paper, we address this issue by asking the question: Can Walker-Wang models describe 3D twisted gauge theories with nontrivial three-loop braiding statistics? We will give an affirmative answer to this question by solving the following two problems:
\begin{enumerate}
\item{How do we choose the input data of the Walker-Wang models?}
\item{How do we determine the topological order of the output theory?}
\end{enumerate}

%

In particular, we study the examples of 3D $\mathbb{Z}_2 \times \mathbb{Z}_2$ gauge theories. There are 4 inequivalent such theories: one untwisted gauge theory and three twisted gauge theories. We find that if we choose the input data of a Walker-Wang model to be the $\mathbb{Z}_2 \times \mathbb{Z}_2$ symmetry charges and the anyons in the non-anomalous (resp. anomalous) projective semion states studied in Ref. \onlinecite{Chen2014}, the output theory is a 3D untwisted (resp. twisted) $\mathbb{Z}_2 \times \mathbb{Z}_2$ gauge theory. As we will see, there are 1 non-anomalous and 3 anomalous projective semion states, corresponding precisely to the 1 untwisted and 3 twisted $\mathbb{Z}_2 \times \mathbb{Z}_2$ gauge theories, respectively.

To determine the topological order in our Walker-Wang models, we perform 3D modular transformations to their ground space on a three-torus and calculate the resulting nonabelian Berry phases. By a dimensional reduction argument, we are able to obtain the three-loop braiding statistics, which distinguish the 3D $\mathbb{Z}_2 \times \mathbb{Z}_2$ gauge theories.

%
The remainder of the paper is organized as follows: In Section \ref{WW}, we review the Walker-Wang construction. In Section \ref{result}, we present the input data of the Walker-Wang models that describe the 3D $\mathbb{Z}_2 \times \mathbb{Z}_2$ gauge theories. We also explain the physical intuition of why such input data is chosen. In Section \ref{method}, we introduce the methods we use to deduce the topological order in our Walker-Wang models. In Section \ref{sum}, we summarize the results and discuss future directions. We also discuss some subtleties involved in doing 3D modular transformations on the Walker-Wang wave function. Some technical details involved in the arguments and calculations can be found in the appendices.

\section{Review of Walker-Wang models}
\label{WW}

\begin{figure}
\centering
\includegraphics[width=0.70\linewidth]{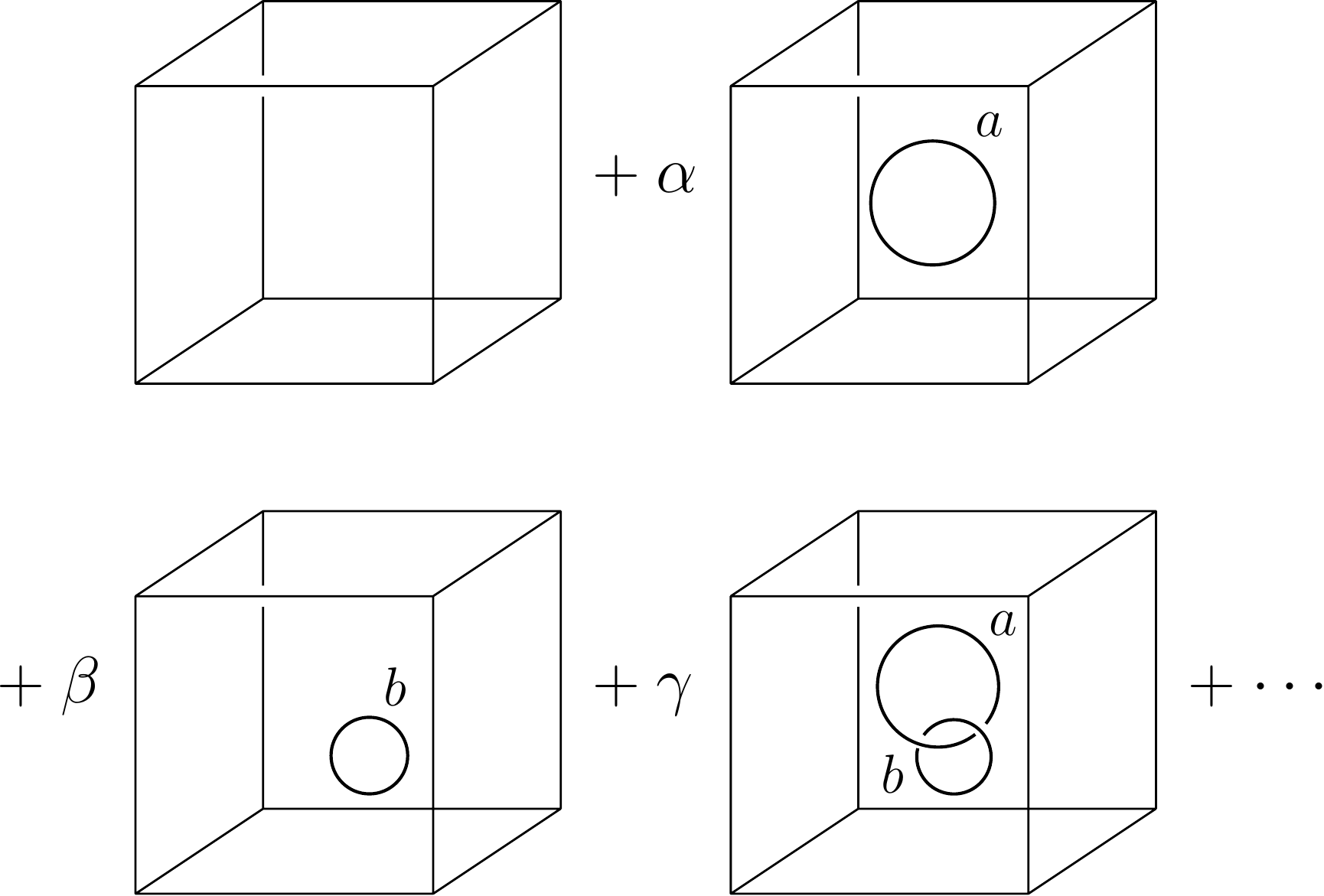}\caption{An example of the ground state wave function of a Walker-Wang model. $a$ and $b$ here label the quasiparticle types.}
\label{wf}
\end{figure}

\begin{figure}
\centering
\includegraphics[width=0.65\linewidth]{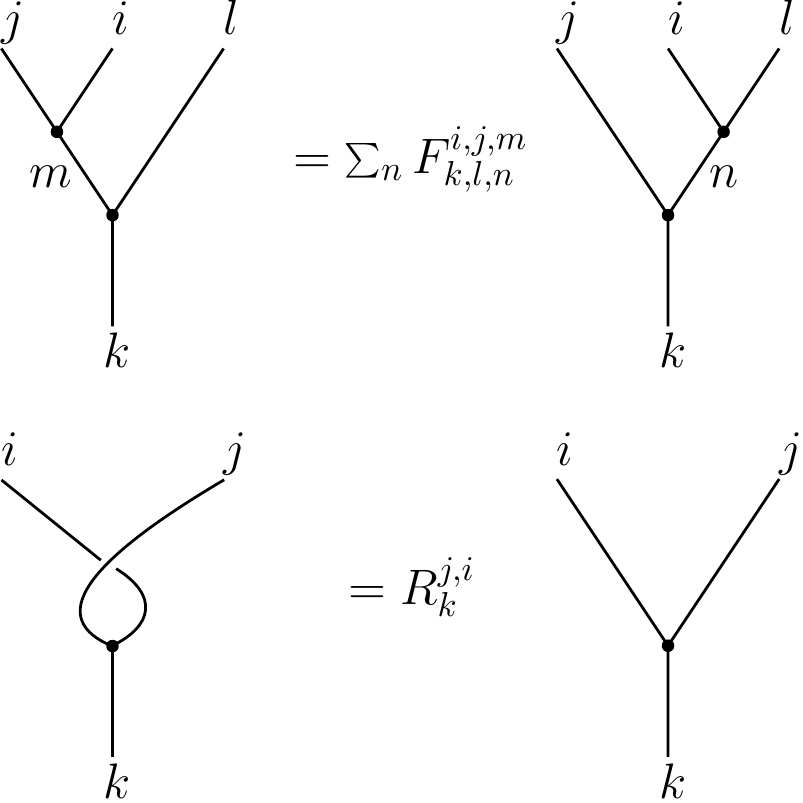}\caption{Graphical definition of $F$ and $R$ symbols.}
\label{fr}
\end{figure}

\begin{figure}
\centering
\includegraphics[width=0.70\linewidth]{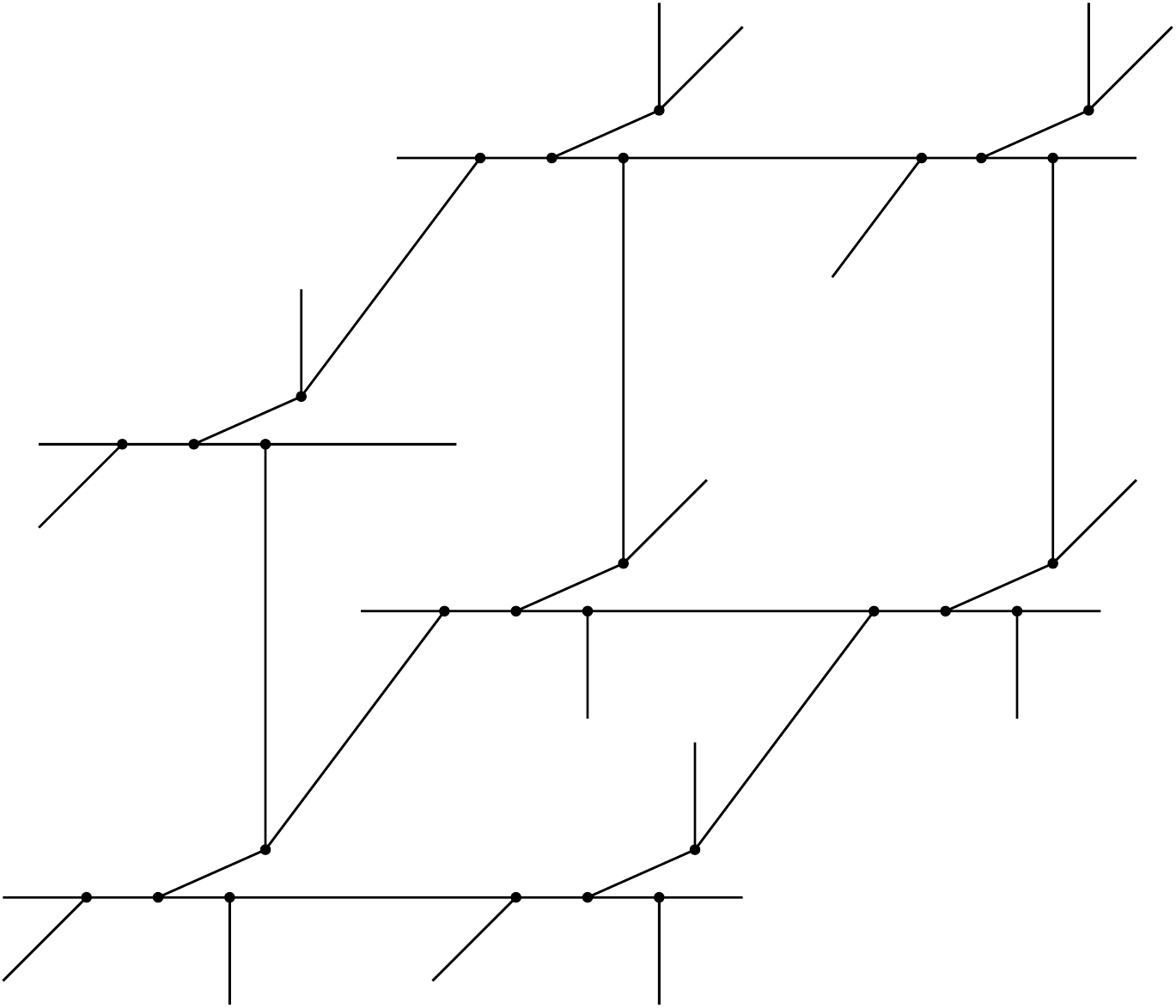}\caption{Planar projection of a trivalent resolution of the cublic lattice.}
\label{lattice}
\end{figure}

\begin{figure}
\centering
\includegraphics[width=0.80\linewidth]{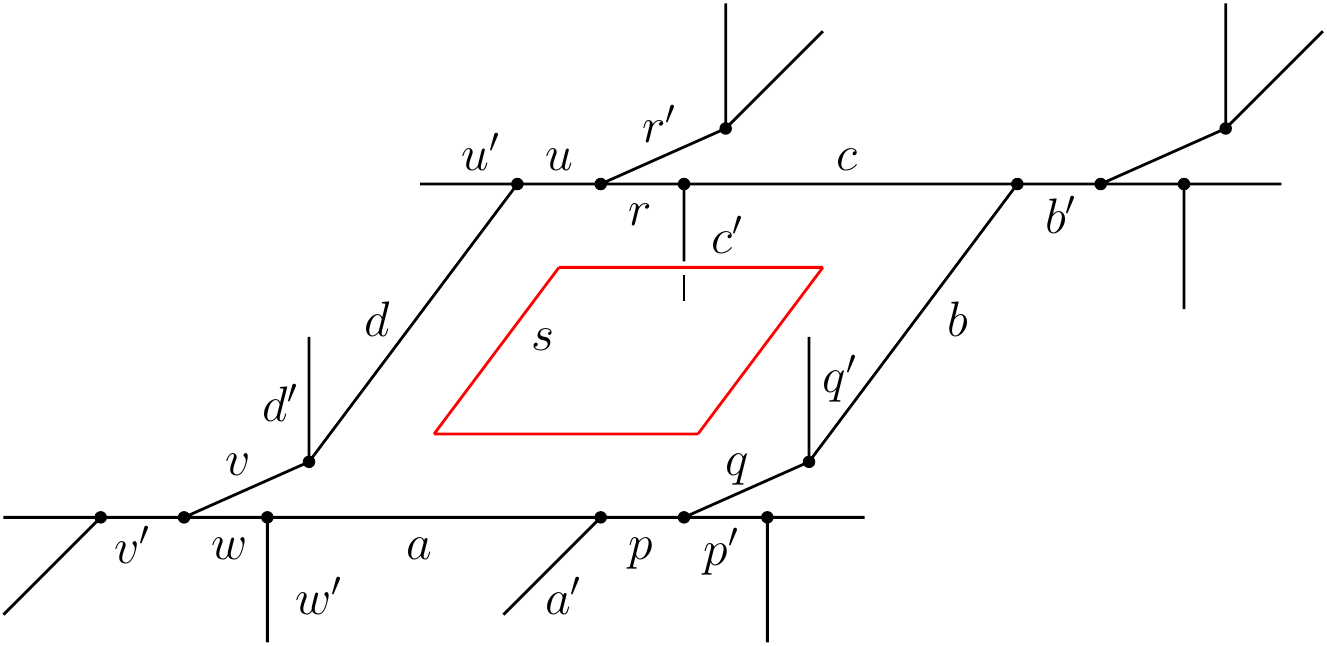}\caption{Plaquette term in the Walker-Wang Hamiltonian.}
\label{plaquette}
\end{figure}

The Walker-Wang models are a class of exactly solvable models for 3D topological orders. The basic intuition behind the Walker-Wang construction is simple. Given a 2D anyon theory, the model is constructed such that the ground state wave function is a superposition of ``3D string-nets'' labeled by the anyon types, which describe the 2+1D space time trajectories of the anyons (Fig. \ref{wf}). The coefficient in front of each configuration is equal to the topological amplitude of the corresponding anyon process. It can be evaluated by using the graphical rules depicting the algebraic data of the anyon theory, captured essentially by the $F$ and $R$ symbols defined in Fig. \ref{fr}, which specify the fusion and braiding rules of the anyons, respectively. The bulk-boundary correspondence described above is similar in spirit to the correspondence between quantum Hall wave functions and the edge conformal field theories. There the systems are in one dimension lower, and the bulk wave function is expressed as a correlator in the boundary CFT.

Mathematically, the input anyon theory of a Walker-Wang model is described by a braided fusion category $\mathcal{A}$. If $\mathcal{A}$ is modular, which means that the only quasiparticle that braids trivially with itself and all other quasiparticles in $\mathcal{A}$ is the vacuum, the output theory would have a trivial bulk and a surface with topological order described by $\mathcal{A}$, and the model belongs to the first type of Walker-Wang models we introduced in the previous section. On the other hand,  if $\mathcal{A}$ is non-modular, the output theory would have a nontrivial bulk, and the model belongs to the second type of Walker-Wang models. The surface theory in this case is more complicated because it contains not only the quasiparticles in $\mathcal{A}$, but also the endpoints of bulk loop excitations that are cut open by the system boundary.

To be more concrete, let us illustrate with two examples. First, we consider the simplest nontrivial input $\mathcal{A}$ possible, which consists of only the vacuum $I$ and a boson $e$. $\mathcal{A}$ is non-modular because $e$ is distinct from the vacuum but braids trivially with everything in $\mathcal{A}$. A Walker-Wang model with such input describes the 3D $\mathbb{Z}_2$ gauge theory with $e$ being the $\mathbb{Z}_2$ gauge charge, which is deconfined in the bulk and on the boundary.\cite{Keyserlingk2012} Next, we modify $\mathcal{A}$ a bit by replacing the boson $e$ with a semion $s$. $\mathcal{A}$ becomes modular in this case, because $s$ braids nontrivially with itself. A Walker-Wang model with the modified input has a trivial bulk and a deconfined semion excitation $s$ on the boundary.\cite{Keyserlingk2012}

In general, deconfined bulk quasiparticle excitations of a Walker-Wang model correspond to quasiparticles in the symmetric center $\mathcal{Z}(\mathcal{A})$ of the input braided fusion category $\mathcal{A}$. A quasiparticle belongs to $\mathcal{Z}(\mathcal{A})$ if it has trivial braiding with itself and all other quasiparticles in $\mathcal{A}$. If $\mathcal{A}$ is modular, $\mathcal{Z}(\mathcal{A})$ is trivial, which is consistent with the fact that a Walker-Wang model with modular input has a trivial bulk. If $\mathcal{A}$ is non-modular, it is known that there are two possibilities for $\mathcal{Z}(\mathcal{A})$\cite{Doplicher1989}: (1) $\mathcal{Z}(\mathcal{A})$ contains only bosons. In this case, it can be identified with the set of irreducible representations of some finite group $G$; (2) $\mathcal{Z}(\mathcal{A})$ contains at least one fermion. In this case, it can also be identified with the set of irreducible representations of some finite group $G$, but each representation comes with a parity, and the set is split into even and odd sectors, corresponding to the bosons and fermions in $\mathcal{Z}(\mathcal{A})$, respectively. Thus, the deconfined bulk quasiparticle excitations of a Walker-Wang model with non-modular input correspond to the irreducible representations of some finite group, and it is plausible that the bulk topological order of the model is a gauge theory of the corresponding group.

Before delving into the exploration of the above possibility, let us review some details of Walker-Wang models. We closely follow Ref. \onlinecite{Walker2011} and refer the reader there for further details. Walker-Wang models are defined on a fixed planar projection of a trivalent resolution of the cublic lattice as shown in Fig. \ref{lattice}. The Hilbert space of a model defined on the lattice is spanned by all labelings of the edges by the input anyon types. The Hamiltonian is of the form
\begin{equation}
H = -\sum_v A_v - \sum_p B_p,
\end{equation}
where $A_v$ is a vertex term which enforces the fusion rules at $v$ by giving an energy penalty to string configurations that violate the fusion rules at $v$, and $B_p$ is a plaquette term of the form $B_p = \sum_s d_s B_p^s$, where the summation is over all the input anyon types $s$, weighted by the quantum dimension of $s$. Each $B_p^s$ acts on the anyon labels of the edges around plaquette $p$, in a way determined by the anyon labels of the edges adjoining $p$. More explicitly, the matrix element of $B_p^s$ sandwiched between states with plaquette edges $(a^{\prime\prime},b^{\prime\prime},c^{\prime\prime},d^{\prime\prime},p^{\prime\prime},q^{\prime\prime},r^{\prime\prime},u^{\prime\prime},v^{\prime\prime},w^{\prime\prime})$ and $(a,b,c,d,p,q,r,u,v,w)$ is given by
\begin{align}
&(B_p^s)^{(a^{\prime\prime},b^{\prime\prime},c^{\prime\prime},d^{\prime\prime},p^{\prime\prime},q^{\prime\prime},r^{\prime\prime},u^{\prime\prime},v^{\prime\prime},w^{\prime\prime})}_{(a,b,c,d,p,q,r,u,v,w)}=
R^{bq^{\prime}}_{q}(R^{rc^{\prime}}_{c})^*(R^{b^{\prime\prime}q^{\prime}}_{q^{\prime\prime}})^* \times \nonumber \\
&R^{r^{\prime\prime}c^{\prime}}_{c^{\prime\prime}}F^{sa^{\prime\prime}p^{\prime\prime}}_{a^{\prime}pa}F^{sp^{\prime\prime}q^{\prime\prime}}_{p^{\prime}qp}
F^{sq^{\prime\prime}b^{\prime\prime}}_{q^{\prime}bq}F^{sb^{\prime\prime}c^{\prime\prime}}_{b^{\prime}cb}
F^{sc^{\prime\prime}r^{\prime\prime}}_{c^{\prime}rc}F^{sr^{\prime\prime}u^{\prime\prime}}_{r^{\prime}us}
F^{su^{\prime\prime}d^{\prime\prime}}_{u^{\prime}du}\times \nonumber \\
&F^{sd^{\prime\prime}v^{\prime\prime}}_{d^{\prime}vd}F^{sv^{\prime\prime}w^{\prime\prime}}_{v^{\prime}wv}F^{sw^{\prime\prime}a^{\prime\prime}}_{w^{\prime}aw},
\end{align}
The above expression looks rather complicated, but there is a simple graphical way of understanding the action of $B_p^s$. Namely, $B_p^s$ temporarily displaces certain links ($c^{\prime}$ and $q^{\prime}$ in Fig. \ref{plaquette}) and fuses a loop with anyon label $s$ to the skeleton of $p$. One can check that all terms in the Hamiltonian commute, and the model is exactly solvable.

To be able to discuss point and loop excitations in Walker-Wang models, we also need to define string operators and membrane operators in these models. The string operators have a graphical definition analogous to that of the plaquette operators. Namely, to create a pair of quasiparticle excitations $\alpha \in \mathcal{A}$ at two points, we just need to lay an $\alpha$-string connecting the two points, and then fuse it to the edges of the lattice. Furthermore, one can show that the string operator commutes (resp. fails to commute) with the plaquette operators threaded by the string if $\alpha \in \mathcal{Z}(\mathcal{A})$ (resp. $\alpha \notin \mathcal{Z}(\mathcal{A})$), and the corresponding quasiparticles are deconfined (resp. confined) in the bulk. On the other hand, all quasiparticles in $\mathcal{A}$ are deconfined on the boundary, because string operators restricted to the boundary do not thread any plaquettes and hence there is no energy penalty associated with them. Unlike the string operators, in general, we do not know how to implement membrane operators in Walker-Wang models, but as we will show below, we can deduce the statistics of the loop excitations without explicitly writing down the membrane operators.

\section{$\mathbb{Z}_2 \times \mathbb{Z}_2$ gauge theories in the Walker-Wang models}
\label{result}

In this section, we discuss how the 3D $\mathbb{Z}_2 \times \mathbb{Z}_2$ gauge theories can be described by Walker-Wang models. In particular, we ask the question: How do we find the input data of the Walker-Wang models that will generate the twisted gauge theories? Our insight into solving this problem comes from the study of 3D $\mathbb{Z}_2 \times \mathbb{Z}_2$ symmetry protected topological (SPT) phases, which are a class of gapped short-range entangled phases of matter protected by a global symmetry. A nontrivial SPT phase has the interesting property that its surface state is anomalous,\cite{Kapustin2014Mar,Kapustin2014Apr,Chen2014,Cho2014} meaning that it cannot exist on its own and must be realized as the boundary of some system in one dimension higher. This implies that a gapped symmetric surface of a nontrivial SPT phase must have nontrivial topological order, and that the symmetry must fractionalize on the anyons in an anomalous way. Specifically to the $\mathbb{Z}_2 \times \mathbb{Z}_2$ SPTs, we will first review a particular kind of gapped symmetric surface states of these SPTs, called the projected semion states. We will introduce its anyon content, and the symmetry fractionalization pattern of $\mathbb{Z}_2 \times \mathbb{Z}_2$ on the anyons. Next, we couple the systems to a $\mathbb{Z}_2 \times \mathbb{Z}_2$ gauge field, and study the surface theories of the gauged systems. Finally, it is known that upon gauging, a trivial (resp. nontrivial) SPT becomes an untwisted (resp. twisted) gauge theory,\cite{Levin2012} which leads us to propose a Walker-Wang model description of the $\mathbb{Z}_2 \times \mathbb{Z}_2$ gauge theories based on the surface anyon content of the gauged $\mathbb{Z}_2 \times \mathbb{Z}_2$ SPTs.

\subsection{Projective semion states and 3D $\mathbb{Z}_2 \times \mathbb{Z}_2$ SPTs}
The projective semion states are 2D symmetry fractionalization patterns with a semion and a $\mathbb{Z}_2 \times \mathbb{Z}_2$ symmetry, first introduced and analyzed in Ref. \onlinecite{Chen2014} and Ref. \onlinecite{Kapustin2014Apr}. They may be considered as variants of the Kalmeyer-Laughlin chiral spin liquid (CSL)\cite{Kalmeyer1987}.

We first give a brief review of the Kalmeyer-Laughlin CSL. The topological order of the theory is the same as that of the $\nu=1/2$ bosonic fractional quantum Hall state. The only nontrivial quasiparticle is a semion, which has topological spin $i$ and fuses into the vacuum with another semion. Moreover, the semion carries a spin-1/2, transforming projectively under the $SO(3)$ symmetry. The CSL can thus be understood as a symmetry fractionalization pattern of $SO(3)$ on a semion. The theory is non-anomalous, because it can be realized in a purely 2D system with the explicit construction in Ref. \onlinecite{Kalmeyer1987}.

To describe the projective semion states, we reduce the $SO(3)$ symmetry to a $\mathbb{Z}_2 \times \mathbb{Z}_2$ subgroup, consisting of rotations along the $x$, $y$, and $z$ axes by $180$ degrees, which we denote by $g_x$, $g_y$, and $g_z$, respectively. By restricting the spin-1/2 representation of $SO(3)$ to this reduced symmetry group, we obtain a projective representation of $\mathbb{Z}_2 \times \mathbb{Z}_2$:
\begin{equation}
\text{CSL:} \ \ \ g_x = i\sigma_x, \ \ \ g_y = i\sigma_y, \ \ \ g_z = i\sigma_z.
\label{CSL}
\end{equation}
The CSL is therefore a symmetry fractionalization pattern of $\mathbb{Z}_2 \times \mathbb{Z}_2$ on a semion where the semion carries a half charge under all nontrivial group elements, because acting a nontrivial group element twice on a spin-1/2 is equivalent to rotating the spin-1/2 by 360 degrees along the corresponding axis, which results in a phase factor of $-1$.

However, the CSL is not the only possible symmetry fractionalization pattern of $\mathbb{Z}_2 \times \mathbb{Z}_2$ on a semion. The semion can also transform under other projective representations of $\mathbb{Z}_2 \times \mathbb{Z}_2$. More specifically, the semion can carry either integral or half-integral charges under the nontrivial elements of $\mathbb{Z}_2 \times \mathbb{Z}_2$, and we have 3 variants of the CSL, called the ``anomalous projective semion'' (APS) states, where the symmetry action on the semion can be represented as 
\begin{align}
&\text{APS-X}: \ \ \ g_x = i\sigma_x, \ \ \ g_y = \sigma_y, \ \ \ g_z = \sigma_z, \nonumber \\
&\text{APS-Y}:\negphantom{\text{APS-Y}}\phantom{\text{APS-X}}\ \ \ g_x = \sigma_x, \phantom{i}\ \ \ g_y = i\sigma_y, \negphantom{i}\ \ \ g_z = \sigma_z, \nonumber \\
&\text{APS-Z}:\negphantom{\text{APS-Z}}\phantom{\text{APS-X}}\ \ \ g_x = \sigma_x, \phantom{i}\ \ \ g_y = \sigma_y, \ \ \ g_z = i\sigma_z.
\label{APS}
\end{align}

If we take $g_x$ and $g_y$ to be the two generators of $\mathbb{Z}_2 \times \mathbb{Z}_2$, the APS-X, APS-Y, and APS-Z theories correspond to the cases where the semion carries a half charge under either the first, second, or both generators, respectively. It was argued in Ref. \onlinecite{Chen2014} and Ref. \onlinecite{Kapustin2014Apr} that the addition of such half charges to the CSL, though compatible with the fusion rules of the semion, leads to anomalies in the theory. This can be seen via the violation of the pentagon equations for the symmetry defects\cite{Chen2014} or the failure in promoting the global symmetry to a gauge symmetry \cite{Kapustin2014Apr} in the effective field theory (dubbed the 't Hooft anomaly\cite{tHooft1980}). The anomalous projective semion theories are therefore not realizable in purely 2D systems. Nevertheless, they can be realized on the boundary of some nontrivial 3D $\mathbb{Z}_2 \times \mathbb{Z}_2$ SPT phases. Exactly solvable models for such 3D SPT phases based on the ``decorated'' Walker-Wang models were constructed in Ref. \onlinecite{Chen2014}. More specifically, the semion Walker-Wang model studied in Ref. \onlinecite{Keyserlingk2012} is decorated with unitary linear representations of $\mathbb{Z}_2 \times \mathbb{Z}_2$, such that the ground state wave function is a loop gas of semion lines dressed with $\mathbb{Z}_2 \times \mathbb{Z}_2$ Haldane chains. The endpoints of open semion lines, which are deconfined semion excitations on the boundary, carry projective representations of $\mathbb{Z}_2\times \mathbb{Z}_2$ as in Eq. \eqref{APS}. Therefore, the boundary of the 3D SPT phases are precisely the anomalous projective semion states. Similarly, one can construct a trivial 3D SPT phase which realizes the non-anomalous projective semion state on its boundary. With a slight modification of the $\mathbb{Z}_2 \times \mathbb{Z}_2$ Haldane chains, the boundary semion excitations can be made to transform under $\mathbb{Z}_2 \times \mathbb{Z}_2$ as in Eq. \eqref{CSL}, as desired for a CSL. We will not delve into the details of the construction. The interested reader may refer to Ref. \onlinecite{Chen2014} for more information.

\subsection{Gauging the $\mathbb{Z}_2 \times \mathbb{Z}_2$ symmetry}
\label{sec:gauging}

\begin{table}
\renewcommand{\arraystretch}{1.2}{
\caption{Berry phases associated with the bulk three-loop braiding processes in the 3D $\mathbb{Z}_2 \times \mathbb{Z}_2$ gauge theories. For simplicity, we use CSL, APS-X, APS-Y, and APS-Z to label the $\mathbb{Z}_2 \times \mathbb{Z}_2$ gauge theories obtained by gauging the corresponding SPT models.}
\begin{tabular}{cccc}
\hline\hline
$\ $& $\ \ $ $\theta_{x,y}$ $ \ \ $ & $\ \ $ $\theta_{y,x}$ $ \ \ 
$\\
\hline
CSL $\ $  & $0$ & $0$\\
APS-X $\ $ & $0$ & $\pi/2$ \\
APS-Y $\ $ & $\pi/2$ & $0$  \\
APS-Z $\ $ & $\pi/2$ & $\pi/2$ \\
\hline
\end{tabular}
\label{tab1}
}
\end{table}

Now, suppose we couple the models discussed above to a $\mathbb{Z}_2 \times \mathbb{Z}_2$ gauge field. We obtain a 3D untwisted (resp. twisted) $\mathbb{Z}_2 \times \mathbb{Z}_2$ gauge theory if the system is in a trivial (resp. nontrivial) SPT phase. These $\mathbb{Z}_2 \times \mathbb{Z}_2$ gauge theories can be distinguished by the following three-loop braiding processes in the bulk: (1) Two $g_x$-flux loops exchanged while both linked with a $g_y$-flux loop; (2) Two $g_y$-flux loops exchanged while both linked with a $g_x$-flux loop. We denote the associated Berry phases by $\theta_{x,y}$ and $\theta_{y,x}$, respectively. The numerical values of $\theta_{x,y}$ and $\theta_{y,x}$ for the various 3D $\mathbb{Z}_2 \times \mathbb{Z}_2$ gauge theories are listed in Table \ref{tab1}.

\begin{figure}
\centering
\includegraphics[width=0.50\linewidth]{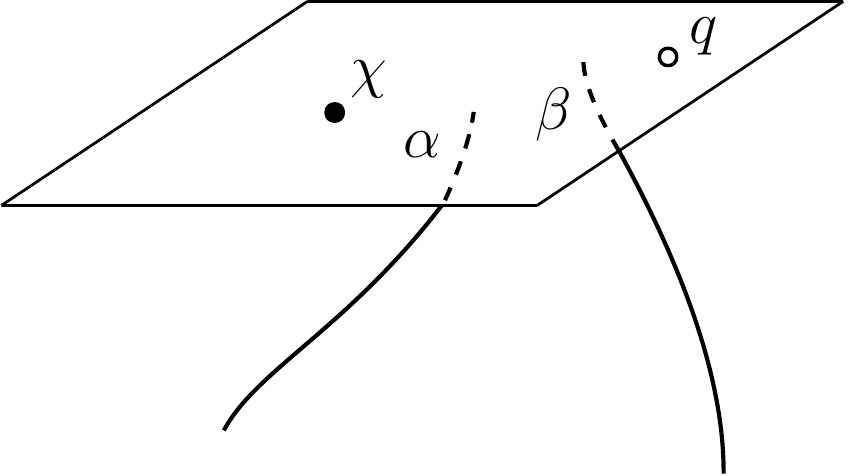}\caption{Sketch of surface excitations in a gauged SPT model. The bulk is below the plane. $q$ represents a gauge charge that can appear in the bulk and on the surface. $\chi$ represents an anyon that can appear only on the surface. $\alpha$ and $\beta$ represent the flux lines in the bulk that end on the surface.}
\label{anyon}
\end{figure}

It is shown in Ref. \onlinecite{WangLevin2015} that the gauged systems host three types of excitations on or near the surface (Fig. \ref{anyon}): (1) gauge charges that can appear in the bulk and on the boundary; (2) flux loops in the bulk that become open flux lines when ending on the boundary; (3) anyons that are pinned to the boundary. It is further argued in the same reference that each surface anyon $\tilde{\chi}$ in the gauged model is naturally associated with a surface anyon $\chi$ in the ungauged model. $\chi$ is referred to as the ``anyonic flux'' carried by $\tilde{\chi}$. In our case, the only nontrivial surface anyon in the ungauged model is a semion $s$. Correspondingly, there is one and only one nontrivial surface anyon in the gauged model, which is $\tilde{s}$. The set of excitations in the gauged model therefore consists of $\tilde{s}$ and the gauge charges and flux loops of $\mathbb{Z}_2 \times \mathbb{Z}_2$.

Having understood the excitations in the 3D $\mathbb{Z}_2 \times \mathbb{Z}_2$ gauge theories, let us try to incorporate them into the Walker-Wang construction of the 3D $\mathbb{Z}_2 \times \mathbb{Z}_2$ gauge theories. Since the deconfined quasiparticle excitations on the surface include the $\mathbb{Z}_2 \times \mathbb{Z}_2$ gauge charges and the anyon with anyonic flux $s$, based on the physical picture that the bulk wave function of a Walker-Wang model is the space time trajectories of the quasiparticles on the surface, we expect that if we use the $\mathbb{Z}_2 \times \mathbb{Z}_2$ gauge charges and the surface anyon $\tilde{s}$ to write a Walker-Wang model, we should get a 3D $\mathbb{Z}_2 \times \mathbb{Z}_2$ gauge theory. In the next subsection, we will make this idea more concrete by explicitly constructing the input data for the Walker-Wang models that describe the $\mathbb{Z}_2 \times \mathbb{Z}_2$ gauge theories.

\subsection{Input data of the Walker-Wang models}
\label{sec:input}

\begin{figure*}
\includegraphics[width=0.70\textwidth]{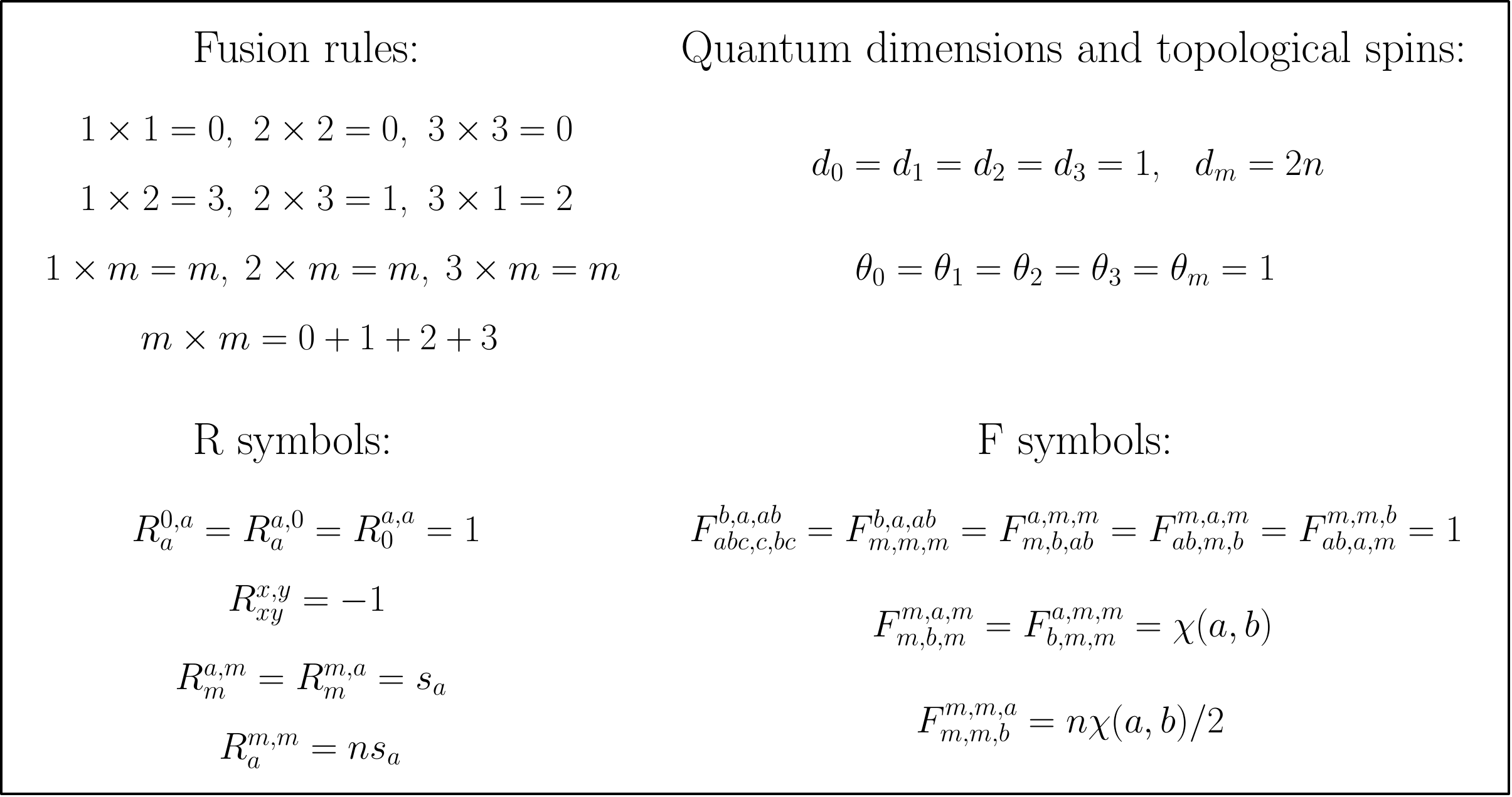}\caption{\label{fig:collected_data} Data for Rep($Q_8$) and Rep($D_4$). The (simple) objects are the irreducible representations (charges) of either $Q_8$ or $D_4$, defined in Eq. \eqref{eqn:1d} and Eq. \eqref{eqn:2d}. They are all self-dual. Here $a,b=0,1,2,3$, and $x,y = 1,2,3$, $x \neq y$. $\chi$ is defined by $\chi(0,a)=\chi(a,0)=1$, $\chi(x,x)=1$, and $\chi(x,y)=-1$. The quantum dimension of a charge is defined to be the product of its Frobenius-Schur indicator and dimension. The $F$ symbols with $n=-1$ (resp. $n=1$) are the $6j$ symbols of $Q_8$ (resp. $D_4$). The $R$ symbols are solutions to the hexagon equations (given the $F$ symbols) with the constraint that all charges are bosonic and have trivial mutual statistics. For $Q_8$, there exists a unique solution, $n=-1$, $s_0=1$, $s_1 = s_2 = s_3 = -1$. For $D_4$, there exists $3$ solutions, $n=1$, $s_0=1$, $s_1=-1$, $s_2=1$, $s_3=1$, and also the ones resulting from the permutations $1 \leftrightarrow 2$ and $1 \leftrightarrow 3$, respectively.}
\label{data}
\end{figure*}


We start by introducing some notations. We denote by $Q_8$ the quaternion group, and $D_4$ the dihedral group of order $8$, defined by the presentations
\begin{align}
&Q_8 = \langle x,y \vert x^2 = y^2 = (xy)^2, x^4=1 \rangle, \nonumber \\
&D_4 = \langle x,y \vert x^4 = y^2 = (xy)^2=1 \rangle.
\end{align}
We denote by $G$ either $Q_8$ or $D_4$. The irreducible representations (charges) of $G$ consist of four 1-dimensional charges, given by
\begin{align}
&\rho_0(x) = 1,\phantom{-} \ \ \ \rho_0(y) = 1, \nonumber \\
&\rho_1(x) = 1,\phantom{-}\ \ \ \rho_1(y) = -1, \nonumber \\
&\rho_2(x) = -1,\ \ \ \rho_2(y) = 1, \nonumber \\
&\rho_3(x) = -1,\ \ \ \rho_3(y) = -1,
\label{eqn:1d}
\end{align}
and one 2-dimensional charge, given by
\begin{align}
&m(x) = i\sigma_z,\ \ \ m(y) = i\sigma_y, \ \ \ \text{for}\ G = Q_8, \nonumber \\
&m(x) = i\sigma_z,\ \ \ m(y) = \sigma_x, \phantom{i}\ \ \ \text{for}\ G = D_4.
\label{eqn:2d}
\end{align}
For simplicity, we denote the 1-dimensional charges $\rho_a$ of $G$ by $a$ ($a=0,1,2,3$), which form a $\mathbb{Z}_2 \times \mathbb{Z}_2$ group under the tensor product of representations. The  charges of $G$ form a braided fusion category Rep($G$) with the fusion and braiding data presented in Fig. \ref{data}. It is known that a Walker-Wang model with input data Rep($G$) describes a 3D untwisted $G$-gauge theory.\cite{LevinWen2004}

Next, we construct the input data for the Walker-Wang models that describe the $\mathbb{Z}_2 \times \mathbb{Z}_2$ gauge theories. It is useful to first study the fusion rules satisfied by the quasiparticles on the surface. From representation theory, we know that the tensor product of the projective representation of $\mathbb{Z}_2 \times \mathbb{Z}_2$ carried by the semion $s$ (Eq. \eqref{CSL} or Eq. \eqref{APS}) with itself gives a reducible linear representation of $\mathbb{Z}_2 \times \mathbb{Z}_2$, which can be further decomposed into a direct sum of the four 1-dimensional representations of $\mathbb{Z}_2 \times \mathbb{Z}_2$. After the gauging procedure, the symmetry charges are promoted to gauge charges, which are deconfined quasiparticle excitations, and the fusion rule of representations becomes the fusion rule of quasiparticles:
\begin{equation}
\tilde{s} \times \tilde{s} = 0 + 1 + 2 + 3,
\label{eqn:fusion}
\end{equation}
where we have identified the charges of $\mathbb{Z}_2 \times \mathbb{Z}_2$ with the 1-dimensional charges of $G$ on the right hand side. Eq. \eqref{eqn:fusion} is identical to the fusion between two 2-dimensional charges $m$ of $G$ in Fig. \ref{data}, provided that we further identify $\tilde{s}$ with $m$. However, the topological spin of $\tilde{s}$ (resp. $m$) is $i$ (resp. $1$), so $m$ needs to be ``twisted'' by a semion before we can make the identification. The precise meaning of this is that we multiply all the $F$ symbols $F^{mma}_{mmb}$ ($a,b = 0,1,2,3$) in Fig. \ref{data} by $-1$, and all the $R$ symbols $R^{mm}_{a}$ ($a = 0,1,2,3$) in Fig. \ref{data} by $i$. One can check that Rep($G$) remains a consistent braided fusion category after the modifications, i.e., the pentagon equations and hexagon equations are satisfied. For convenience, we will denote the modified category by $\text{Rep}_s(G)$.\footnote{Fusion categories with fusion rules that of $\text{Rep}(G)$ and $\text{Rep}_s(G)$ are actually examples of the Tambara-Yamagami categories\cite{Tambara1998} based on the group $\mathbb{Z}_2 \times \mathbb{Z}_2$. Not all Tambara-Yamagami categories admit consistent braiding as $\text{Rep}(G)$ and $\text{Rep}_s(G)$ do.} Furthermore, comparison between Eq. \eqref{eqn:2d} and Eq. \eqref{CSL} (resp. Eq. \eqref{APS}) shows that we should take $G$ to be $Q_8$ (resp. $D_4$) if $s$ carries the projective representation in Eq. \eqref{CSL} (resp. Eq. \eqref{APS}).\footnote{It is a mathematical fact that given a projective representation of a group $G$, one can lift it to a linear representation of a different group $C$, which is a central extension of $G$. In our cases, one can actually show that the projective representation in Eq. \eqref{CSL} (resp. Eq. \eqref{APS}) can be lifted to a linear representation of $Q_8$ (resp. $D_4$), which is a central extension of $\mathbb{Z}_2 \times \mathbb{Z}_2$ by $\mathbb{Z}_2$.} Therefore, after the identification above, the fusion and braiding information of the quasiparticles on the surface of a 3D untwisted (resp. twisted) $\mathbb{Z}_2 \times \mathbb{Z}_2$ gauge theory are neatly captured by the braided fusion category $\text{Rep}_s(Q_8)$ (resp. $\text{Rep}_s(D_4)$), which leads us to the proposal that a Walker-Wang model with input  $\text{Rep}_s(Q_8)$ (resp. $\text{Rep}_s(D_4)$) describes a 3D untwisted (resp. twisted) $\mathbb{Z}_2 \times \mathbb{Z}_2$ gauge theory.

Physically, the semion-twisted 2-dimensional charge of $G$ is an anyon, which is confined in the bulk and deconfined on the boundary because it braids nontrivially with itself. The 1-dimensional charges of $G$ remain deconfined in the bulk and on the boundary. The set of quasiparticle excitations in the Walker-Wang models therefore agrees with that in the $\mathbb{Z}_2 \times \mathbb{Z}_2$ gauge theories described in Section \ref{sec:gauging}, provided that our identification between the twisted 2-dimensional charge of $G$ and the surface anyon $\tilde{s}$ is correct. The identification between the 1-dimensional charges of $G$ and the $\mathbb{Z}_2 \times \mathbb{Z}_2$ gauge charges are natural because their fusion and braiding data are identical. In the next section, we will give a more direct verification that the Walker-Wang models we proposed indeed describe the $\mathbb{Z}_2 \times \mathbb{Z}_2$ gauge theories. More specifically, we will compute the three-loop braiding statistics in our models and check that they agree with those listed in Table \ref{tab1}.

\section{Detecting the topological order in the Walker-Wang models}
\label{method}

In this section, we verify that the Walker-Wang models we proposed in the previous section describe the 3D $\mathbb{Z}_2 \times \mathbb{Z}_2$ gauge theories. Our approach is to do 3D modular transformations to the ground space of our Walker-Wang models on a three-torus and calculate the resulting nonabelian Berry phases. Similar methods have been used to determine the topological order in chiral spin liquid,\cite{Zhang2011} 2D topological orders represented by tensor networks,\cite{Cincio2013,He2014}2D string-net models,\cite{Liu2013} and untwisted or twisted quantum double models in 2D\cite{Hu2013} and 3D.\cite{Moradi2015,Jiang2014,WangWen2015} Furthermore, by making a dimensional reduction argument, we are able to deduce the three-loop braiding statistics of our models. We can compare them with the data listed in Table \ref{tab1} to determine which $\mathbb{Z}_2 \times \mathbb{Z}_2$ gauge theory a particular model is describing.

\subsection{$S$ and $T$ matrices from 3D modular transformations}
\label{snt}

\begin{figure}
\centering
\includegraphics[width=0.75\linewidth]{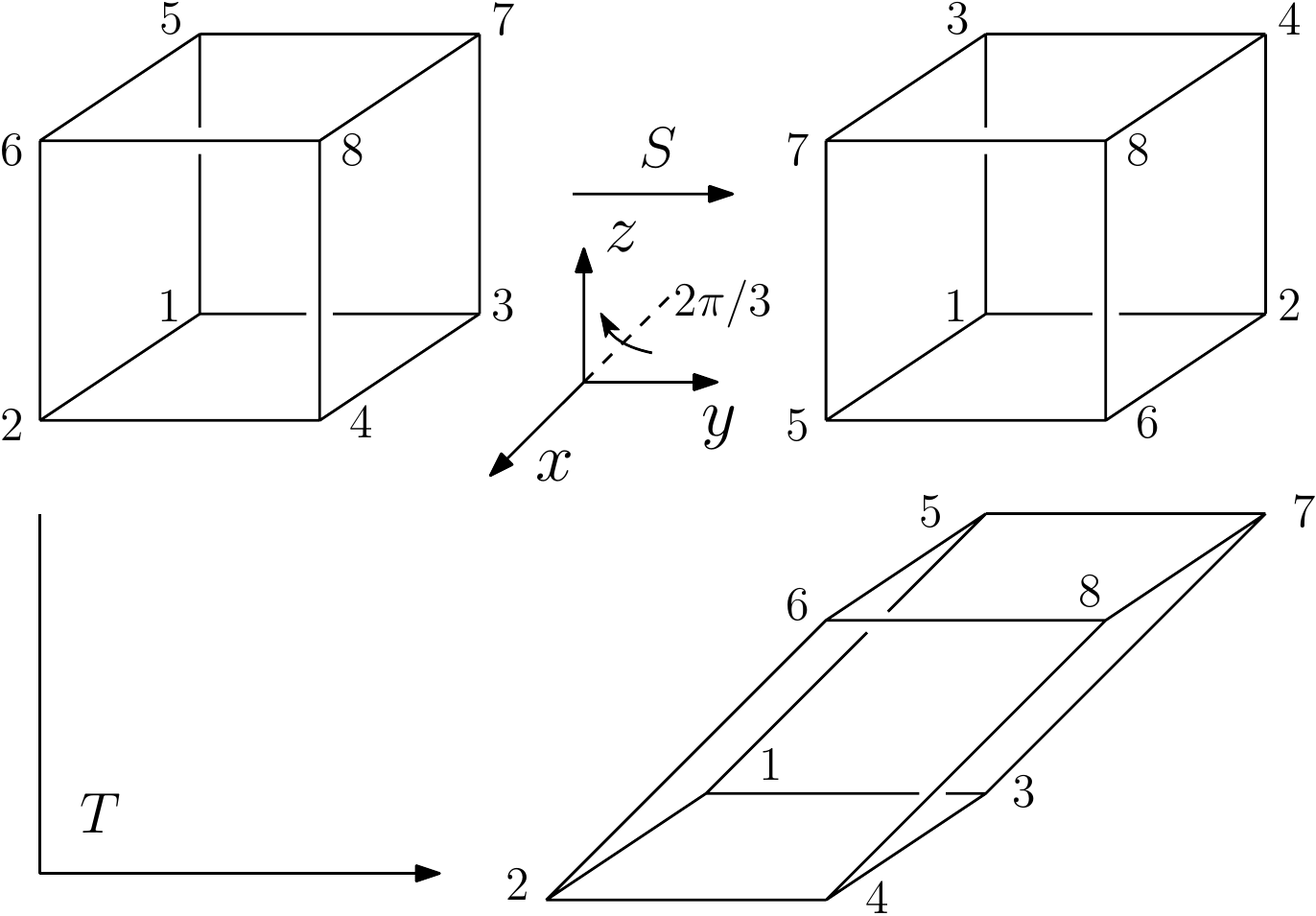}\caption{\label{fig:cube}$S$ and $T$ transformations on a three-torus.}
\end{figure}

\begin{figure}
\centering
\includegraphics[width=0.75\linewidth]{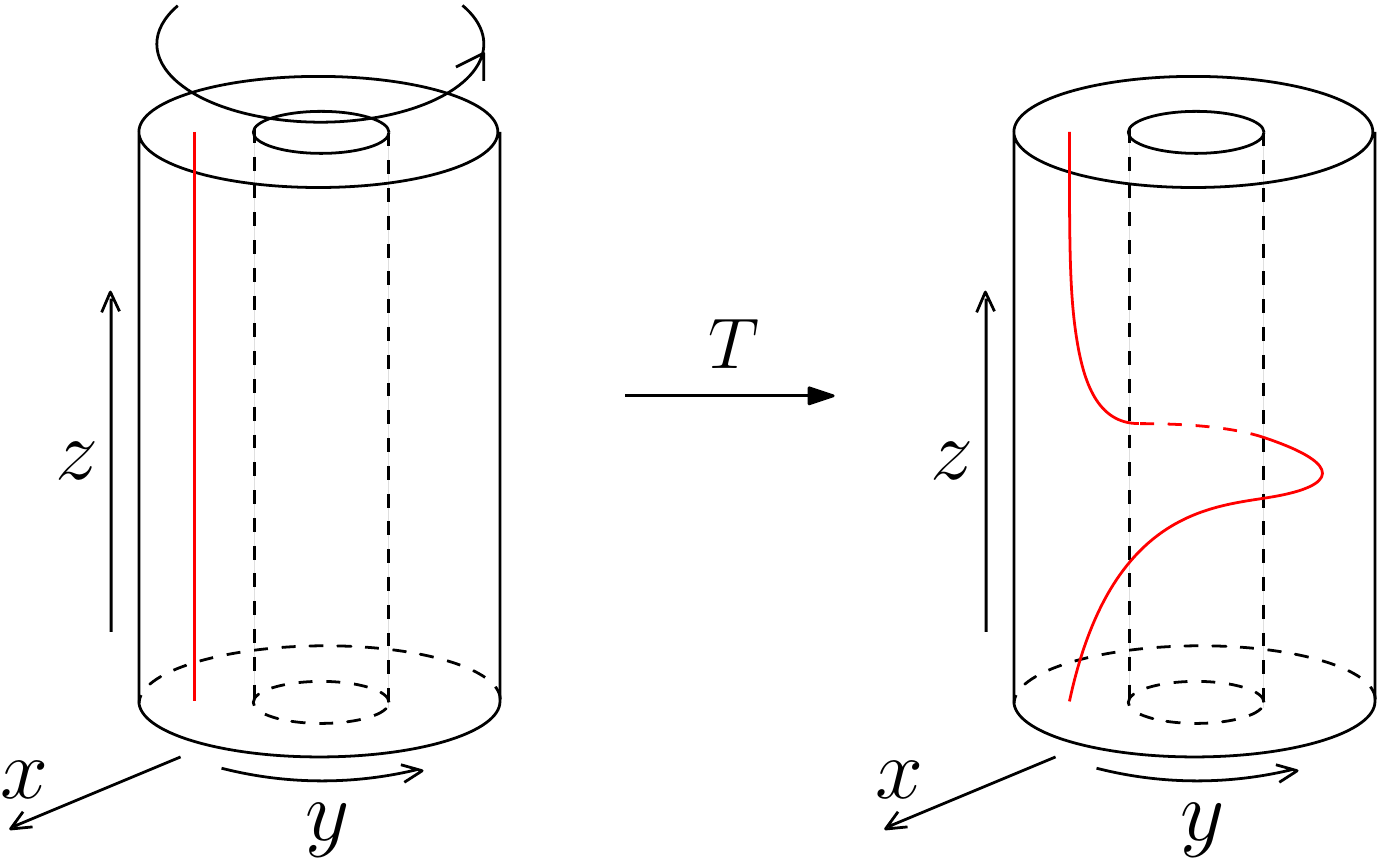}\caption{\label{fig:dehn}$T$ transformation as the Dehn twist of a hollow cylinder.}
\end{figure}

\begin{figure}
\centering
\includegraphics[width=0.4\linewidth]{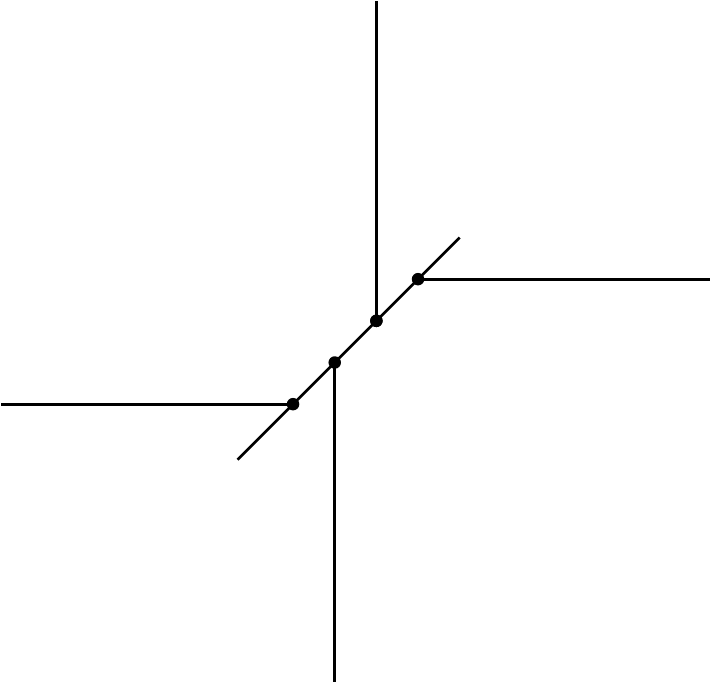}\caption{\label{fig:lat}A ``minimal'' trivalent lattice on the three-torus, which consists of 4 vertices, 6 edges, and 3 plaquettes.}
\end{figure}

The 3D modular transformations are elements of the mapping class group of the three-torus $\text{MCG}(\mathbb{T}^3) = SL(3,\mathbb{Z})$. The group has two generators, $S$ and $T$, which are of the form
\begin{equation}\label{eq:SL3ZGenerators}
	S = \begin{pmatrix}
				0 & 0 & 1 \\
				1 & 0 & 0 \\
				0 & 1 & 0
			 \end{pmatrix},
			 \quad \text{and} \quad
	T= \begin{pmatrix}
				1 & 0 & 0 \\
				0 & 1 & 1 \\
				0 & 0 & 1
			 \end{pmatrix}.
\end{equation}
If we represent the three-torus as a cube with opposite faces identified, and if we draw the cube in a right-handed coordinate frame as in Fig. \ref{fig:cube}, then $S$ is a clockwise rotation of the cube by $120 \degree$ along a diagonal, and $T$ is a shear transformation in the $yz$-plane. There is another way to visualize the $T$ transformation. By closing the periodic direction of the cube along the $z$ axis, we can equivalently think of the three-torus as a hollow cylinder with the top and bottom faces identified, and inner and outer faces identified (Fig. \ref{fig:dehn}). The $T$ transformation then becomes the Dehn twist of the hollow cylinder along the $yz$-plane.

Note that the 2D modular transformations in the $yz$-plane, generated by 
\begin{equation}\label{eq:SL2ZGenerators}
	S_{yz} = \begin{pmatrix}
				1 & 0 & 0 \\
				0 & 0 & -1 \\
				0 & 1 & 0
			 \end{pmatrix},
			 \quad \text{and} \quad
	T_{yz}= \begin{pmatrix}
				1 & 0 & 0 \\
				0 & 1 & 1 \\
				0 & 0 & 1
			 \end{pmatrix},
\end{equation}
form an $SL(2,\mathbb{Z})$ subgroup of $SL(3,\mathbb{Z})$, so they can be written as a combination of $S$, $T$, and their inverses. In particular, 
\begin{equation}
S_{yz} = (T^{-1}S)^3(ST)^2ST^{-1}, \quad \text{and} \quad T_{yz} = T.
\label{eqn:styz}
\end{equation}
A presentation of $SL(3,\mathbb{Z})$ is obtained by specifying the relations among the generators\cite{Coxeter1980}:
\begin{align}
&S^3 = S_{yz}^4 = (S_{yz}S)^2 = (T^{-1}SS_{yz}^2S^{-1})^2 = I, \nonumber \\
&S^{-1}TSTS^{-1}T^{-1}ST^{-1} = S_{yz}S^{-1}TSS_{yz}^{-1}, \nonumber \\
&(T^{-1}S_{yz}^{-1})^3=(S_{yz})^2, \quad [S_{yz}TS_{yz}^{-1},STS^{-1}]=I, \nonumber \\
&[S_{yz}TS_{yz}^{-1},S^{-1}TS]=I,
\label{eqn:rels}
\end{align}
where $[A,B] = ABA^{-1}B^{-1}$ denotes the commutator of matrices $A$ and $B$.

Since Walker-Wang models are fixed-point models and are scale-invariant, we can apply the $S$ and $T$ transformations to a model defined on a ``minimal'' trivalent lattice on the three-torus (Fig. \ref{fig:lat}). The Hilbert space of a model defined on the lattice is spanned by all labelings of the edges by the input anyon types that are consistent with the fusion rules. Each such labeling can be denoted by a sextuple $(i,j,k,l,m,n)$, where each entry corresponds to a particular input anyon type. For the modified $Q_8$ and $D_4$ input data discussed in Section \ref{sec:input}, the Hilbert spaces on the lattice are all of dimension 176. The matrix elements of the $S$ and $T$ matrices are derived in Appendix \ref{app_a}, and are given by
\begin{align}
&S^{(k,i,j,\tilde{n},\tilde{m},\tilde{l})}_{(i,j,k,l,m,n)}=F^{jil}_{km\tilde{l}}F^{kmn}_{ji\tilde{n}}F^{j\tilde{l}m}_{j\tilde{n}\tilde{m}}, \nonumber \\
&T^{(i,\tilde{l},k,m,\tilde{n},\tilde{j})}_{(i,j,k,l,m,n)} = F^{k\tilde{j}n}_{km\tilde{n}}F^{k\tilde{l}j}_{in\tilde{j}}(R^{kj}_{\tilde{l}})^*F^{mkl}_{ji\tilde{l}} R^{kl}_m.
\end{align}

Note that there is an additional complication due to the non-abelian fusion rules of the input anyons. By computing the Hamiltonian of our Walker-Wang models on the minimal lattice (see Appendix \ref{app_b}), we find that the ground space of the Hamiltonian is only a 64 dimensional subspace of the 176 dimensional Hilbert space on the lattice, so the ground state degeneracy of our models matches that of the 3D $\mathbb{Z}_2 \times \mathbb{Z}_2$ gauge theories. To restrict the modular transformations to be within the ground space, we need to diagonalize the Hamiltonian, and project the $S$ and $T$ matrices to the ground space of the Hamiltonian, as explained in the appendix. The 64 by 64 $S$ and $T$ matrices thus obtained satisfy the relations in Eq. \eqref{eqn:rels}, so that they form a representation of the $SL(3,\mathbb{Z})$ group. They encode all the braiding statistics of our 3D topological orders, but it takes a bit more work to read them out, which is done in the next subsection.

\subsection{Dimensional reduction and three-loop braiding statistics from the $S$ and $T$ matrices. }
\label{sec:dimred}

\begin{table}
\renewcommand{\arraystretch}{1.2}{
\caption{Summary of dimensional reduction results in the Walker-Wang models. We label the Walker-Wang models by their input data. The first row lists the $\mathbb{Z}_2 \times\mathbb{Z}_2$ gauge fluxes threaded through the ``$x$-hole'' after the $x$-direction is compactified. The resulting 2D $\mathbb{Z}_2 \times\mathbb{Z}_2$ gauge theories on the $yz$-plane are represented by the $\mathbb{Z}_2 \times\mathbb{Z}_2 \times \mathbb{Z}_2$ group elements in the entries.}
\begin{tabular}{cccccc}
\hline\hline
$\ $& $\ \  $ $\text{Trivial}$ $ \ \ $ & $\ \ $ $g_x$ $ \ \ $ & $ \ \ $ $g_y$ $ \ \ $ & $\ \ $ $g_z$ $ \ \ 
$\\
\hline
$\text{Rep}_s$($Q_8$) $\ $  & $1$ & $1$ & $1$ & $1$ \\
$\text{Rep}_s$($D_4$)  $\ $ &$1$ & $\omega_2$ & $\omega_{12}$ & $\omega_2 \omega_{12}$ \\
$\text{Rep}_s$($D_4$) with $1 \leftrightarrow 2$ $\ $ &$1$ & $\omega_{12}$ & $\omega_{1}$ & $\omega_1 \omega_{12}$  \\
$\text{Rep}_s$($D_4$) with $1 \leftrightarrow 3$ $\ $ &$1$ & $\omega_2\omega_{12}$ &   $\omega_{1}\omega_{12}$ & $\omega_1 \omega_2$ \\
\hline
\end{tabular}
\label{tab2}
}
\end{table}

\begin{figure}
\centering
\includegraphics[width=\linewidth]{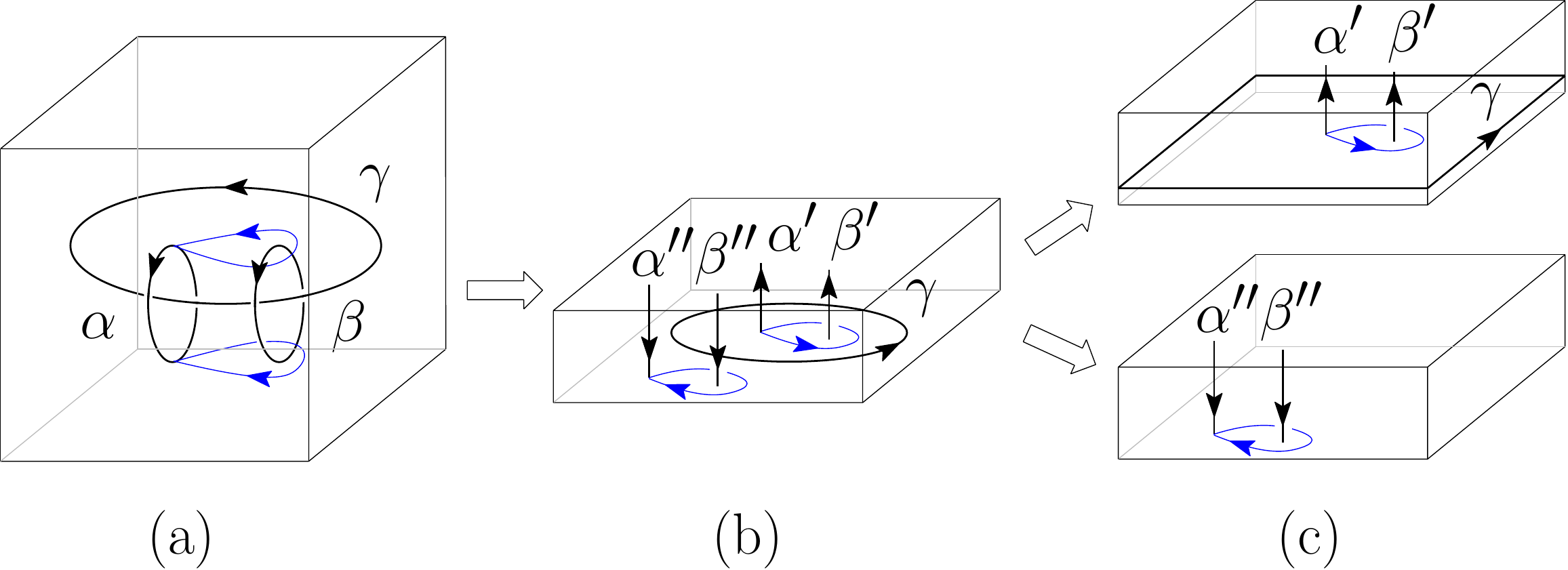}\caption{\label{fig:dimred}Decomposition of a three-loop braiding process into two separate braiding processes in the dimensionally-reduced 2D system.}
\end{figure}

We first review the dimensional reduction phenomenon in 3D discrete gauge theories and its connection to the three-loop braiding processes. For our purposes, it suffices to consider theories with an Abelian gauge group $G$ and Abelian statistics. It was observed in Ref. \onlinecite{Moradi2015} and Ref. \onlinecite{WangWen2015} that the 2D modular matrices $S_{yz}$ and $T_{yz}$ of a 3D $G$-gauge theory $\mathcal{C}^{\text{3D}}$ admit the following direct sum decomposition:
\begin{equation}
S_{yz} = \bigoplus_g S_{yz,g}, \ \ \ T_{yz} = \bigoplus_g T_{yz,g},
\end{equation}
where $g$ runs over all gauge fluxes (group elements) of $G$, and each pair ($S_{yz,g},T_{yz,g}$) describes some particular 2D $G$-gauge theory $\mathcal{C}^{\text{2D}}_{g}$. Furthermore, the basis in which $S_{yz}$ and $T_{yz}$ take the above block diagonal form consists of the simultaneous eigenstates of the charge Wilson loop operators along the $x$-axis.

The above observations can be understood in terms of the dimensional reduction of the 3D $G$-gauge theory $\mathcal{C}^{\text{3D}}$. Physically, it implies that if we put the 3D $G$-gauge theory $\mathcal{C}^{\text{3D}}$ on a three-torus and make one spatial dimension (say the $x$-direction) of the three-torus very small, then $\mathcal{C}^{\text{3D}}$ can be viewed as a direct sum of 2D $G$-gauge theories $\mathcal{C}^{\text{2D}}_{g}$ with degenerate ground state energy:
\begin{equation}
\mathcal{C}^{\text{3D}} = \bigoplus_g \mathcal{C}^{\text{2D}}_{g}.
\end{equation}
The degeneracy is accidental and can be lifted by fixing a $G$-gauge flux $g$ through the hole bound by the $x$-axis (dubbed the ``$x$-hole'' following Ref. \onlinecite{WangLevin2014}). This reduces the 3D $G$-gauge theory $\mathcal{C}^{\text{3D}}$ to the 2D $G$-gauge theory $\mathcal{C}^{\text{2D}}_{g}$. The gauge flux $g$ can be detected by winding the $G$-gauge charges around the ``$x$-hole'' and studying the associated Aharonov-Bohm phases. Therefore, the ground space of each sector $\mathcal{C}^{\text{2D}}_{g}$ is actually the eigenspace of the charge Wilson loop operators along the $x$-axis with a particular set of eigenvalues, which agrees with the observations in the previous paragraph.

To deduce the three-loop braiding statistics in $\mathcal{C}^{\text{3D}}$, we adopt the approach from Ref. \onlinecite{WangLevin2014} to decompose a three-loop braiding process in a 3D system into two separate processes in the dimensionally-reduced 2D systems. More specifically, let us consider the three-loop braiding process depicted in Fig. \ref{loop}, where a flux loop $\alpha$ sweeps out a torus which contains another flux loop $\beta$ while both linked with a ``base'' flux loop $\gamma$. We denote the Berry phase associated with the above braiding process by $\theta_{\alpha\beta,\gamma}$. Without loss of generality, we suppose that $\gamma$ lies in the $yz$-plane, and $\alpha$ and $\beta$ lie in the $xy$-plane. After we compactify the $x$-direction into a small circle, $\alpha$ extends across the $x$-direction, fuses with itself, and splits into two noncontractible loops $\alpha^{\prime}$ and $\alpha^{\prime\prime}$ (Fig. \ref{fig:dimred}(b)). Similarly, $\beta$ fuses with itself and splits into $\beta^{\prime}$ and $\beta^{\prime\prime}$. The three-loop braiding process can then be decomposed into two separate processes in which $\alpha^{\prime}$ is braided around $\beta^{\prime}$ inside the base loop $\gamma$ and $\alpha^{\prime\prime}$ is braided around $\beta^{\prime\prime}$ outside the base loop $\gamma$ (Fig. \ref{fig:dimred}(b)). For the first process, we can stretch $\gamma$ so that it subtends the $yz$-plane (Fig. \ref{fig:dimred}(c)). This leaves a flux line $\gamma$ threaded through the ``$x$-hole'', which reduces the 3D $G$-gauge theory $\mathcal{C}^{\text{3D}}$ to the 2D $G$-gauge theory $\mathcal{C}^{\text{2D}}_{\phi_{\gamma}}$, where $\phi_{\gamma}$ denotes the gauge flux carried by the loop $\gamma$. Similarly, for the second process, we can shrink $\gamma$ till it fuses and annihilates with itself (Fig. \ref{fig:dimred}(c)). This leaves no gauge flux through the ``$x$-hole'', and $\mathcal{C}^{\text{3D}}$ is reduced to $\mathcal{C}^{\text{2D}}_{0}$, where we denote the group identity of $G$ by $0$. In the 2D limit, noncontractible flux loops along the $x$-direction become point-like gauge fluxes in the 2D $G$-gauge theories, because the extent of the $x$-direction is negligible compared with that of the $y$ and $z$-directions. Therefore, the three-loop braiding process in $\mathcal{C}^{\text{3D}}$ that we started with is reduced to two separate braiding processes between gauge fluxes in $\mathcal{C}^{\text{2D}}_{\phi_{\gamma}}$ and $\mathcal{C}^{\text{2D}}_{0}$. This implies the following relation between the associated Berry phases:
\begin{equation}
\theta_{\alpha\beta,\gamma} = \theta^{\text{2D}}_{\alpha^{\prime}\beta^{\prime}}(\gamma) - \theta^{\text{2D}}_{\alpha^{\prime\prime}\beta^{\prime\prime}}(0),
\label{eqn:dimred}
\end{equation}
where the first and second terms on the right hand side are the Berry phases resulting from braiding $\alpha^{\prime}$ around $\beta^{\prime}$ in $\mathcal{C}^{\text{2D}}_{\phi_{\gamma}}$ and $\alpha^{\prime\prime}$ around $\beta^{\prime\prime}$ in $\mathcal{C}^{\text{2D}}_{0}$, respectively. The relative minus sign takes into account the fact that the two pairs of gauge fluxes are braided in opposite directions.

Now, we carry out the above procedure to analyze the dimensional reduction phenomenon and compute the three-loop braiding statistics in our Walker-Wang models. First, we apply the relations in Eq. \eqref{eqn:styz} to obtain the 2D modular matrices $S_{yz}$ and $T_{yz}$ from the 3D modular matrices $S$ and $T$. Then we compute the charge string operators along the non-contractible loops along the $x$, $y$, and $z$ axes (see Appendix \ref{app_c}), which we denote by $W^s_x$, $W^s_y$, and $W^s_z$, respectively. Here $s = 0,1,2,3$ labels the $\mathbb{Z}_2 \times \mathbb{Z}_2$ charges. Without loss of generality, we identify $1$ with $g_x$, and $2$ with $g_y$. Next, we do a basis transformation, and rewrite $S_{yz}$ and $T_{yz}$ in the simultaneous eigenstates of $W^s_x$, $W^s_y$, and $W^s_z$. We find that if we organize the basis states according to the eigenvalues of $W^s_x$ (equivalently the eigenvalues of the pair ($W^1_x,W^2_x$)), $S_{yz}$ and $T_{yz}$ are block diagonal with each block of size 16 by 16. For simplicity, we denote by $S_{a,b}$ and $T_{a,b}$ ($a,b = \pm 1$) the block corresponding to $(W^1_x,W^2_x) = (a,b)$. Since $W^1_x$ (resp. $W^2_x$) detects the $g_x$ (resp. $g_y$) flux through the ``$x$-hole'', ($S_{a,b}$,$T_{a,b}$) describes the 2D topological order obtained by making the $x$-direction of the three-torus into a small circle, and threading some particular $\mathbb{Z}_2 \times \mathbb{Z}_2$ flux $\nu(a,b)$ through the ``$x$-hole'', where $\nu(a,b)$ $=$ $0$, $g_x$, $g_y$, or $g_z$ for $(a,b)$ $=$ $(1,1)$, $(-1,1)$, $(1,-1)$, or $(-1,-1)$, respectively. 

Note that $S_{a,b}$ and $T_{a,b}$ are written in the simultaneous eigenstates of $W^s_y$ and $W^s_z$, and are not yet presented in their canonical form, where the entries of the $S$ and $T$ matrices are the braiding statistics and topological spins of quasiparticles, respectively. It is shown in Ref. \onlinecite{Zhang2011} that by choosing the basis states in the ground space to be the minimum entropy states (MESs), one can put $S$ and $T$ into the canonical form. The MESs are the simultaneous eigenstates of the charge string operators and flux string operators that encircle the two-torus. Without loss of generality, we define our MESs to be the simultaneous eigenstates of $W^s_y$ and $V^s_y$, where we denote the flux string operators along the $y$-axis by $V^s_y$. The flux string operators are the flux membrane operators in the $xy$-plane before we dimensionally reduce our system to the $yz$-plane. In general, we do not know how to implement the membrane operators in Walker-Wang models, so it is hard to write down $V^s_y$ explicitly. However, we do know that $V^s_y$ and $W^s_z$ satisfy the following commutation and anticommutation relations:
\begin{align}
&\{W_z^1, V_y^1\} = 0, \ \ \ [W_z^1,V_y^2] = 0, \ \ \  \{W_z^1, V_y^3\} = 0, \nonumber \\
&[W_z^2,V_y^1] = 0, \negphantom{[W_z^2,V_y^1] = 0,} \phantom{\{W_z^1, V_y^1\} = 0,} \ \ \ \{W_z^2, V_y^2\} = 0, \negphantom{[W_z^2,V_y^1] = 0, \negphantom{[]} \phantom{\{ \}} \ \ \ \{W_z^2, V_y^2\} = 0,}\phantom{\{W_z^1, V_y^1\} = 0, \ \ \ [W_z^1,V_y^2] = 0,}\ \ \  \{W_z^2, V_y^3\} = 0, \nonumber \\
&\{W_z^3, V_y^1\} = 0, \negphantom{\{W_z^3, V_y^1\} = 0,} \phantom{\{W_z^1, V_y^1\} = 0,}\ \ \ \{W_z^3, V_y^2\} = 0, \negphantom{\{W_z^3, V_y^1\} = 0, \negphantom{\{W_z^3, V_y^1\} = 0,} \phantom{\{W_z^1, V_y^1\} = 0,}\ \ \ \{W_z^3, V_y^2\} = 0,} \phantom{\{W_z^1, V_y^1\} = 0, \ \ \ [W_z^1,V_y^2] = 0,}\ \ \  [W_z^3,V_y^3] = 0.
\end{align}
This follows from the Aharonov-Bohm interaction between gauge charges and flux loops in a 3D gauge theory. We are able to deduce from this the basis transformation from the simultaneous eigenstates of $W^s_y$ and $W^s_z$ to the MESs. For details about the basis transformation, we refer the reader to Appendix \ref{app_d}.

After rewriting $S_{a,b}$ and $T_{a,b}$ in the MES basis, we find that they are identical to the 2D modular matrices of the 2D $\mathbb{Z}_2 \times \mathbb{Z}_2$ gauge theories. There are 8 inequivalent such theories: 1 untwisted gauge theory and 7 twisted gauge theories. They can be distinguished by a triple $(\theta_x,\theta_y,\theta_{xy})$, where the first, second and third entries are the Berry phases associated with the exchange of two $g_x$ fluxes, the exchange of two $g_y$ fluxes, and the braiding of a $g_x$ flux around a $g_y$ flux, respectively. $\theta_x$ and $\theta_y$ can take value either $0$ or $\pi/2$, and $\theta_{xy}$ can take value either $0$ or $-\pi/2$, and the $8$ combinations correspond to the 8 different 2D $\mathbb{Z}_2 \times \mathbb{Z}_2$ gauge theories. The 8 theories are classified by the cohomology group $H^3(\mathbb{Z}_2 \times \mathbb{Z}_2 , U(1)) = \mathbb{Z}_2 \times \mathbb{Z}_2 \times \mathbb{Z}_2$. The trivial element of the group corresponds to the untwisted gauge theory, and the $7$ nontrivial elements correspond to the twisted gauge theories. The three generators of the group (written multiplicatively), which we denote by $\omega_1$, $\omega_2$, and $\omega_{12}$, can be taken to be the 2D twisted $\mathbb{Z}_2 \times \mathbb{Z}_2$ gauge theories with $(\theta_x,\theta_y,\theta_{xy})=(\pi/2,0,0)$, $(0,\pi/2,0)$, and $(0,0,-\pi/2)$, respectively. Each dimensionally-reduced 2D $\mathbb{Z}_2 \times \mathbb{Z}_2$ gauge theory can then be represented by a combination of the three generators. The results are summarized in Table \ref{tab2}.

To compute the three-loop braiding statistics in our Walker-Wang models, we follow our earlier discussion to decompose a three-loop braiding process into two separate braiding processes in the dimensionally-reduced 2D systems and find the relation between their associated Berry phases. In particular, the three-loop braiding process considered in Section \ref{sec:gauging}, where two $g_x$-flux loops are exchanged while both linked with a $g_x$-flux loop, can be decomposed into the following two braiding processes in 2D: (1) Two $g_y$ fluxes exchanged inside a $g_y$-flux loop; (2) Two $g_x$ fluxes exchanged outside the $g_y$-flux loop. Therefore, we have the following relation between the associated Berry phases:
\begin{equation}
\theta_{x,y} = \theta_{x}^{\text{2D}}(y) - \theta_{x}^{\text{2D}}(0),
\label{eqn:dimred1}
\end{equation}
where the first and second terms on the right hand side are the Berry phases resulting from exchanging two $g_x$ fluxes either inside or outside a $g_y$-flux loop, respectively. Similar analysis applies to the case where the roles of $g_x$ and $g_y$ are switched and we have the following expression:
\begin{equation}
\theta_{y,x} = \theta_{y}^{\text{2D}}(x) - \theta_{y}^{\text{2D}}(0).
\label{eqn:dimred2}
\end{equation}

Let us now apply Eq. \eqref{eqn:dimred1} and Eq. \eqref{eqn:dimred2} to two examples. First, we consider a Walker-Wang model with input $\text{Rep}_s(Q_8)$. From Table \ref{tab2}, we know that after dimensional reduction, we get the 2D untwisted $\mathbb{Z}_2 \times \mathbb{Z}_2$ gauge theory both inside and outside each $\mathbb{Z}_2 \times \mathbb{Z}_2$ flux loop, which implies that
\begin{align}
&\theta^{\text{2D}}_{x}(0) = \theta^{\text{2D}}_{y}(0) = 0, \nonumber \\
&\theta^{\text{2D}}_{x}(y) = \theta^{\text{2D}}_{y}(x) = 0.
\end{align} 
Hence 
\begin{equation}
\theta_{x,y} =0, \ \ \  \theta_{y,x} = 0,
\end{equation}
and the Walker-Wang model describes the 3D untwisted $\mathbb{Z}_2 \times \mathbb{Z}_2$ gauge theory.

Next, we consider a Walker-Wang model with input $\text{Rep}_s(D_4)$. As in the previous example, we get 2D $\mathbb{Z}_2 \times \mathbb{Z}_2$ gauge theories both inside and outside each $\mathbb{Z}_2 \times \mathbb{Z}_2$ flux loop after dimensional reduction. The only difference is that the $\mathbb{Z}_2 \times \mathbb{Z}_2$ gauge theory is twisted inside a nontrivial flux loop. More specifically, we get the 2D twisted $\mathbb{Z}_2 \times \mathbb{Z}_2$ gauge theory represented by the $\mathbb{Z}_2 \times \mathbb{Z}_2 \times \mathbb{Z}_2$ group element $\omega_2$ (resp. $\omega_{12}$) inside a $g_x$-flux (resp. $g_y$-flux) loop. We can deduce from this that 
\begin{equation}
\theta_{x}^{\text{2D}}(y) = 0, \ \ \ \theta_{y}^{\text{2D}}(x) = \frac{\pi}{2}.
\end{equation}
Together with
\begin{equation}
\theta^{\text{2D}}_{x}(0) = \theta^{\text{2D}}_{y}(0) = 0,
\end{equation}
they imply that 
\begin{equation}
\theta_{x,y} =0, \ \ \  \theta_{y,x} = \frac{\pi}{2},
\end{equation}
and the Walker-Wang model describes a 3D twisted $\mathbb{Z}_2 \times \mathbb{Z}_2$ gauge theory (the APS-X theory).

We can carry out similar computations for Walker-Wang models with the other two sets of input data in Table \ref{tab2}. We find that when we permute the labels $1$ and $2$ (resp. $1$ and $3$) in $\text{Rep}_s(D_4)$, the resulting Walker-Wang model describes the APS-Y (resp. APS-Z) theory.

\section{Summary and Discussion}
\label{sum}

In this paper, we studied in detail the realization of the 3D $\mathbb{Z}_2 \times \mathbb{Z}_2$ gauge theories, both twisted and untwisted, in terms of Walker-Wang models. Our proposal is based on the study of the surface topological order of $\mathbb{Z}_2 \times \mathbb{Z}_2$ gauge theories.\cite{Chen2014, Kapustin2014Apr} We propose that if we take the input data of a Walker-Wang model to be the $\mathbb{Z}_2 \times \mathbb{Z}_2$ symmetry charges and the surface anyon content of a trivial (resp. nontrivial) 3D $\mathbb{Z}_2 \times \mathbb{Z}_2$ SPT (or rather their corresponding excitations in the gauged models), the output theory is a 3D untwisted (resp. twisted) $\mathbb{Z}_2 \times \mathbb{Z}_2$ gauge theory. To check the validity of our proposal, we perform 3D modular transformations to the ground space of our Walker-Wang models on the three-torus and extract the resulting $S$ and $T$ matrices. By making a dimensional reduction argument, we are able to deduce the three-loop braiding statistics from the $S$ and $T$ matrices, which determine the topological order in our models. 

Note that there is a subtlety involved in doing modular transformations in Walker-Wang models. In Walker-Wang models, we work with a fixed planar projection of the 3D trivalent lattice. The strings living on the lattice are actually ribbons with the blackboard framing. Therefore, it is important that we choose to calculate Berry phases associated with the modular transformations that preserve the projection of the 3D lattice (equivalently, the framing of the ribbon graphs). This is solely for the sake of convenience. Otherwise, we need to transform the ribbon graphs back to the original framing after the modular transformations, and this introduces extra factors into the wave function. This is precisely the reason why we did not calculate the Berry phases associated with the 2D modular transformation $S_{yz}$ directly in Section \ref{snt}. $S_{yz}$ changes the framing of the ribbon graphs, whereas $S$ and $T$ do not. Thus, it is easier to first calculate $S$ and $T$, and deduce $S_{yz}$ from the relation Eq. \eqref{eqn:styz}.

So far we have only considered the Walker-Wang construction of 3D $\mathbb{Z}_2 \times \mathbb{Z}_2$ gauge theories. It would be interesting to generalize the construction to other 3D discrete gauge theories. It would also be interesting to find the connection between the Walker-Wang description and the Dijkgraaf-Witten description of these discrete gauge theories. Note that the ground state wave function in the former (resp. latter) description is a condensate of loops (resp. membranes), so the two descriptions should be dual to each other in some sense. It would be nice to make this duality more concrete and study how general it is. Another interesting direction is to generalize the Walker-Wang construction to discrete gauge theories with (at least one) fermionic gauge chareges. The simplest example of this kind is a 3D $\mathbb{Z}_2$ gauge theory with fermionic $\mathbb{Z}_2$ gauge charges. This theory can be described by a Walker-Wang model.\cite{LevinWen2004} One can simply take a Walker-Wang model describing the 3D $\mathbb{Z}_2$ gauge theory, and ``twisting'' the $R$ symbols of the input data by a fermion. More precisely, $R^{11}_0$ takes the value $-1$ in the fermionic case and $1$ in the bosonic case, where $1$ labels the $\mathbb{Z}_2$ gauge charge, and $0$ labels the vacuum. This example is interesting because Dijkgraaf-Witten models fail to describe discrete gauge theories with fermionic gauge charges. It would be nice to have a Walker-Wang description for more of such fermionic discrete gauge theories, especially ones that are twisted. We will leave these problems to future work.

\acknowledgments

We would like to thank Meng Cheng, Lukasz Fidkowski, Yuting Hu, and Zhenghan Wang for valuable discussions, and especially Yichen Huang for discussions and collaborations during various stages of the work. X.C. is supported by the Caltech Institute for Quantum Information and Matter and the Walter Burke Institute for Theoretical Physics.


%

\appendix
\allowdisplaybreaks
\section{3D modular transformations for the Walker-Wang models on the minimal lattice}
\label{app_a}

In this appendix, we calculate the matrix representation of the $S$ and $T$ transformations in the ground space of a Walker-Wang model defined on the minimal lattice. For simplicity, we will use a labeling of the minimal lattice by the input anyons to represent the amplitude of the associated string-net configuration in the ground state wave function. The set of all string-net configurations constitute a basis for the ground state Hilbert space. To compute the $S$ and $T$ matrices, we apply the corresponding trasformations to a particular basis vector, and express the resulting vector as a superposition of the basis vectors by applying the $F$ and $R$ moves. The coefficients in front of the superposition are nothing but the matrix elements of $S$ and $T$.

\subsection{$S$ matrix}
\begin{align*}
&\vcenter{\hbox{\includegraphics[width=0.45\linewidth]{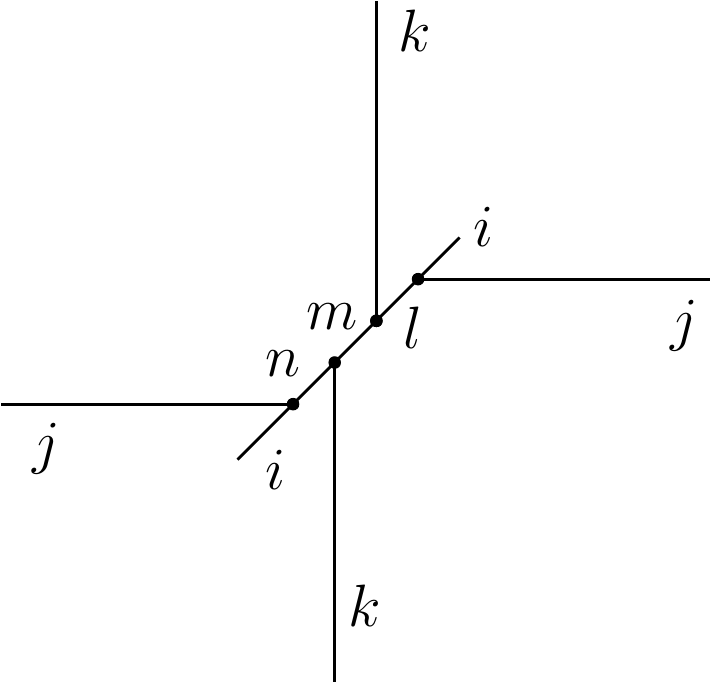}}}\xrightarrow[S]{\quad }\vcenter{\hbox{\includegraphics[width=0.45\linewidth]{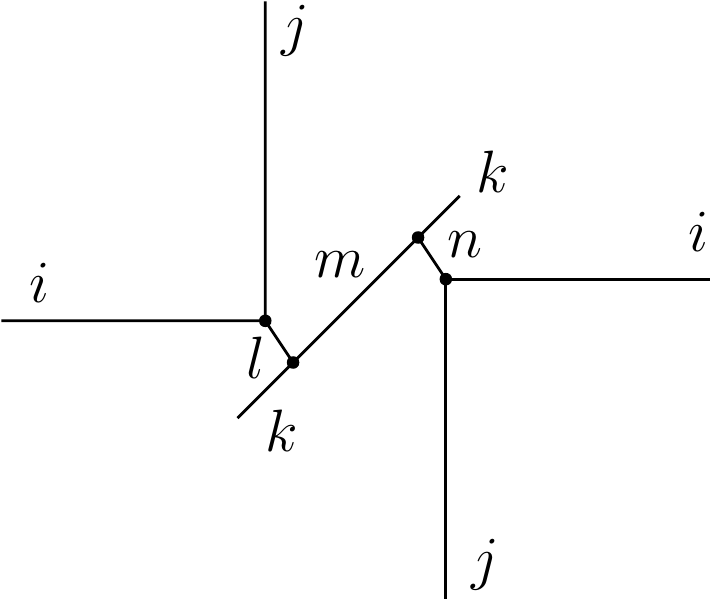}}} \\
&\xrightarrow[F \ \rm move]{\sum_{\tilde{l}}F^{jil}_{km\tilde{l}}} \ \vcenter{\hbox{\includegraphics[width=0.45\linewidth]{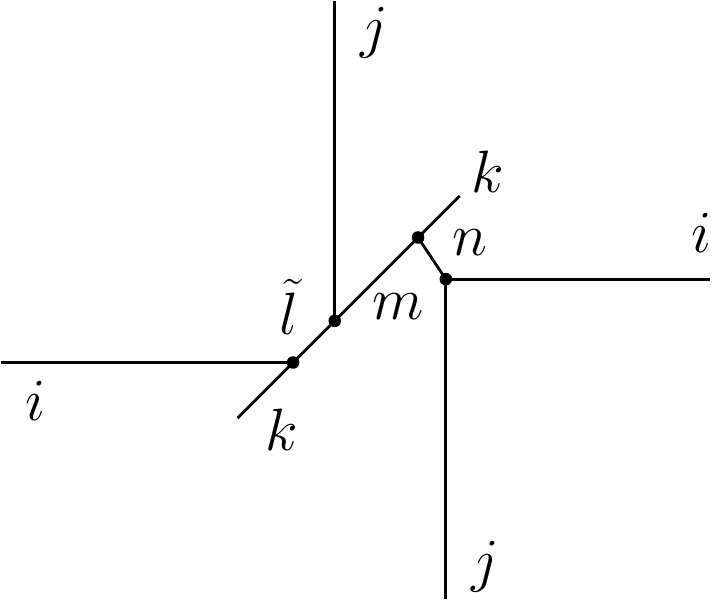}}} \\
&\xrightarrow[F \ \rm move]{\sum_{\tilde{n}}F^{kmn}_{ji\tilde{n}}}\ \vcenter{\hbox{\includegraphics[width=0.45\linewidth]{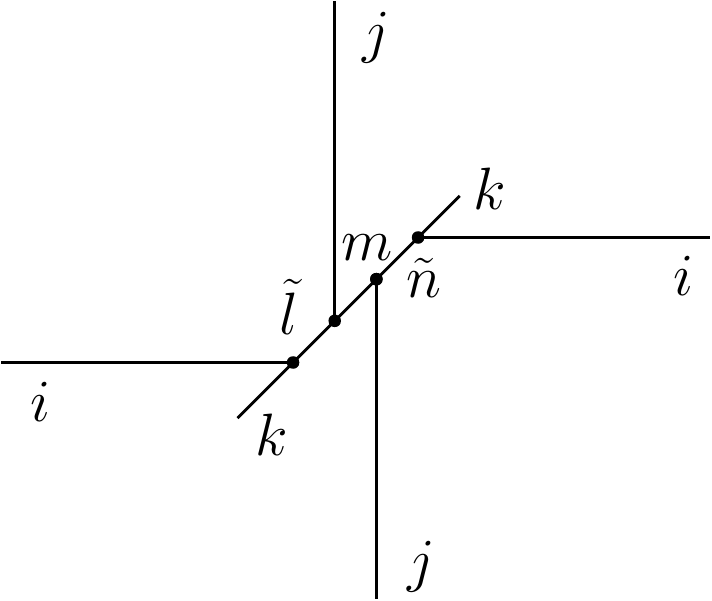}}} \\
&\xrightarrow[F \ \rm move]{\sum_{\tilde{m}}F^{j\tilde{l}m}_{j\tilde{n}\tilde{m}}}\ \vcenter{\hbox{\includegraphics[width=0.45\linewidth]{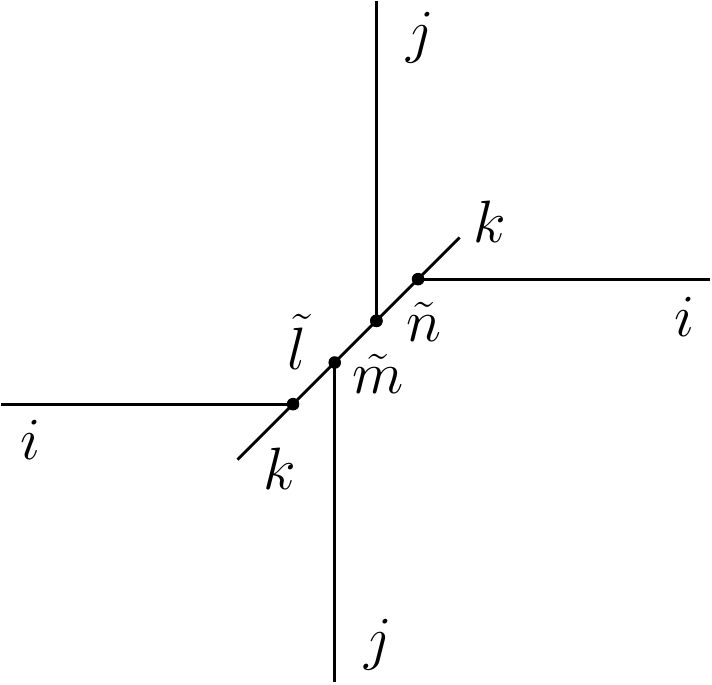}}}
\end{align*}

Collecting the coefficients from each step, we obtain 
\begin{equation}
S^{(k,i,j,\tilde{n},\tilde{m},\tilde{l})}_{(i,j,k,l,m,n)}=F^{jil}_{km\tilde{l}}F^{kmn}_{ji\tilde{n}}F^{j\tilde{l}m}_{j\tilde{n}\tilde{m}}.
\end{equation}

\subsection{$T$ matrix}
\begin{align*}
&\vcenter{\hbox{\includegraphics[width=0.45\linewidth]{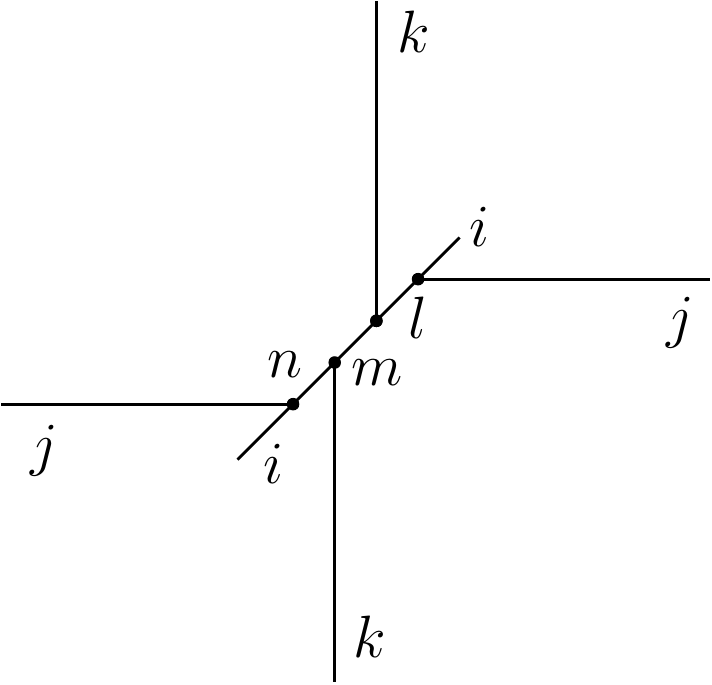}}}\xrightarrow[T]{\quad}\vcenter{\hbox{\includegraphics[width=0.45\linewidth]{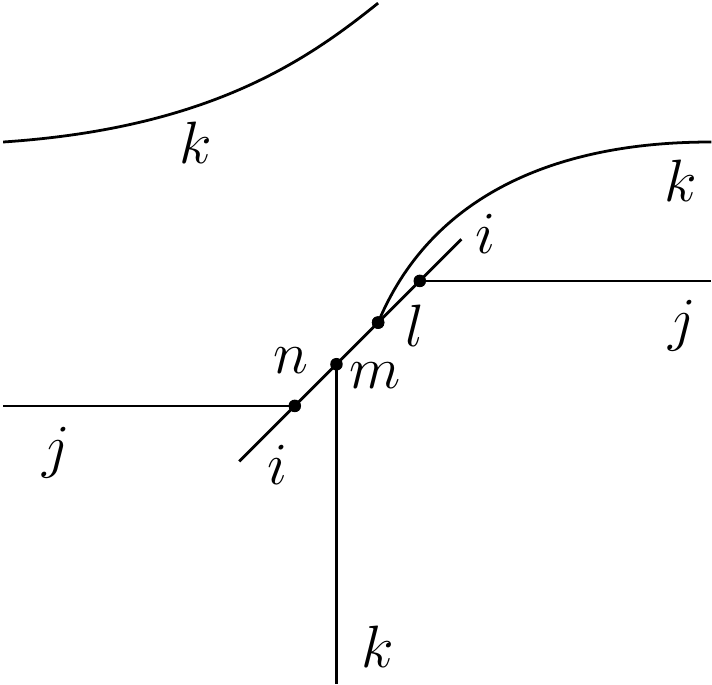}}} \nonumber \\
&\xrightarrow[R \ \rm move]{R^{kl}_m} \ \vcenter{\hbox{\includegraphics[width=0.45\linewidth]{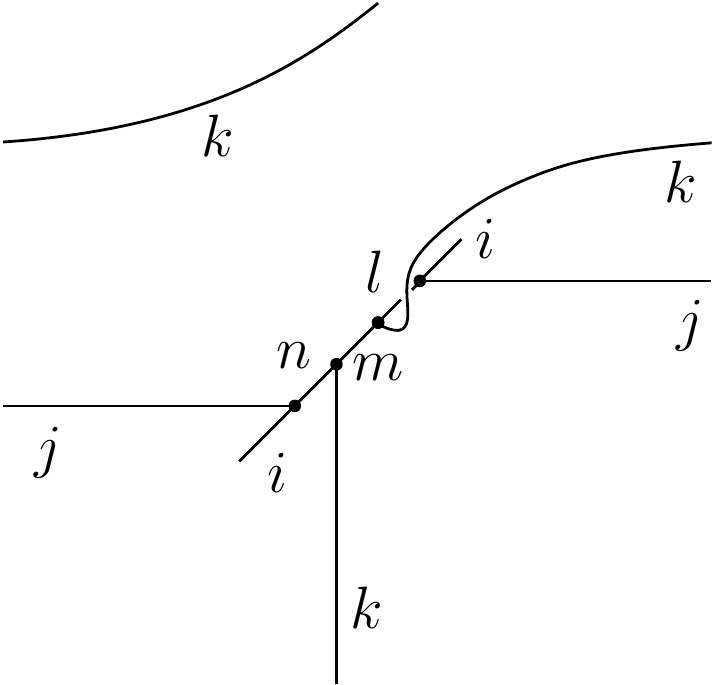}}} \\
&\xrightarrow[\text{Commute $F$ and $R$ moves}]{} \ \vcenter{\hbox{\includegraphics[width=0.45\linewidth]{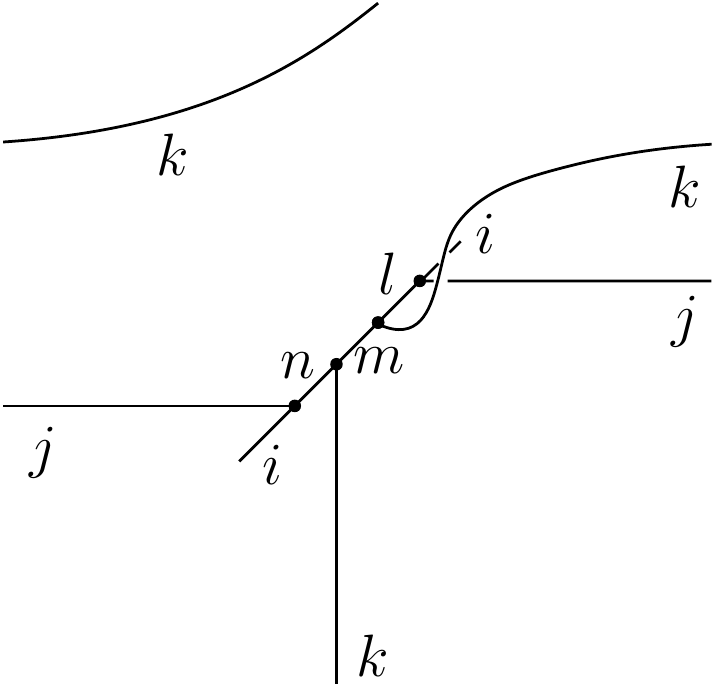}}} \\
&\xrightarrow[F \ \rm move]{\sum_{\tilde{l}}F^{mkl}_{ji\tilde{l}}}\ \vcenter{\hbox{\includegraphics[width=0.45\linewidth]{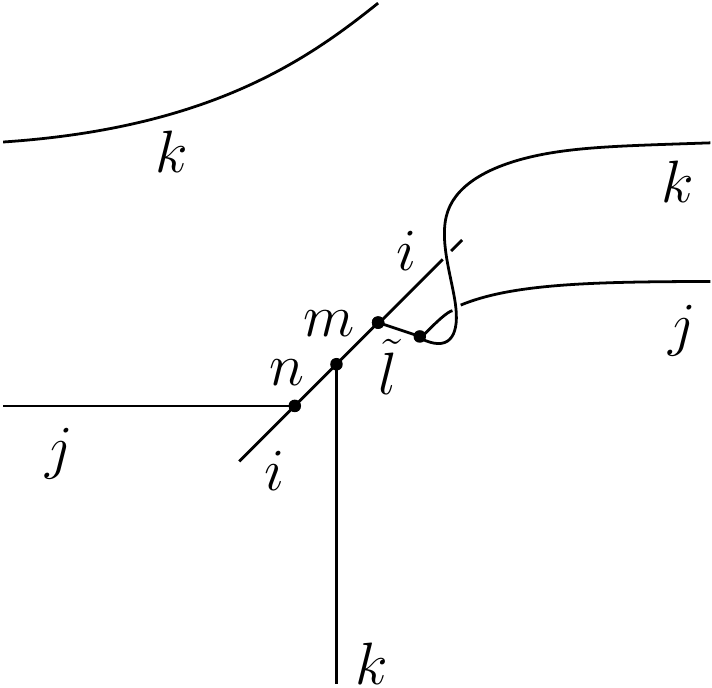}}}\\
&\xrightarrow[R \ \rm move]{(R^{kj}_{\tilde{l}})^*} \ \vcenter{\hbox{\includegraphics[width=0.45\linewidth]{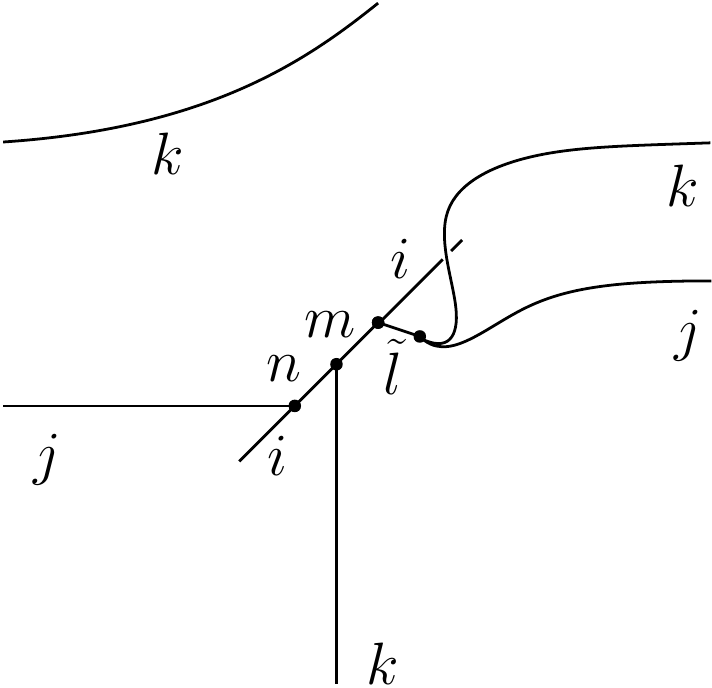}}}\\
&\xrightarrow[\rm Deformation]{}\ \vcenter{\hbox{\includegraphics[width=0.45\linewidth]{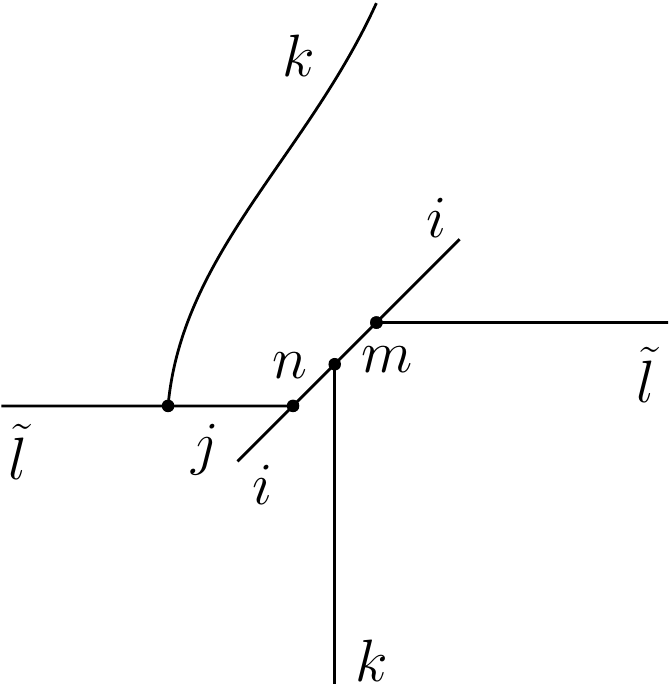}}}\\
&\xrightarrow[F \ \rm move]{\sum_{\tilde{j}}F^{k\tilde{l}j}_{in\tilde{j}}}\ \vcenter{\hbox{\includegraphics[width=0.45\linewidth]{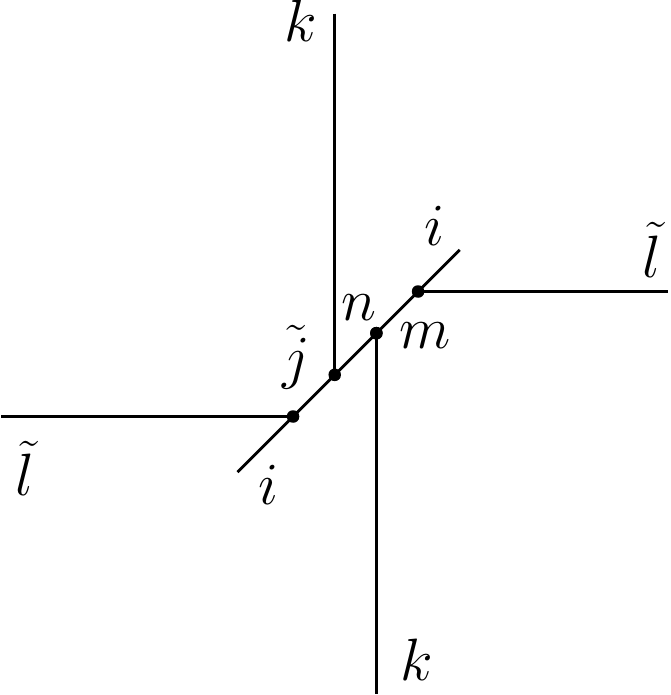}}}\\
&\xrightarrow[F \ \rm move]{\sum_{\tilde{n}}F^{k\tilde{j}n}_{km\tilde{n}}}\ \vcenter{\hbox{\includegraphics[width=0.45\linewidth]{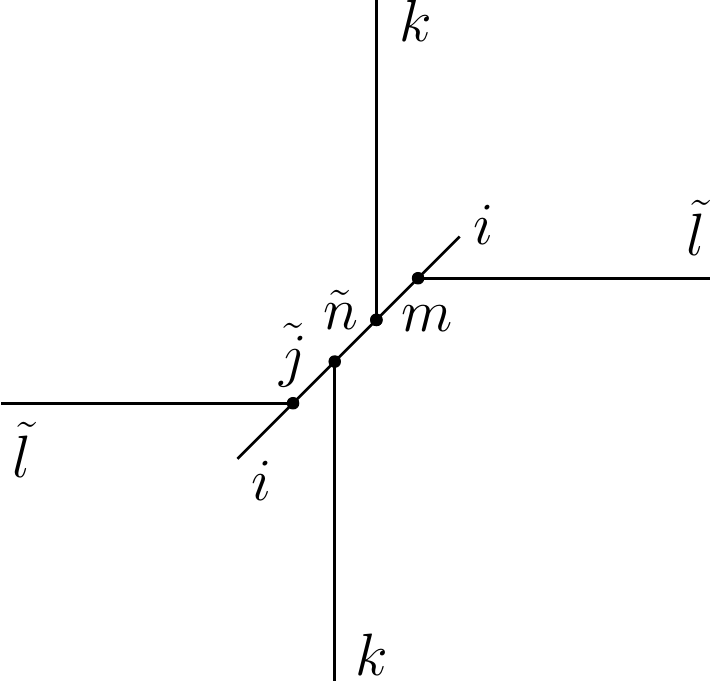}}}
\end{align*}

Collecting the coefficients from each step, we obtain 
\begin{equation}
T^{(i,\tilde{l},k,m,\tilde{n},\tilde{j})}_{(i,j,k,l,m,n)}=R^{kl}_mF^{mkl}_{ji\tilde{l}}(R^{kj}_{\tilde{l}})^*F^{k\tilde{l}j}_{in\tilde{j}}F^{k\tilde{j}n}_{km\tilde{n}}.
\end{equation}

\section{Hamiltonian for the Walker-Wang models on the minimal lattice}
\label{app_b}

In this appendix, we calculate the plaquette operators of a Walker-Wang model defined on the minimal lattice. For simplicity, we will calculate $B_p^s$ for a particular anyon label $s$, and the full plaquette operator can be obtained by summing over $s$ weighted by the quantum dimension of $s$: $B_p = \sum_s d_s B_p^s$.

\subsection{Plaquette operator in the $xy$-plane}
\begin{align*}
&\vcenter{\hbox{\includegraphics[width=0.45\linewidth]{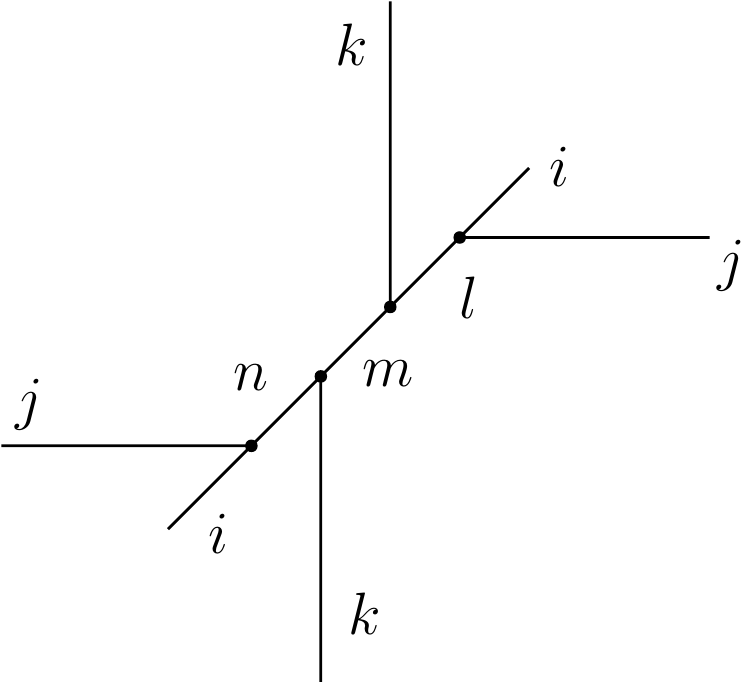}}}\xrightarrow[R\ \rm move]{(R^{nk}_{m})^*}\vcenter{\hbox{\includegraphics[width=0.45\linewidth]{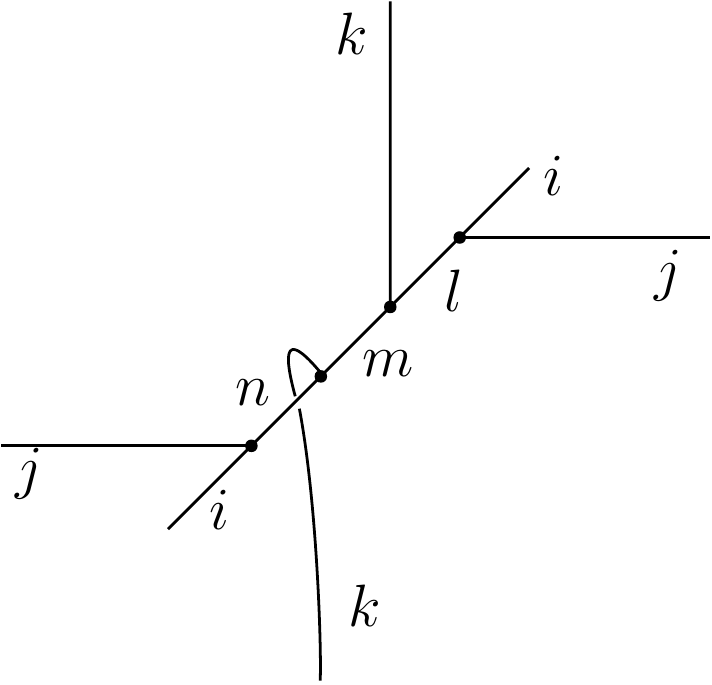}}}\\
&\xrightarrow[\text{add an $s$-loop in the $xy$-plane}]{} \ \vcenter{\hbox{\includegraphics[width=0.45\linewidth]{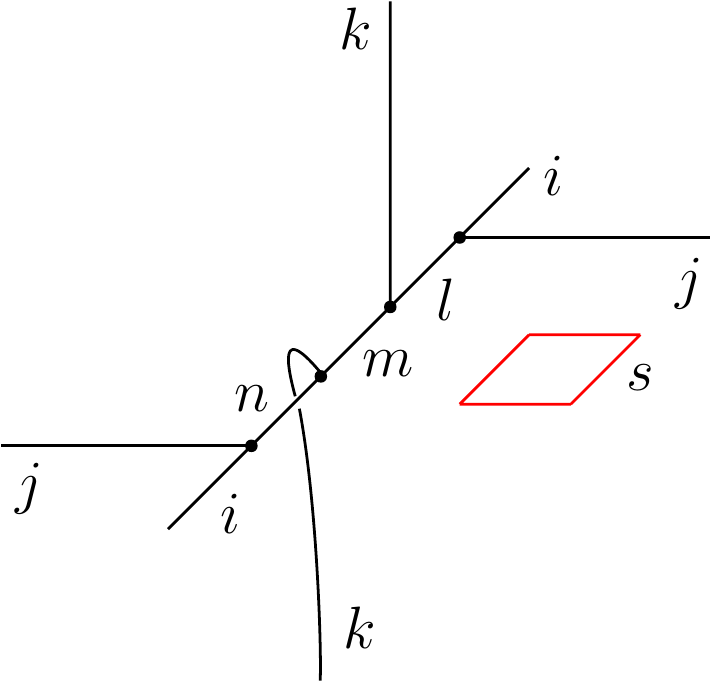}}}\\
&\xrightarrow[\rm Deformation]{}\ \vcenter{\hbox{\includegraphics[width=0.45\linewidth]{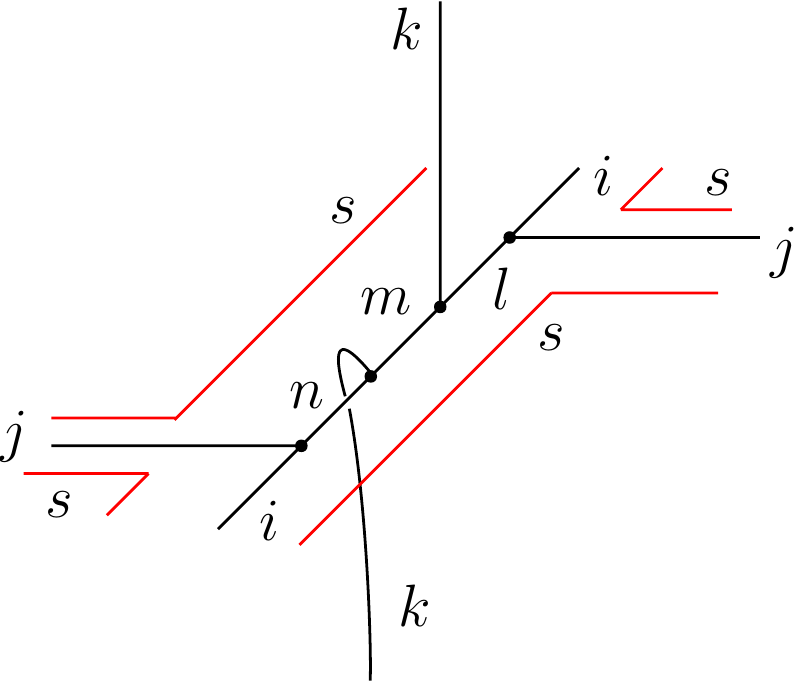}}}\\
&\xrightarrow[F \ \rm move]{\sum_{j^{\prime}}F^{jj0}_{ssj^{\prime}}}\ \vcenter{\hbox{\includegraphics[width=0.45\linewidth]{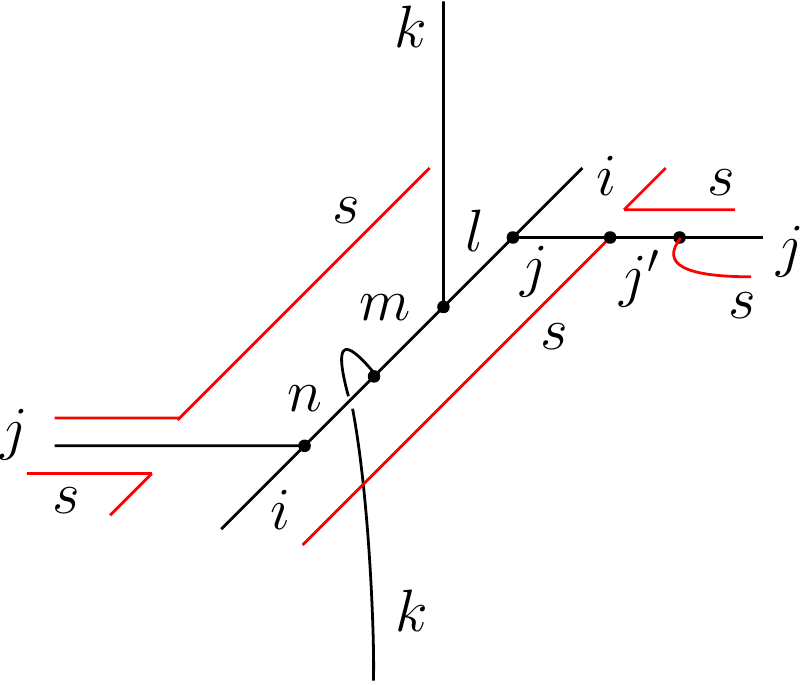}}}\\
&\xrightarrow[F \ \rm move]{\sum_{l^{\prime}}F^{ilj}_{sj^{\prime}l^{\prime}}}\ \vcenter{\hbox{\includegraphics[width=0.45\linewidth]{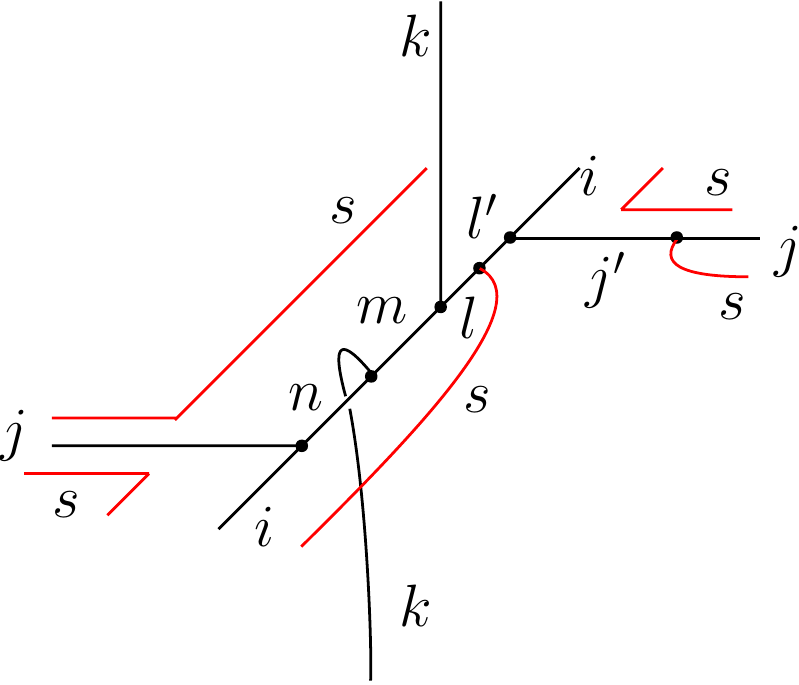}}}\\
&\xrightarrow[F \ \rm move]{\sum_{m^{\prime}}F^{kml}_{sl^{\prime}m^{\prime}}}\ \vcenter{\hbox{\includegraphics[width=0.45\linewidth]{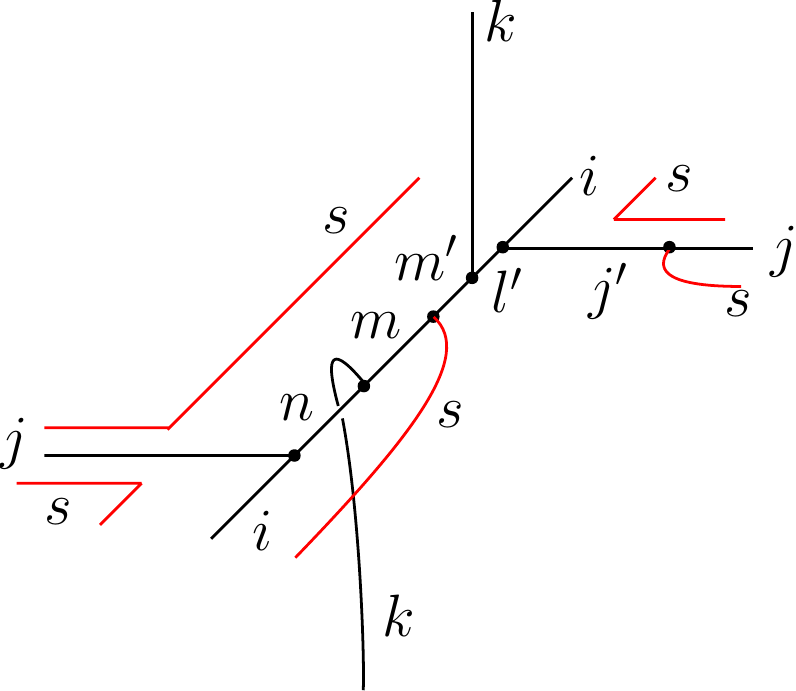}}}\\
&\xrightarrow[F\ \rm move]{\sum_{n^{\prime}}F^{knm}_{sm^{\prime}n^{\prime}}}\ \vcenter{\hbox{\includegraphics[width=0.45\linewidth]{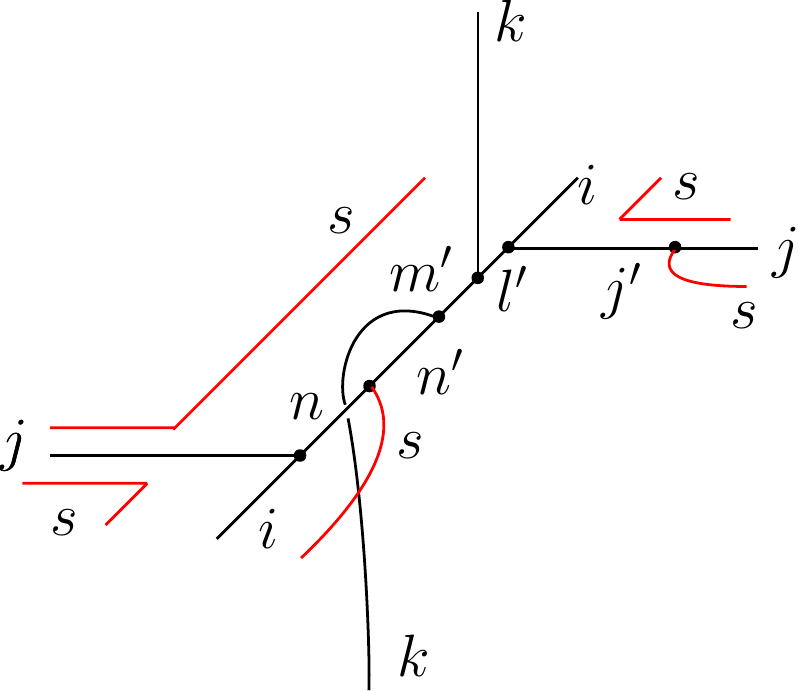}}}\\
&\xrightarrow[\text{Commute $F$ and $R$ moves}]{}\ \vcenter{\hbox{\includegraphics[width=0.45\linewidth]{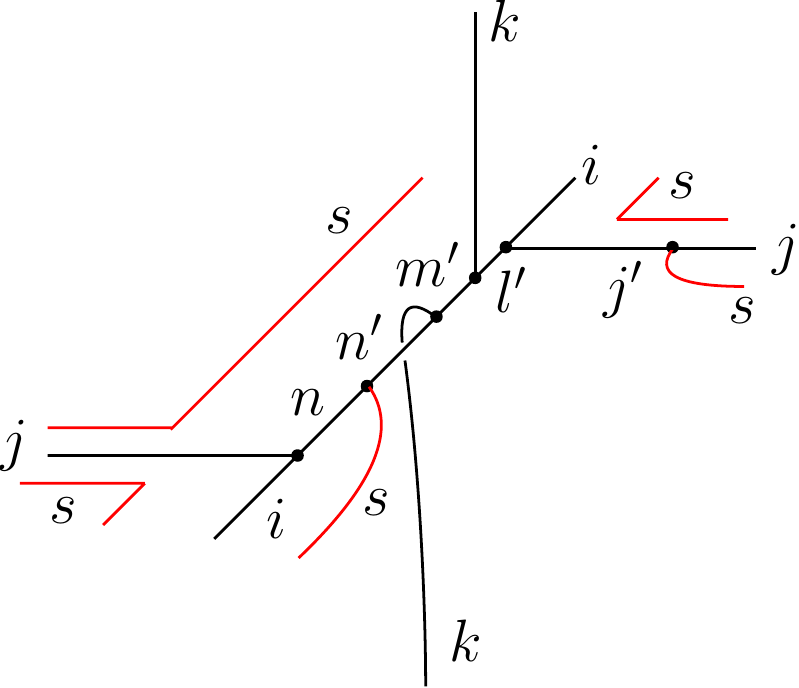}}}\\
&\xrightarrow[R \ \rm move]{R^{n^{\prime}k}_{m^{\prime}}}\ \vcenter{\hbox{\includegraphics[width=0.45\linewidth]{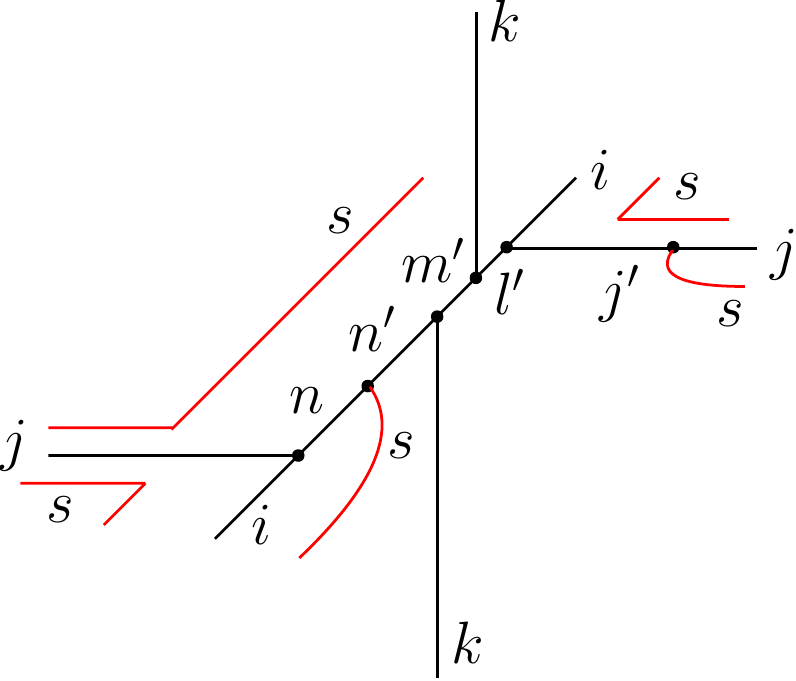}}}\\
&\xrightarrow[R \ \rm move]{R^{kl^{\prime}}_{m^{\prime}}}\ \vcenter{\hbox{\includegraphics[width=0.45\linewidth]{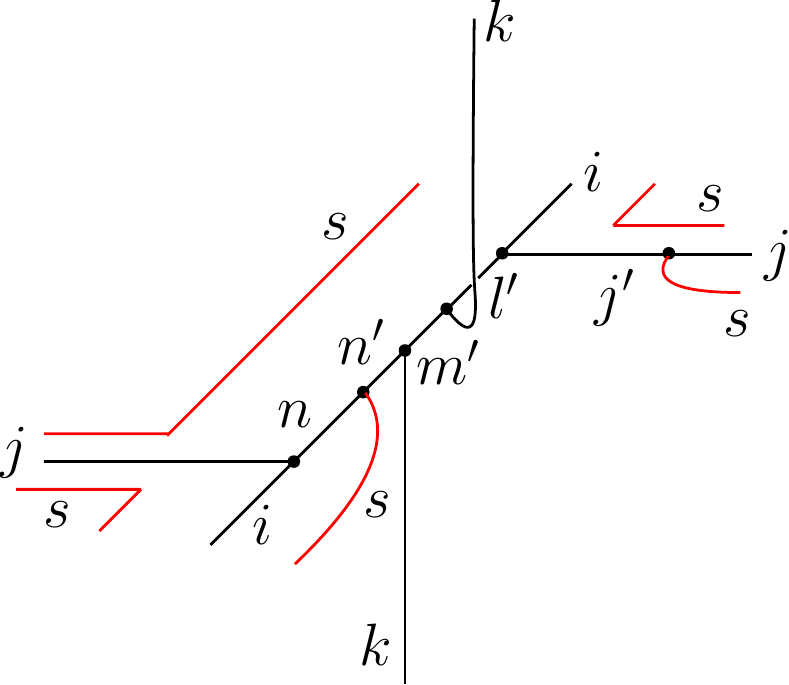}}}\\
&\xrightarrow[F \ \rm move]{\sum_{i^{\prime}}F^{jin}_{sn^{\prime}i^{\prime}}}\ \vcenter{\hbox{\includegraphics[width=0.45\linewidth]{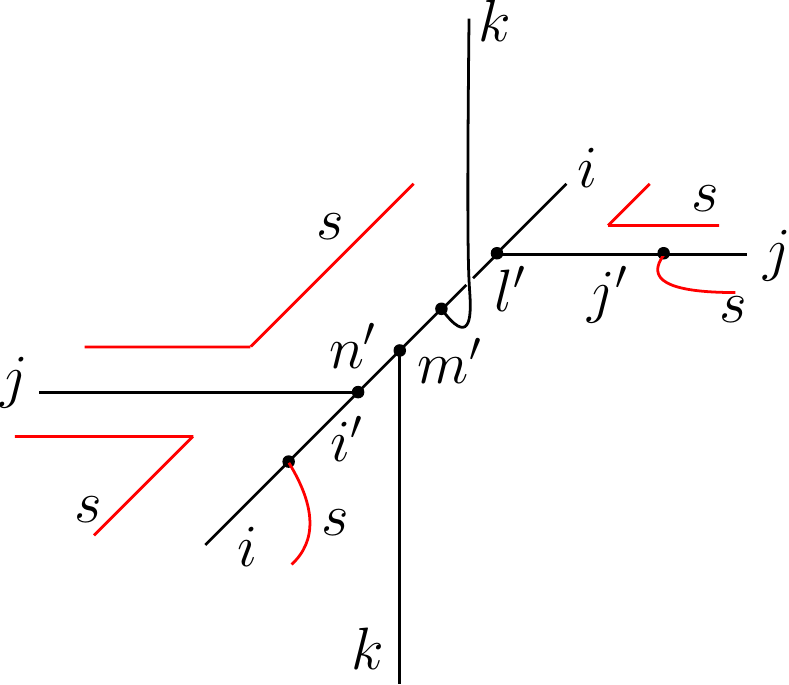}}}\\
&\xrightarrow[F \ \rm move]{\sum_{l^{\prime\prime}}F^{l^{\prime}j^{\prime}i}_{si^{\prime}l^{\prime\prime}}}\ \vcenter{\hbox{\includegraphics[width=0.45\linewidth]{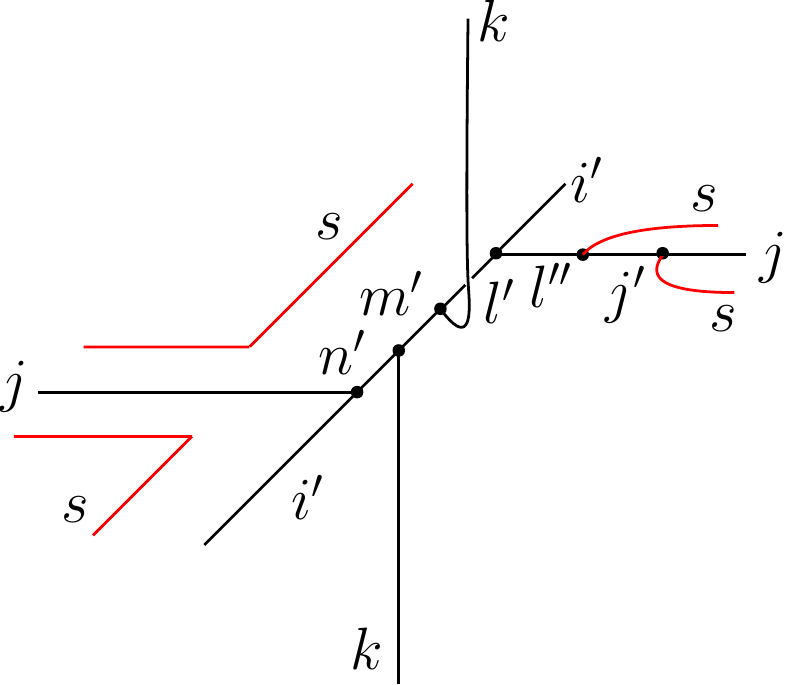}}}\\
&\xrightarrow[F \ \rm move]{\sum_{j^{\prime\prime}}F^{j^{\prime}sj}_{i^{\prime}n^{\prime}j^{\prime\prime}}}\ \vcenter{\hbox{\includegraphics[width=0.45\linewidth]{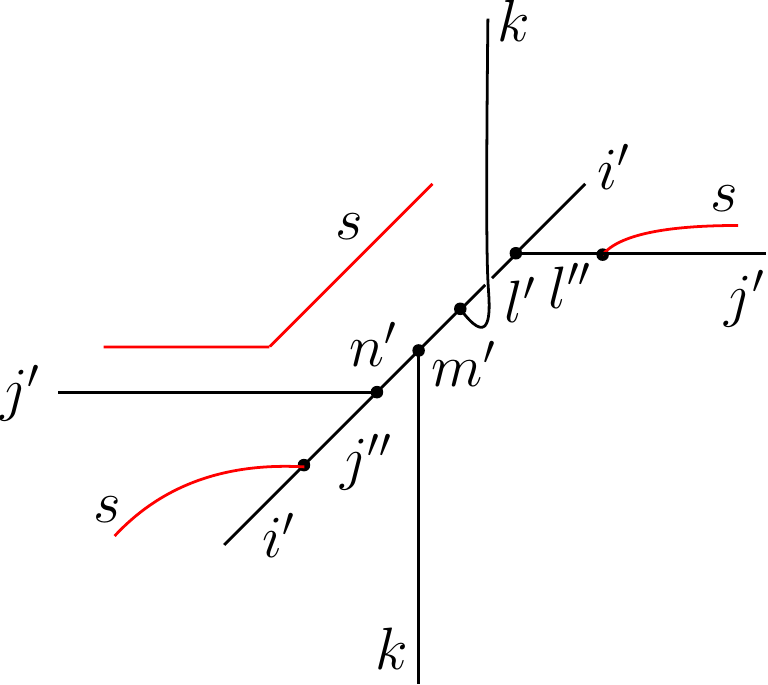}}}\\
&\xrightarrow[F \ \rm move]{\sum_{n^{\prime\prime}}F^{sl^{\prime\prime}j^{\prime}}_{j^{\prime\prime}n^{\prime}n^{\prime\prime}}}\ \vcenter{\hbox{\includegraphics[width=0.45\linewidth]{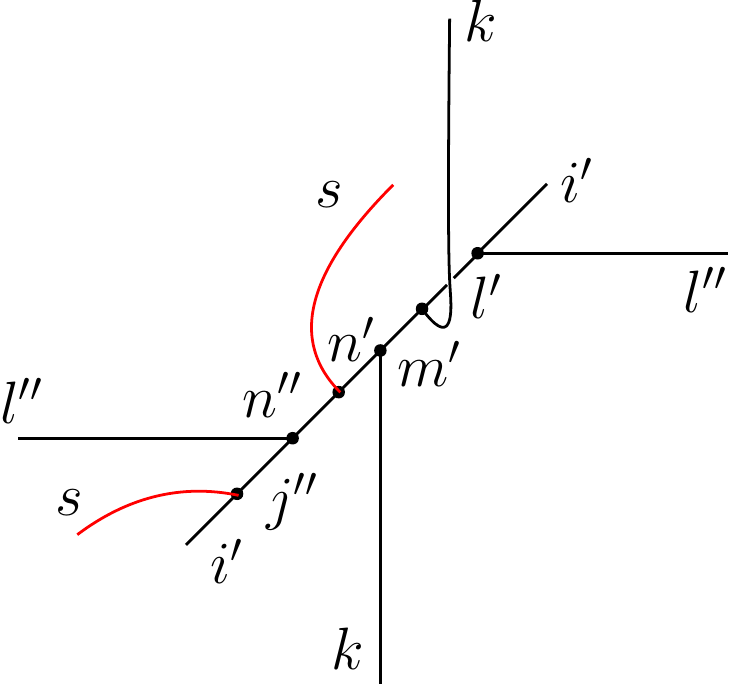}}}\\
&\xrightarrow[F \ \rm move]{\sum_{i^{\prime\prime}}F^{j^{\prime\prime}si^{\prime}}_{l^{\prime}l^{\prime\prime}i^{\prime \prime}}}\ \vcenter{\hbox{\includegraphics[width=0.45\linewidth]{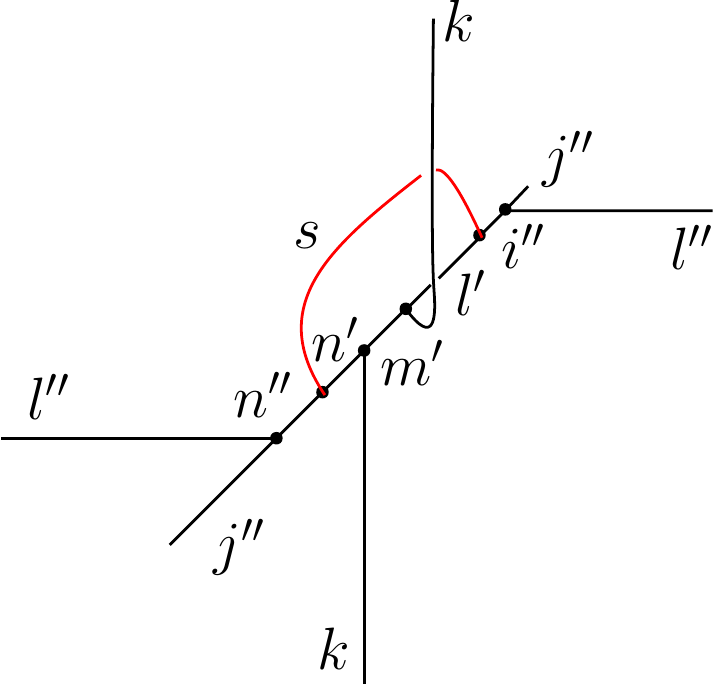}}}\\
&\xrightarrow[F \ \rm move]{\sum_{m^{\prime\prime}}F^{sn^{\prime\prime}n^{\prime}}_{km^{\prime}m^{\prime\prime}}}\ \vcenter{\hbox{\includegraphics[width=0.45\linewidth]{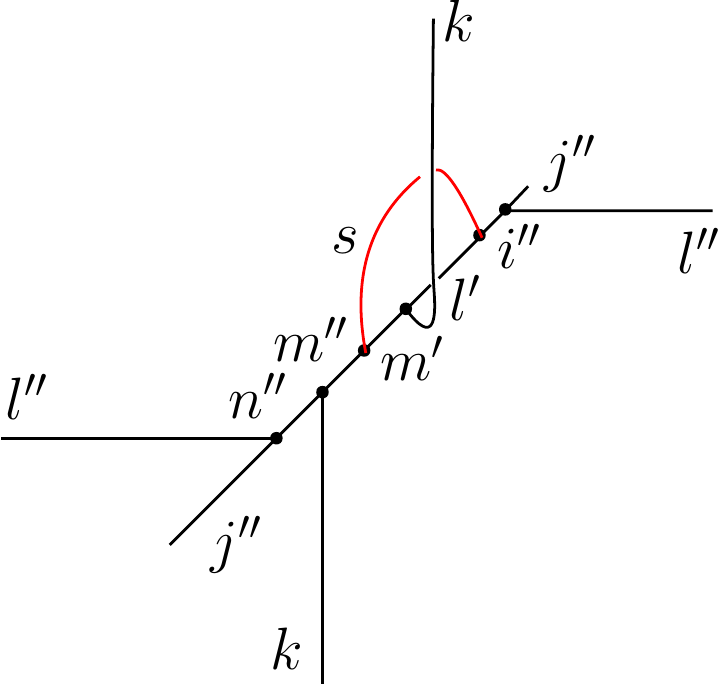}}}\\
&\xrightarrow[F \ \rm move]{\sum_{a}F^{sm^{\prime\prime}m^{\prime}}_{kl^{\prime}a}}\ \vcenter{\hbox{\includegraphics[width=0.45\linewidth]{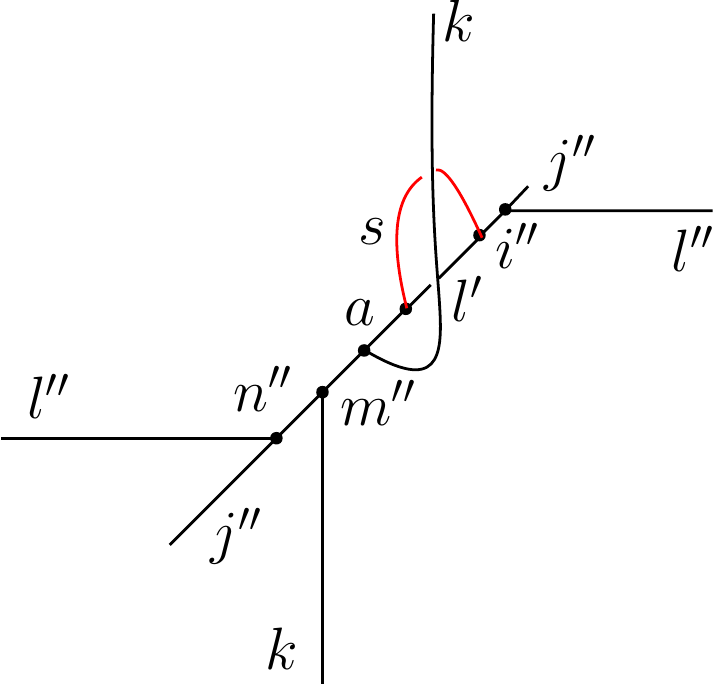}}}\\
&\xrightarrow[\text{Commute $F$ and $R$ moves}]{}\ \vcenter{\hbox{\includegraphics[width=0.45\linewidth]{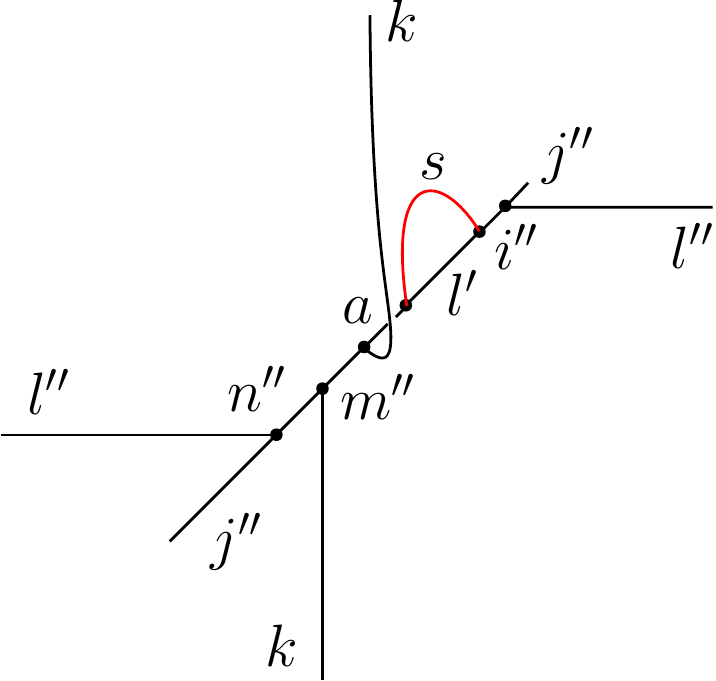}}}\\
&\xrightarrow[R \ \rm move]{(R^{ka}_{m^{\prime\prime}})^*}\ \vcenter{\hbox{\includegraphics[width=0.45\linewidth]{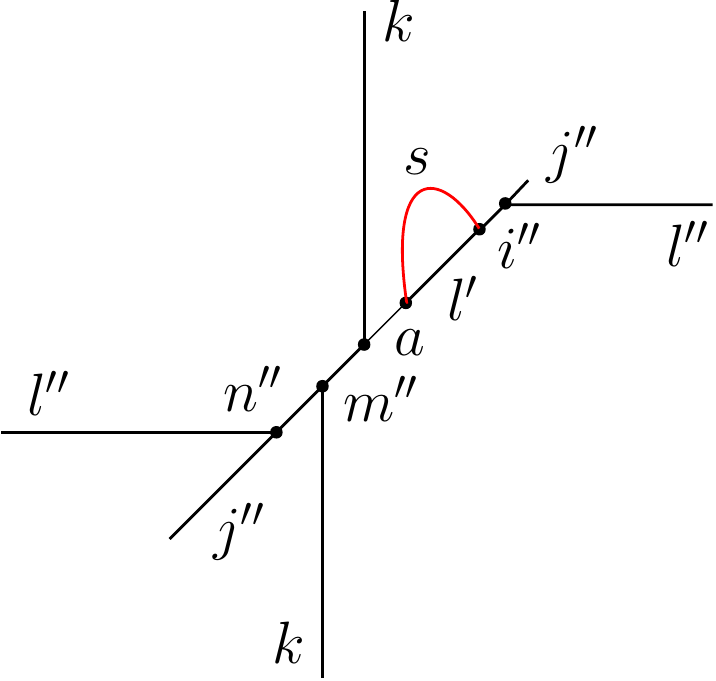}}}\\
&\xrightarrow[\text{squeezing bubbles}]{F^{si^{\prime\prime}l^{\prime}}_{as0}d_s\delta_{ai^{\prime\prime}}}\ \vcenter{\hbox{\includegraphics[width=0.45\linewidth]{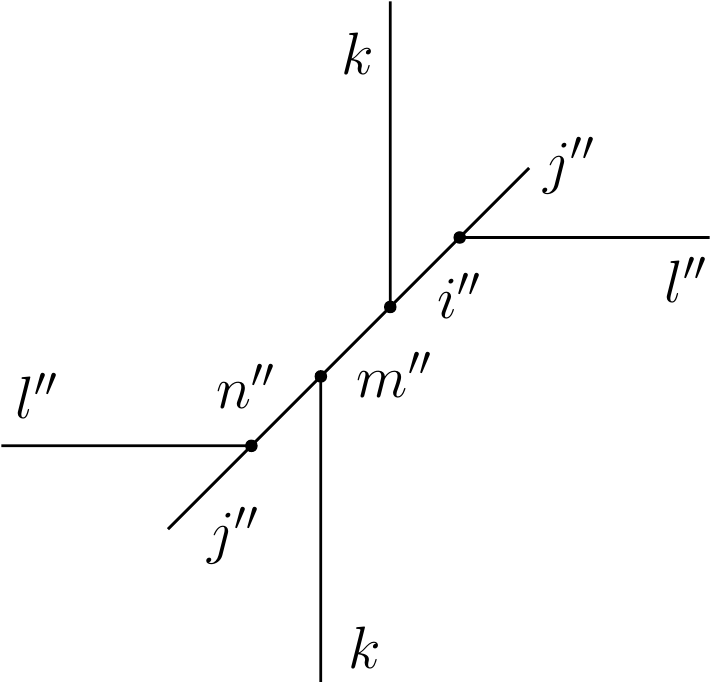}}}
\end{align*}

Collecting the coefficients from each step, we obtain 
\begin{align}
&(B^s_{xy})^{(j^{\prime\prime},l^{\prime\prime},k,i^{\prime\prime},m^{\prime\prime},n^{\prime\prime})}_{(i,j,k,l,m,n)}=\sum_{j^{\prime},l^{\prime},m^{\prime},n^{\prime},i^{\prime}}(R^{nk}_{m})^*F^{jj0}_{ssj^{\prime}}
\times \nonumber \\
&F^{ilj}_{sj^{\prime}l^{\prime}}F^{kml}_{sl^{\prime}m^{\prime}}F^{knm}_{sm^{\prime}n^{\prime}}R^{n^{\prime}k}_{m^{\prime}}R^{kl^{\prime}}_{m^{\prime}}F^{jin}_{sn^{\prime}i^{\prime}}F^{l^{\prime}j^{\prime}i}_{si^{\prime}l^{\prime\prime}}F^{j^{\prime}sj}_{i^{\prime}n^{\prime}j^{\prime\prime}} F^{sl^{\prime\prime}j^{\prime}}_{j^{\prime\prime}n^{\prime}n^{\prime\prime}}\times \nonumber \\
&F^{j^{\prime\prime}si^{\prime}}_{l^{\prime}l^{\prime\prime}i^{\prime \prime}}F^{sn^{\prime\prime}n^{\prime}}_{km^{\prime}m^{\prime\prime}}F^{sm^{\prime\prime}m^{\prime}}_{kl^{\prime}i^{\prime\prime}}(R^{ki^{\prime\prime}}_{m^{\prime\prime}})^*F^{si^{\prime\prime}l^{\prime}}_{i^{\prime\prime}s0}d_s.
\end{align}

\subsection{Plaquette operator in the $xz$-plane}
\begin{align*}
&\vcenter{\hbox{\includegraphics[width=0.45\linewidth]{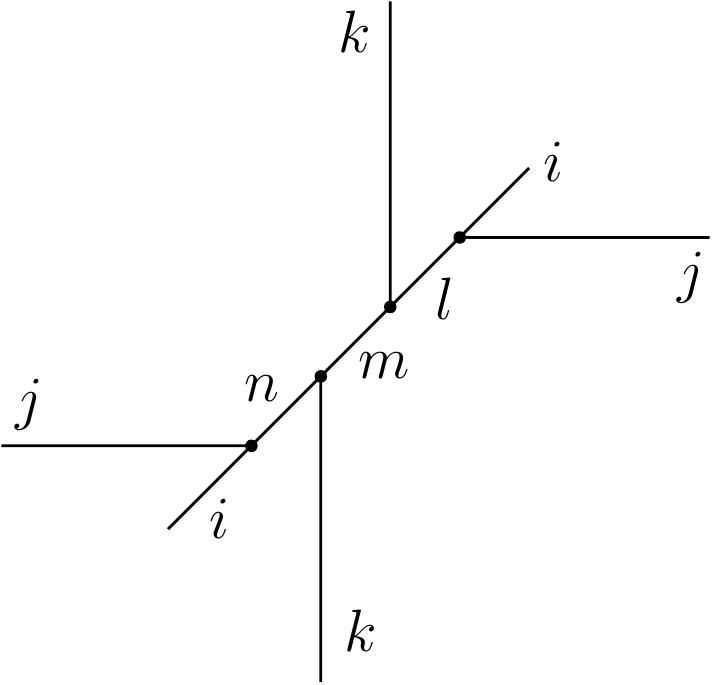}}}\xrightarrow[R\ \rm move]{(R^{ji}_{l})^*}\vcenter{\hbox{\includegraphics[width=0.45\linewidth]{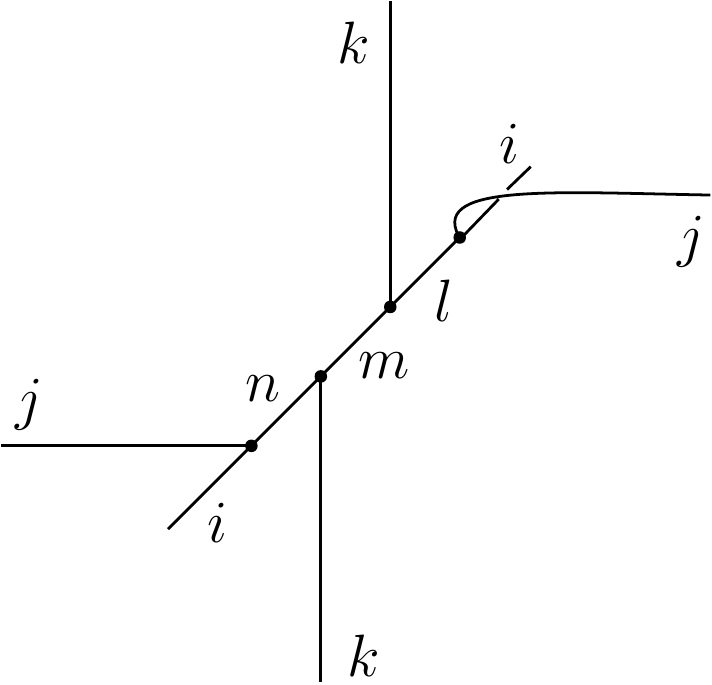}}}\\
&\xrightarrow[\text{add an $s$-loop in the $xz$-plane}]{} \ \vcenter{\hbox{\includegraphics[width=0.45\linewidth]{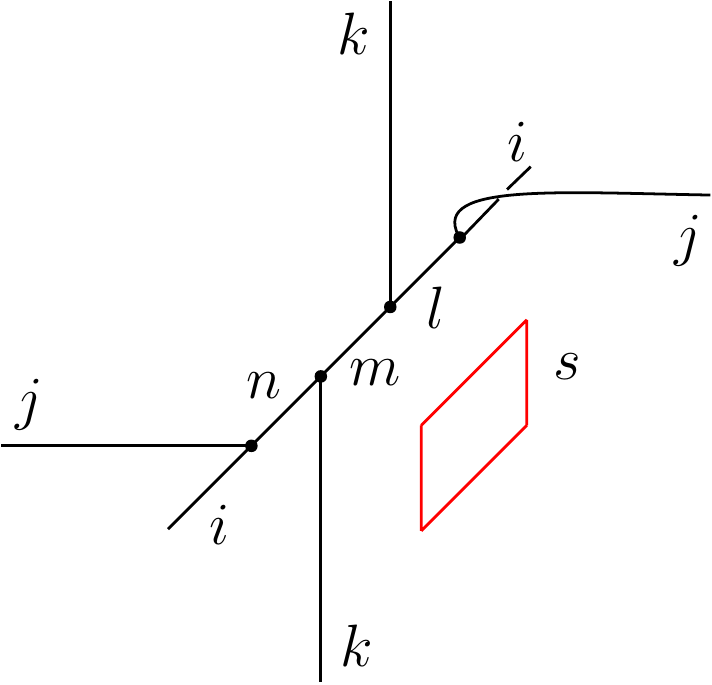}}}\\
&\xrightarrow[\rm Deformation]{}\ \vcenter{\hbox{\includegraphics[width=0.45\linewidth]{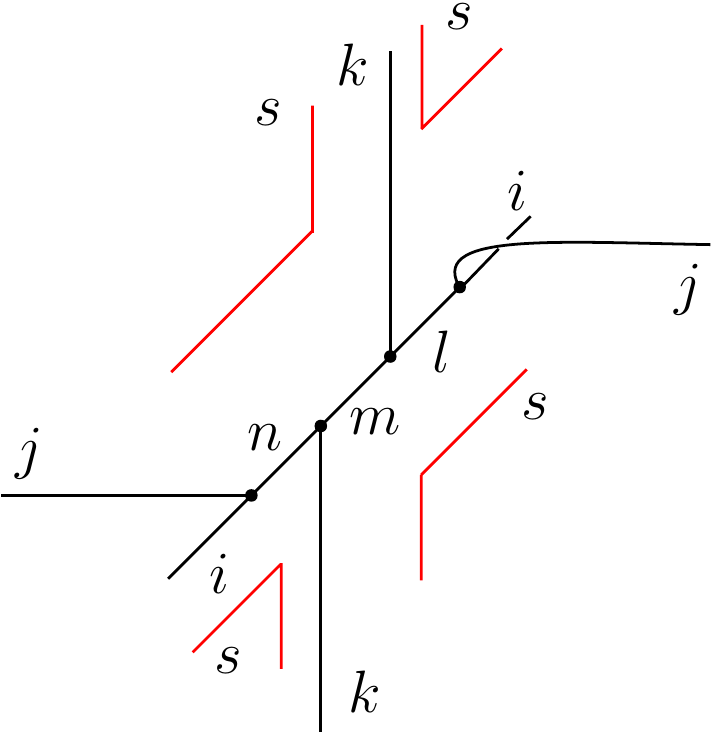}}}\\
&\xrightarrow[F \ \rm move]{\sum_{k^{\prime}}F^{kk0}_{ssk^{\prime}}}\ \vcenter{\hbox{\includegraphics[width=0.45\linewidth]{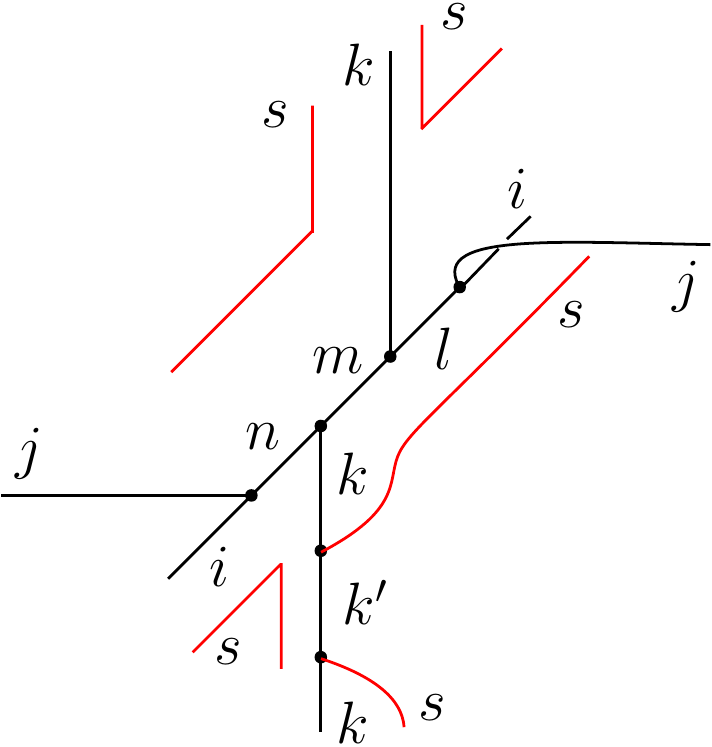}}}\\
&\xrightarrow[F \ \rm move]{\sum_{m^{\prime}}F^{k^{\prime}sk}_{mnm^{\prime}}}\ \vcenter{\hbox{\includegraphics[width=0.45\linewidth]{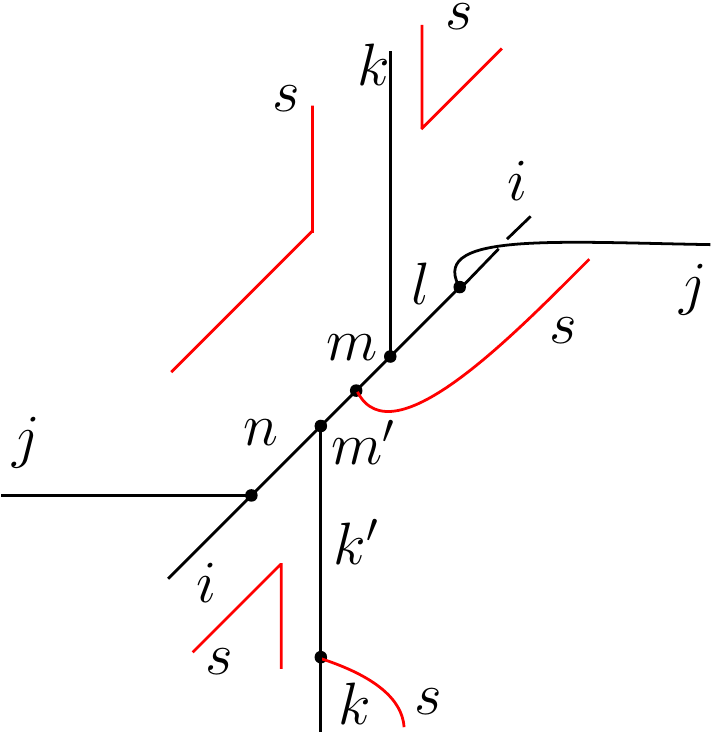}}}\\
&\xrightarrow[F \ \rm move]{\sum_{l^{\prime}}F^{m^{\prime}sm}_{lkl^{\prime}}}\ \vcenter{\hbox{\includegraphics[width=0.45\linewidth]{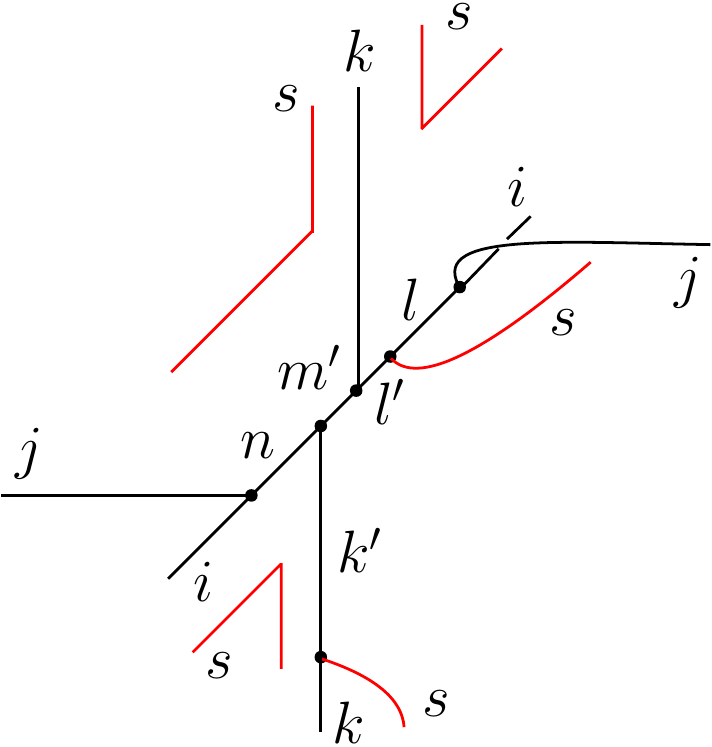}}}\\
&\xrightarrow[F\ \rm move]{\sum_{i^{\prime}}F^{l^{\prime}sl}_{iji^{\prime}}}\ \vcenter{\hbox{\includegraphics[width=0.45\linewidth]{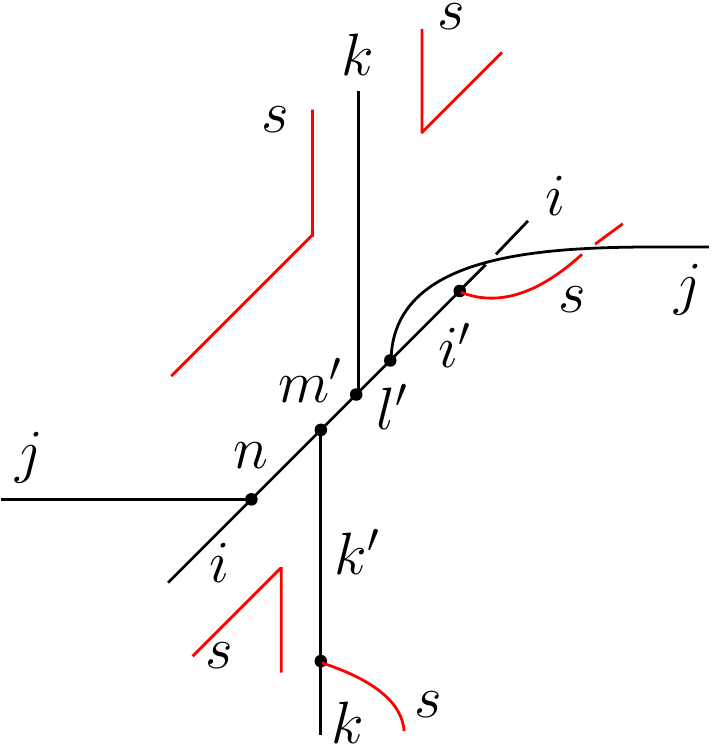}}}\\
&\xrightarrow[\text{Commute $F$ and $R$ moves}]{}\ \vcenter{\hbox{\includegraphics[width=0.45\linewidth]{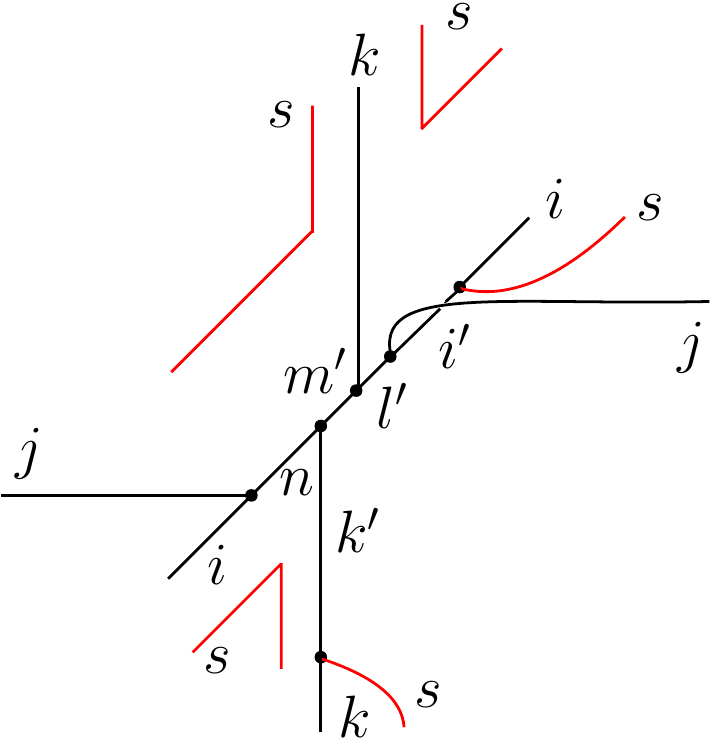}}}\\
&\xrightarrow[R \ \rm move]{R^{ji^{\prime}}_{l^{\prime}}}\ \vcenter{\hbox{\includegraphics[width=0.45\linewidth]{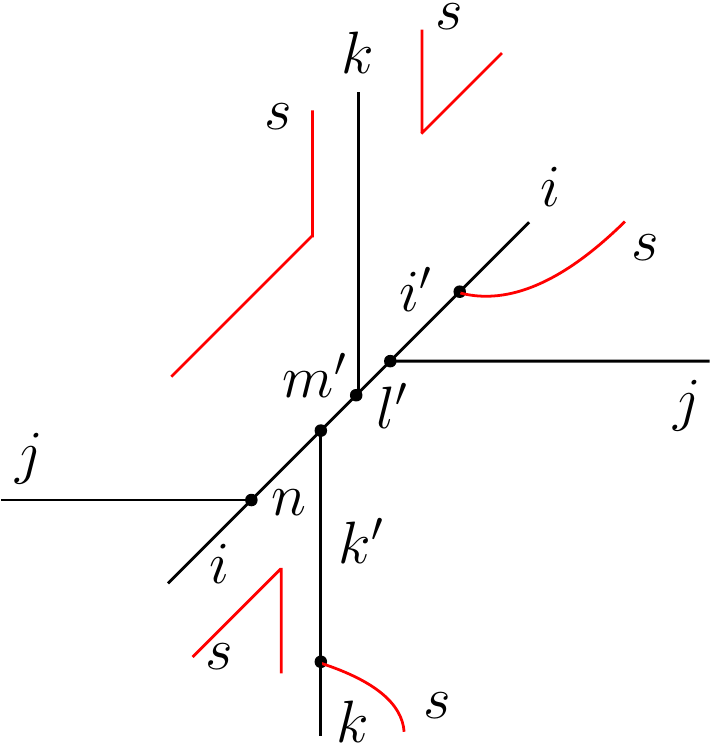}}}\\
&\xrightarrow[F \ \rm move]{\sum_{n^{\prime}}F^{i^{\prime}si}_{njn^{\prime}}}\ \vcenter{\hbox{\includegraphics[width=0.45\linewidth]{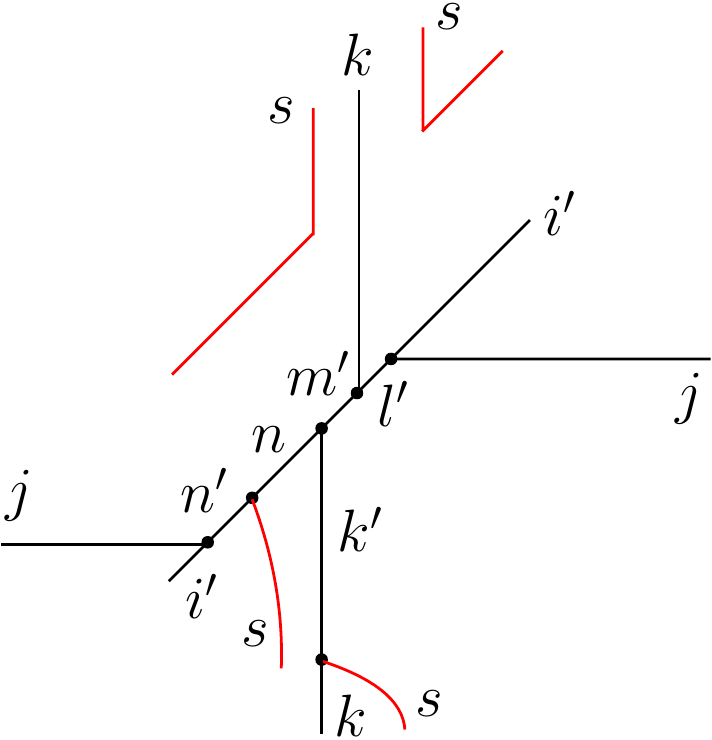}}}\\
&\xrightarrow[F \ \rm move]{\sum_{k^{\prime\prime}}F^{n^{\prime}sn}_{k^{\prime}m^{\prime}k^{\prime\prime}}}\ \vcenter{\hbox{\includegraphics[width=0.45\linewidth]{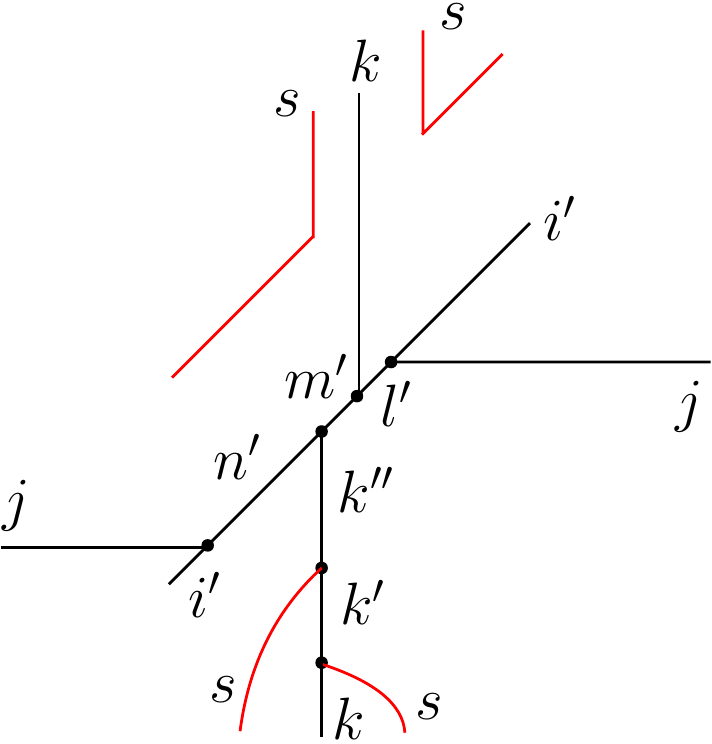}}}\\
&\xrightarrow[R \ \rm move]{(R^{n^{\prime}j}_{i^{\prime}})^*}\ \vcenter{\hbox{\includegraphics[width=0.45\linewidth]{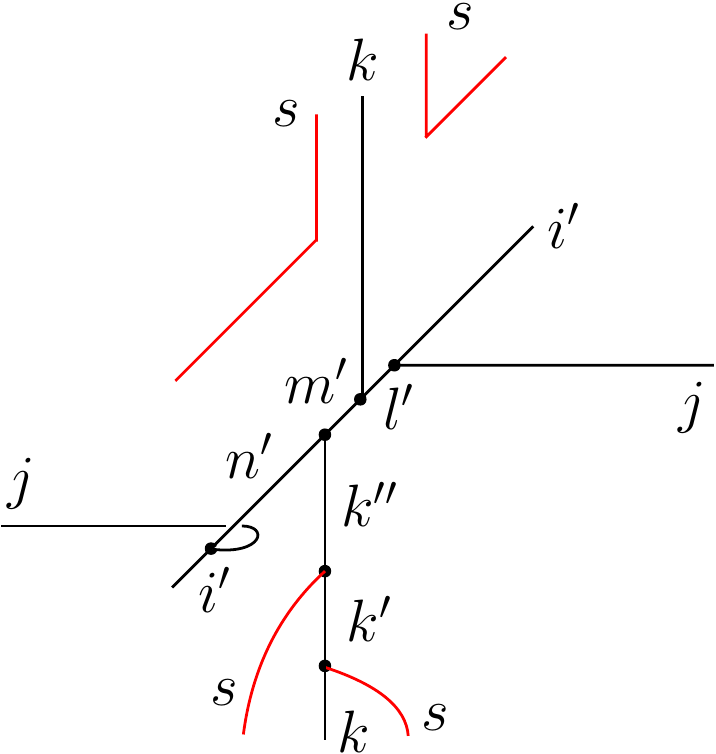}}}\\
&\xrightarrow[F \ \rm move]{\sum_{l^{\prime\prime}}F^{m^{\prime}l^{\prime}k}_{sk^{\prime}l^{\prime\prime}}}\ \vcenter{\hbox{\includegraphics[width=0.45\linewidth]{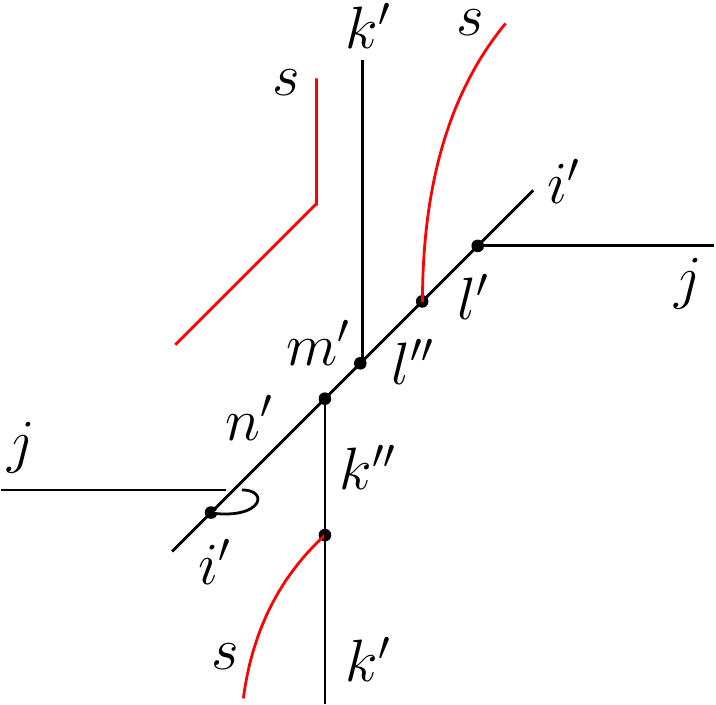}}}\\
&\xrightarrow[F \ \rm move]{\sum_{i^{\prime\prime}}F^{sl^{\prime\prime}l^{\prime}}_{ji^{\prime}i^{\prime\prime}}}\ \vcenter{\hbox{\includegraphics[width=0.45\linewidth]{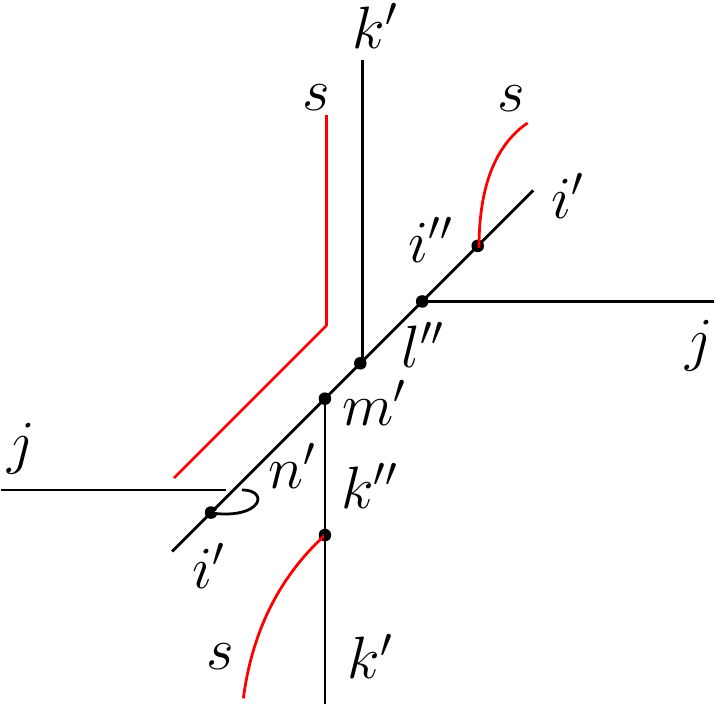}}}\\
&\xrightarrow[F \ \rm move]{\sum_{n^{\prime\prime}}F^{si^{\prime\prime}i^{\prime}}_{jn^{\prime}n^{\prime\prime}}}\ \vcenter{\hbox{\includegraphics[width=0.45\linewidth]{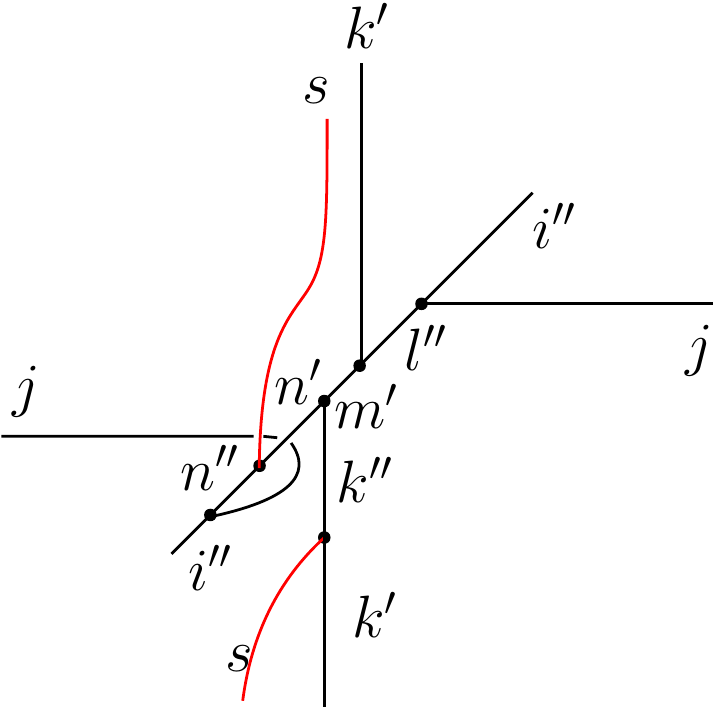}}}\\
&\xrightarrow[\text{Commute $F$ and $R$ moves}]{}\ \vcenter{\hbox{\includegraphics[width=0.45\linewidth]{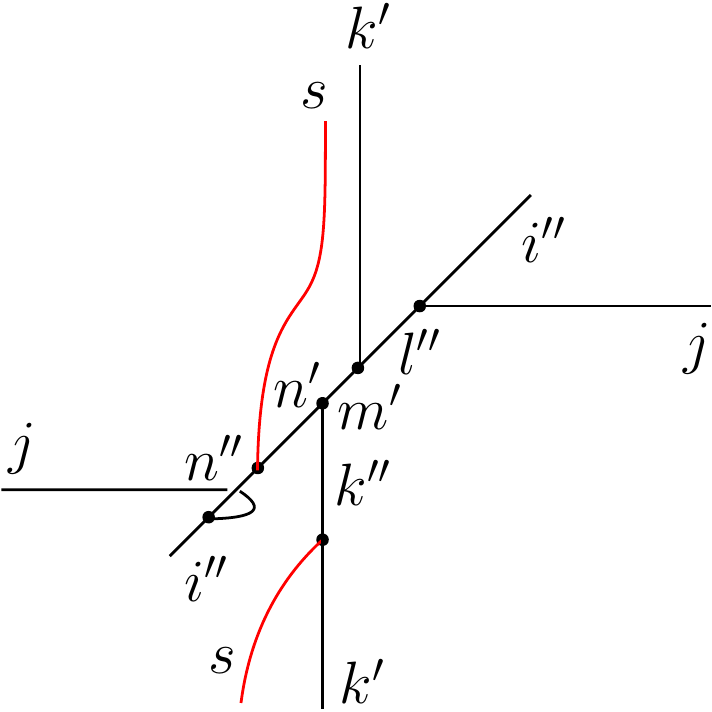}}}\\
&\xrightarrow[R \ \rm move]{R^{n^{\prime\prime}j}_{i^{\prime\prime}}}\ \vcenter{\hbox{\includegraphics[width=0.45\linewidth]{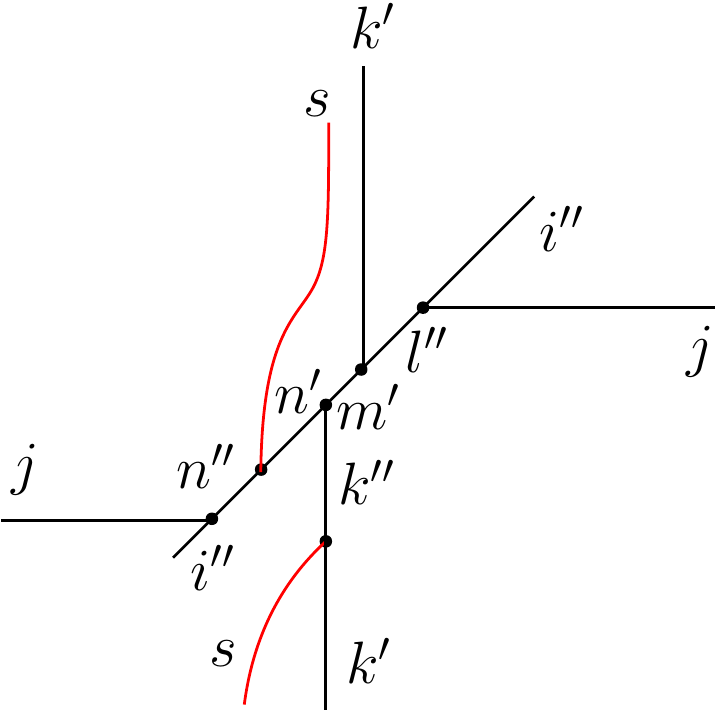}}}\\
&\xrightarrow[F \ \rm move]{\sum_{m^{\prime\prime}}F^{k^{\prime\prime}sk^{\prime}}_{m^{\prime}l^{\prime\prime}m^{\prime\prime}}}\ \vcenter{\hbox{\includegraphics[width=0.45\linewidth]{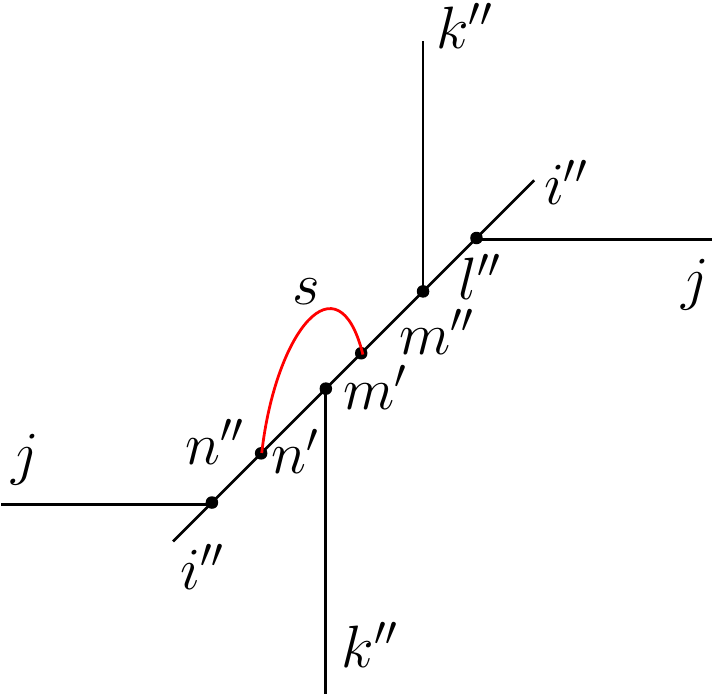}}}\\
&\xrightarrow[F \ \rm move]{\sum_{a}F^{n^{\prime}k^{\prime\prime}m^{\prime}}_{m^{\prime\prime}sa}}\ \vcenter{\hbox{\includegraphics[width=0.45\linewidth]{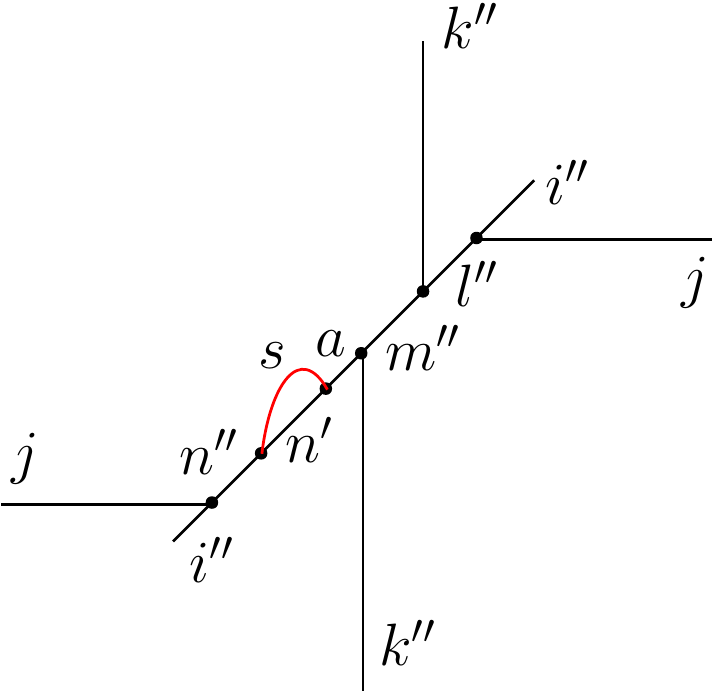}}}\\
&\xrightarrow[\text{squeezing bubbles}]{F^{san^{\prime}}_{n^{\prime\prime}s0}d_s\delta_{an^{\prime\prime}}}\ \vcenter{\hbox{\includegraphics[width=0.45\linewidth]{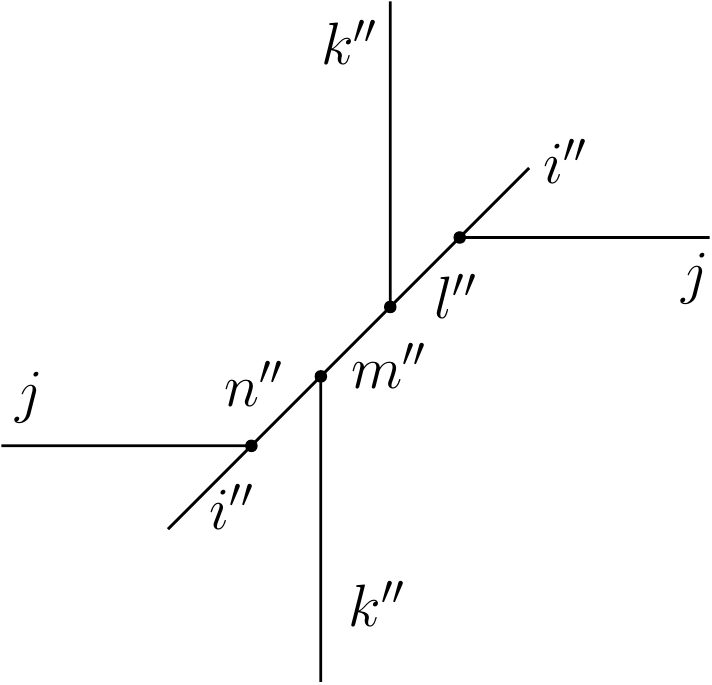}}}
\end{align*}

Collecting the coefficients from each step, we obtain 
\begin{align}
&(B^s_{xz})^{(i^{\prime\prime},j,k^{\prime\prime},l^{\prime\prime},m^{\prime\prime},n^{\prime\prime})}_{(i,j,k,l,m,n)}=
\sum_{k^{\prime},m^{\prime},l^{\prime},i^{\prime},n^{\prime}}(R^{ji}_{l})^*
F^{kk0}_{ssk^{\prime}} \times \nonumber \\
&F^{k^{\prime}sk}_{mnm^{\prime}}
F^{m^{\prime}sm}_{lkl^{\prime}}
F^{l^{\prime}sl}_{iji^{\prime}}
R^{ji^{\prime}}_{l^{\prime}}
F^{i^{\prime}si}_{njn^{\prime}}
F^{n^{\prime}sn}_{k^{\prime}m^{\prime}k^{\prime\prime}}
(R^{n^{\prime}j}_{i^{\prime}})^* F^{m^{\prime}l^{\prime}k}_{sk^{\prime}l^{\prime\prime}}
F^{sl^{\prime\prime}l^{\prime}}_{ji^{\prime}i^{\prime\prime}}\times \nonumber \\
&F^{si^{\prime\prime}i^{\prime}}_{jn^{\prime}n^{\prime\prime}}
R^{n^{\prime\prime}j}_{i^{\prime\prime}}
F^{k^{\prime\prime}sk^{\prime}}_{m^{\prime}l^{\prime\prime}m^{\prime\prime}}
F^{n^{\prime}k^{\prime\prime}m^{\prime}}_{m^{\prime\prime}sn^{\prime\prime}}
F^{sn^{\prime\prime}n^{\prime}}_{n^{\prime\prime}s0}d_s.
\end{align}

\subsection{Plaquette operator in the $yz$-plane}
\begin{align*}
&\vcenter{\hbox{\includegraphics[width=0.45\linewidth]{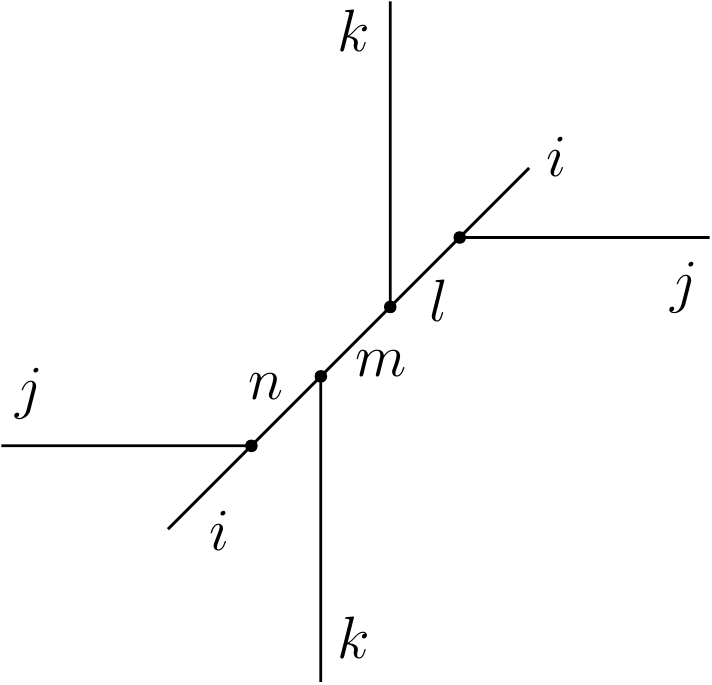}}}\xrightarrow[R\ \rm move]{(R^{ji}_{l})^*}\vcenter{\hbox{\includegraphics[width=0.45\linewidth]{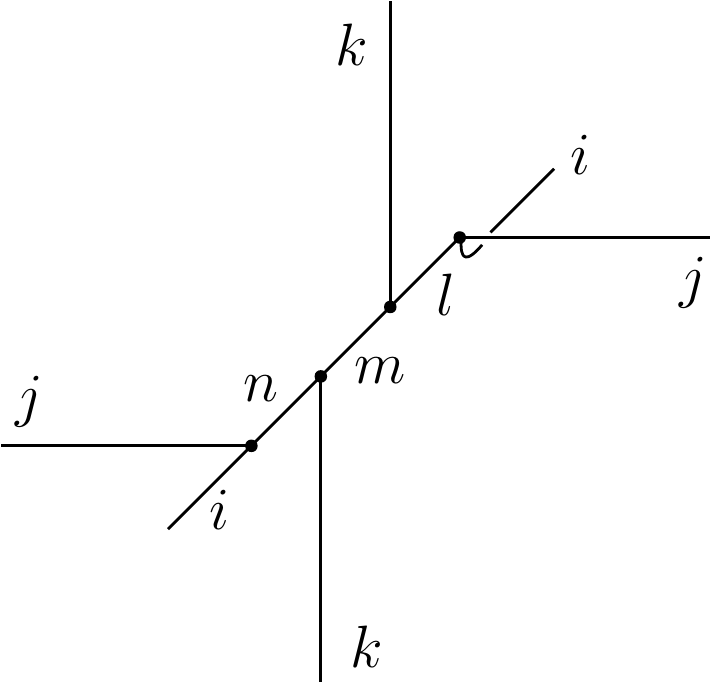}}}\\
&\xrightarrow[\text{add an $s$-loop in the $yz$-plane}]{} \ \vcenter{\hbox{\includegraphics[width=0.45\linewidth]{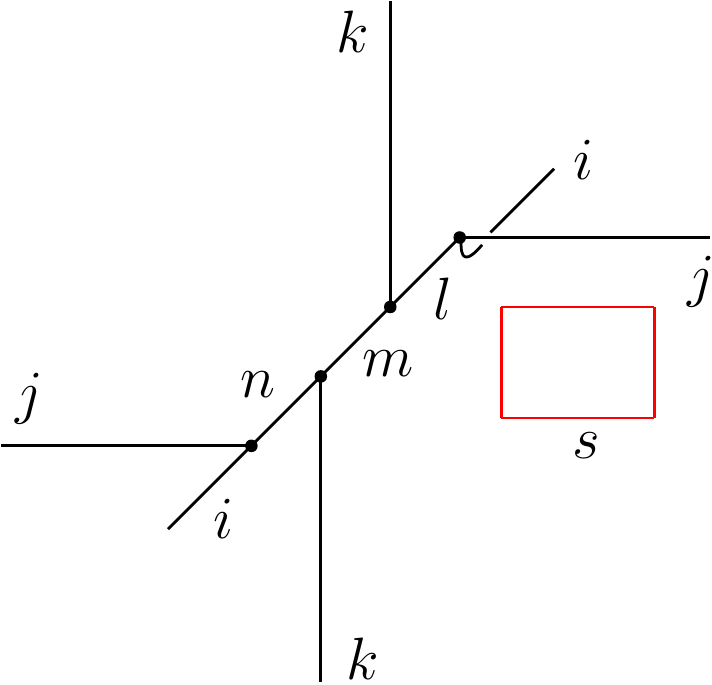}}}\\
&\xrightarrow[\rm Deformation]{}\ \vcenter{\hbox{\includegraphics[width=0.45\linewidth]{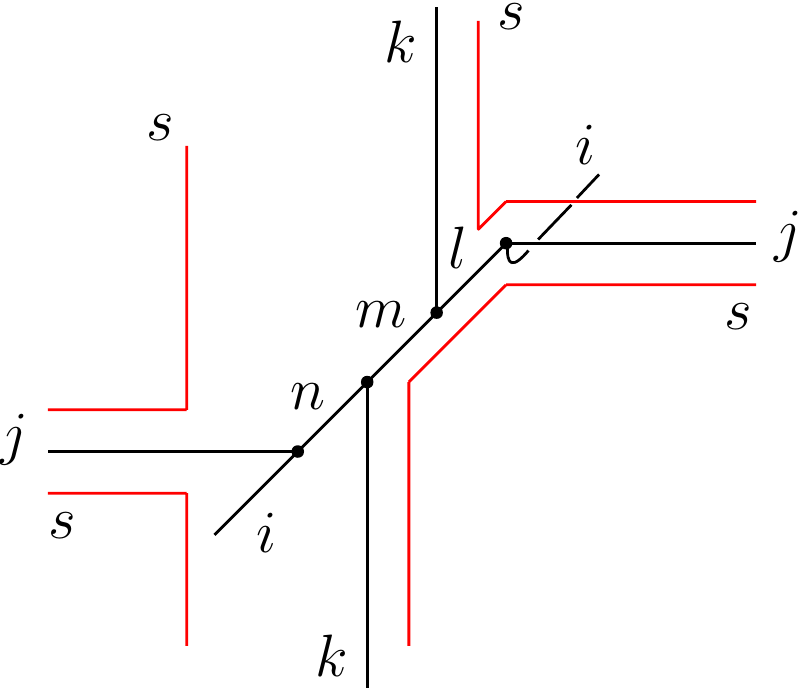}}}\\
&\xrightarrow[F \ \rm move]{\sum_{k^{\prime}}F^{kk0}_{ssk^{\prime}}}\ \vcenter{\hbox{\includegraphics[width=0.45\linewidth]{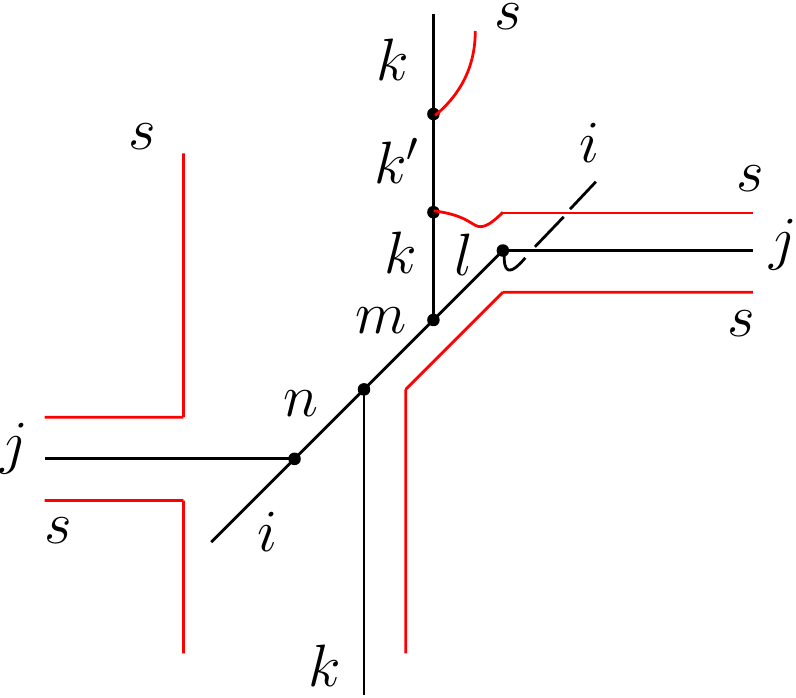}}}\\
&\xrightarrow[F \ \rm move]{\sum_{l^{\prime}}F^{mlk}_{sk^{\prime}l^{\prime}}}\ \vcenter{\hbox{\includegraphics[width=0.45\linewidth]{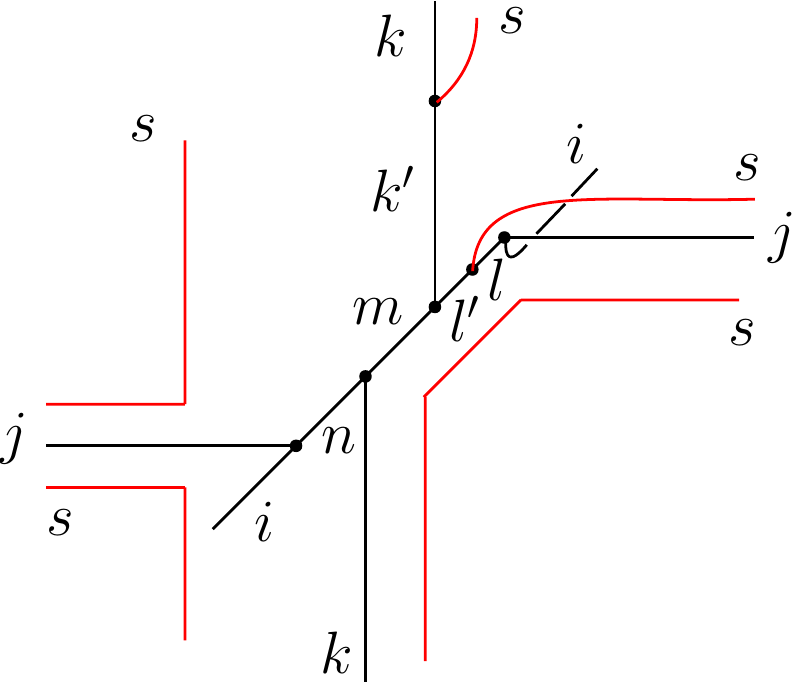}}}\\
&\xrightarrow[F \ \rm move]{\sum_{j^{\prime}}F^{sl^{\prime}l}_{ijj^{\prime}}}\ \vcenter{\hbox{\includegraphics[width=0.45\linewidth]{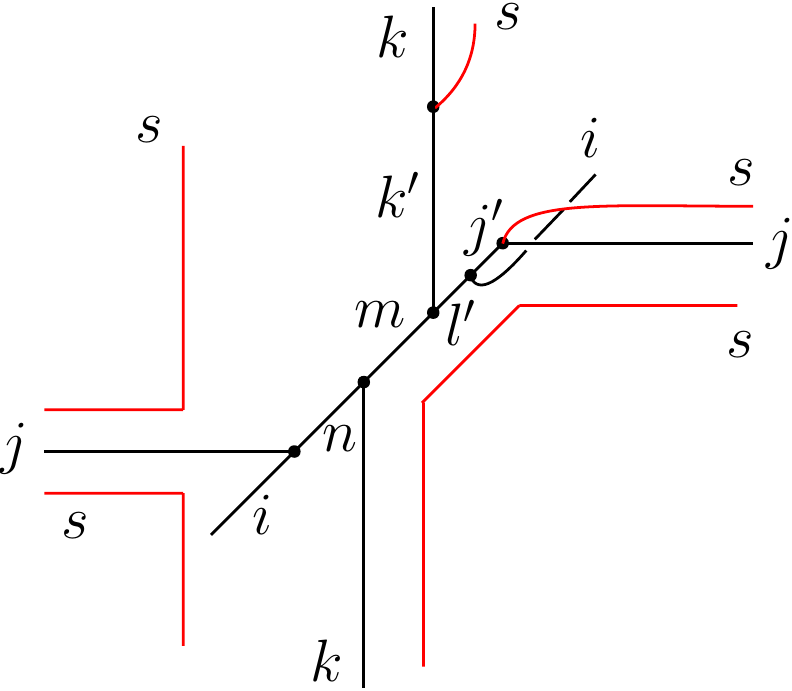}}}\\
&\xrightarrow[\text{Commute $F$ and $R$ moves}]{}\ \vcenter{\hbox{\includegraphics[width=0.45\linewidth]{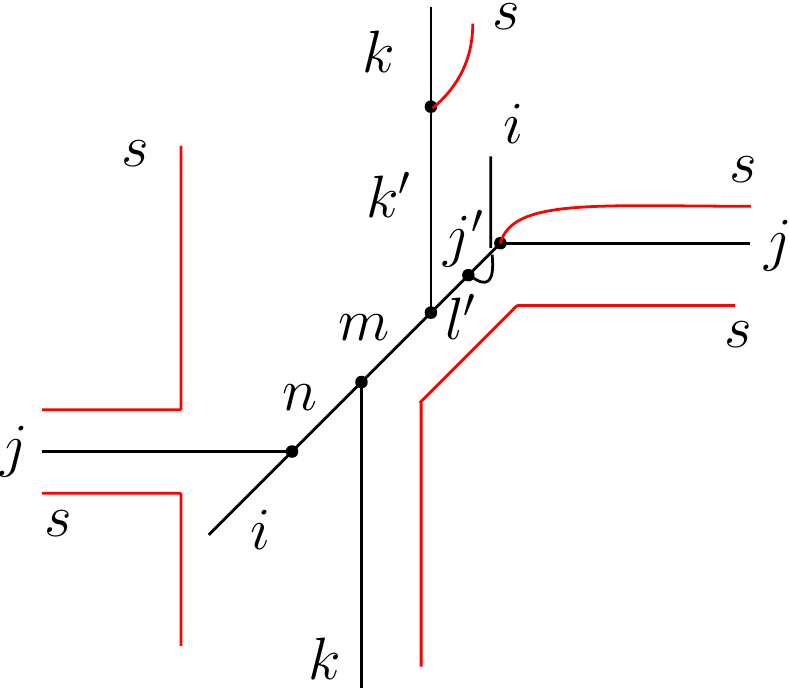}}}\\
&\xrightarrow[R\ \rm move]{R^{j^{\prime}i}_{l^{\prime}}}\ \vcenter{\hbox{\includegraphics[width=0.45\linewidth]{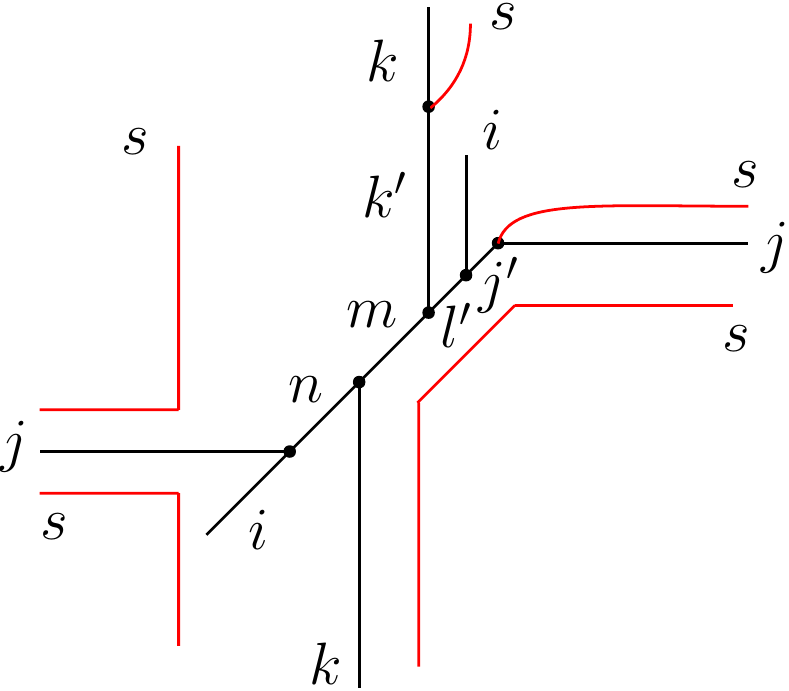}}}\\
&\xrightarrow[F \ \rm move]{\sum_{n^{\prime}}F^{sj^{\prime}j}_{inn^{\prime}}}\ \vcenter{\hbox{\includegraphics[width=0.45\linewidth]{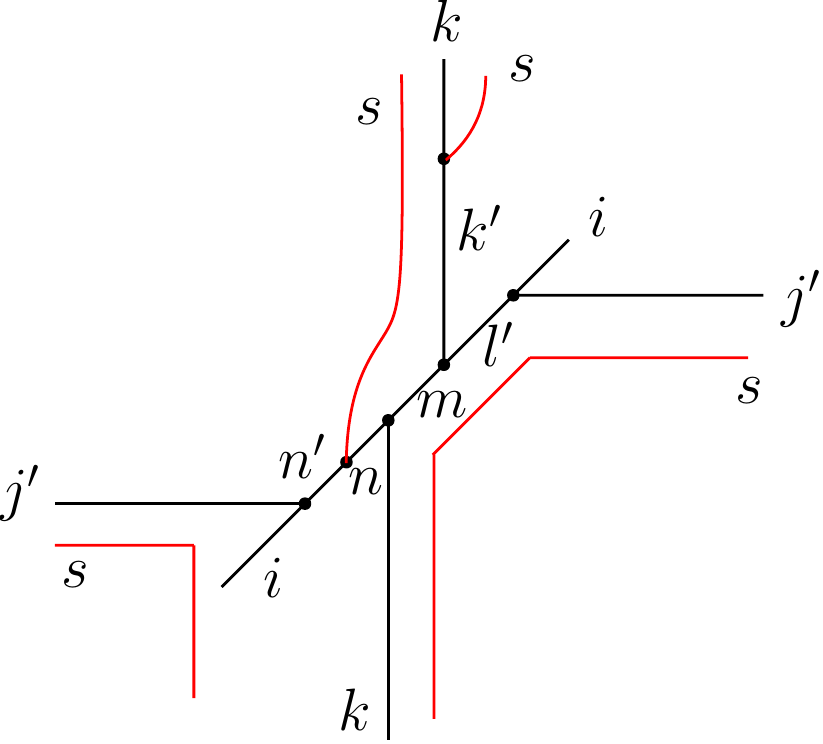}}}\\
&\xrightarrow[R \ \rm move]{R^{ij^{\prime}}_{n^{\prime}}}\ \vcenter{\hbox{\includegraphics[width=0.45\linewidth]{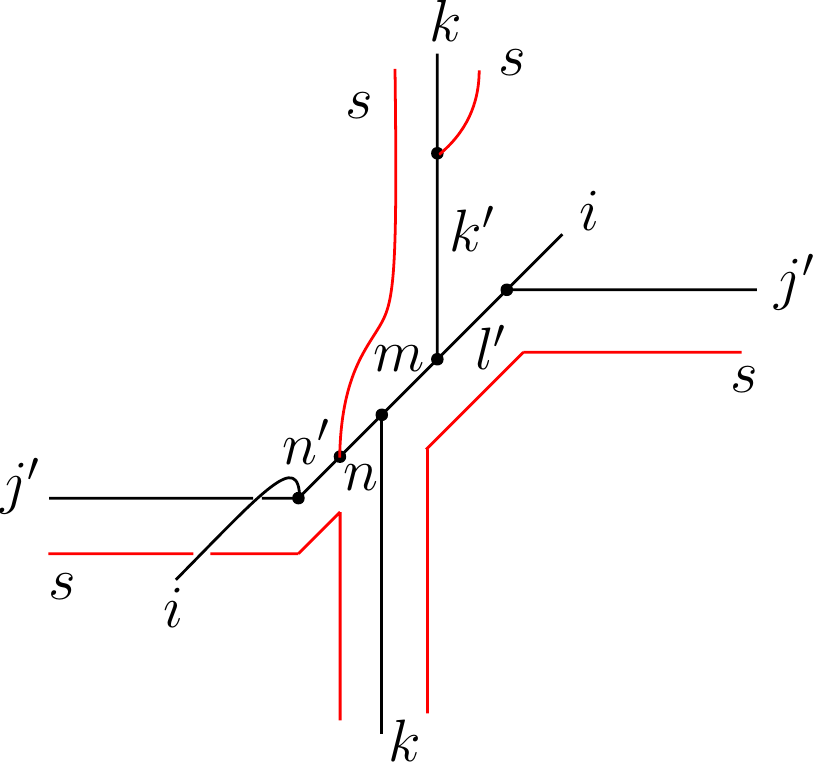}}}\\
&\xrightarrow[F \ \rm move]{\sum_{m^{\prime}}F^{sn^{\prime}n}_{kmm^{\prime}}}\ \vcenter{\hbox{\includegraphics[width=0.45\linewidth]{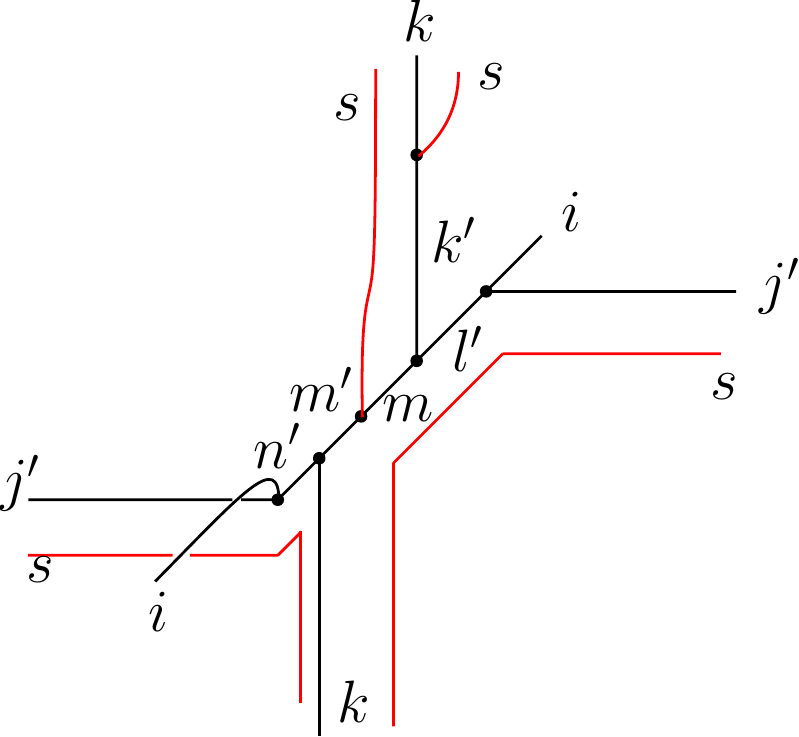}}}\\
&\xrightarrow[F \ \rm move]{\sum_{k^{\prime\prime}}F^{sm^{\prime}m}_{l^{\prime}k^{\prime}k^{\prime\prime}}}\ \vcenter{\hbox{\includegraphics[width=0.45\linewidth]{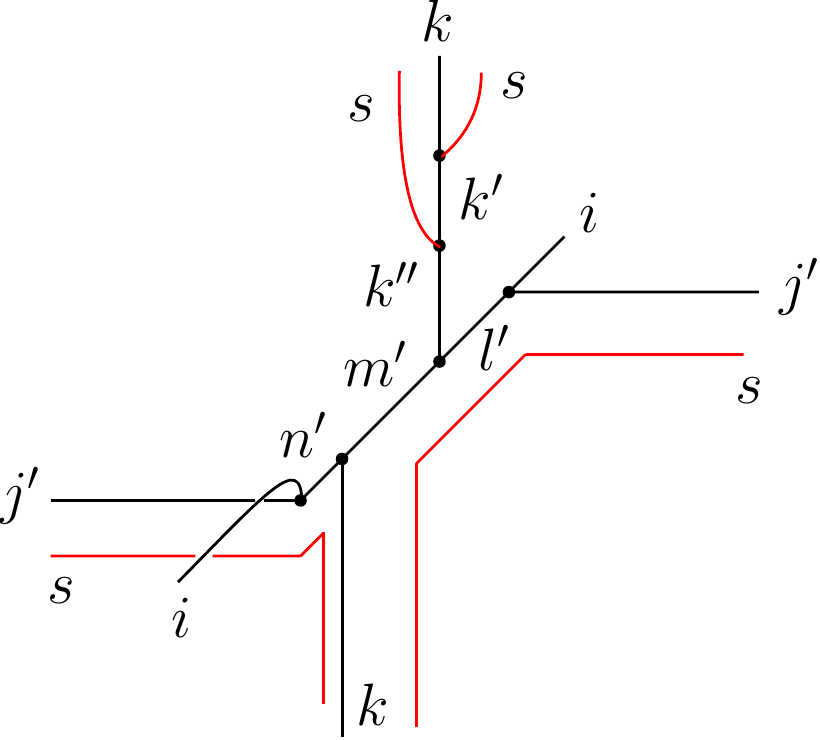}}}\\
&\xrightarrow[F \ \rm move]{\sum_{m^{\prime\prime}}F^{k^{\prime}sk}_{m^{\prime}n^{\prime}m^{\prime\prime}}}\ \vcenter{\hbox{\includegraphics[width=0.45\linewidth]{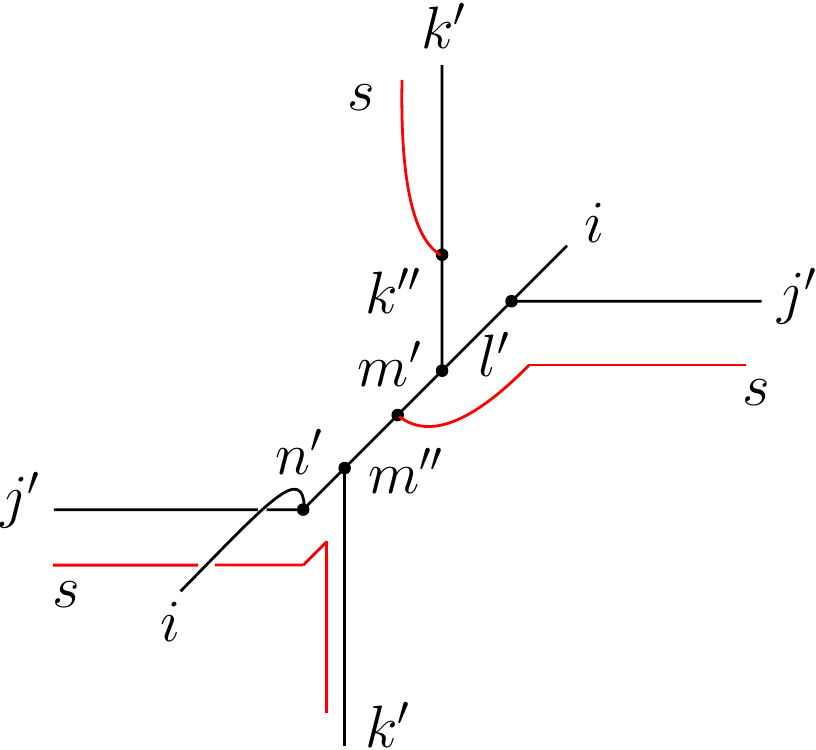}}}\\
&\xrightarrow[F \ \rm move]{\sum_{l^{\prime\prime}}F^{m^{\prime\prime}sm^{\prime}}_{l^{\prime}k^{\prime\prime}l^{\prime\prime}}}\ \vcenter{\hbox{\includegraphics[width=0.45\linewidth]{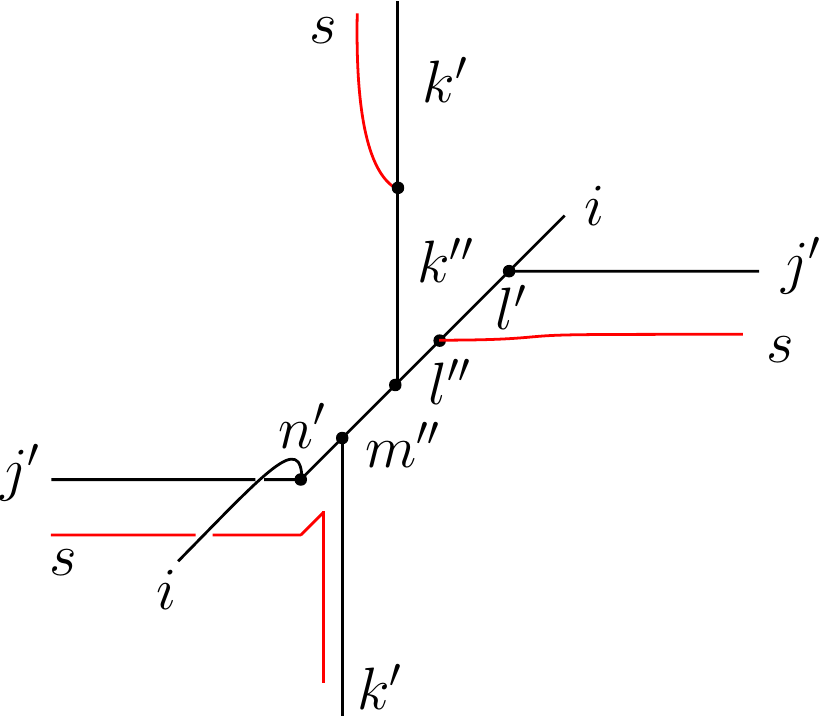}}}\\
&\xrightarrow[F \ \rm move]{\sum_{j^{\prime\prime}}F^{l^{\prime\prime}sl^{\prime}}_{j^{\prime}ij^{\prime\prime}}}\ \vcenter{\hbox{\includegraphics[width=0.45\linewidth]{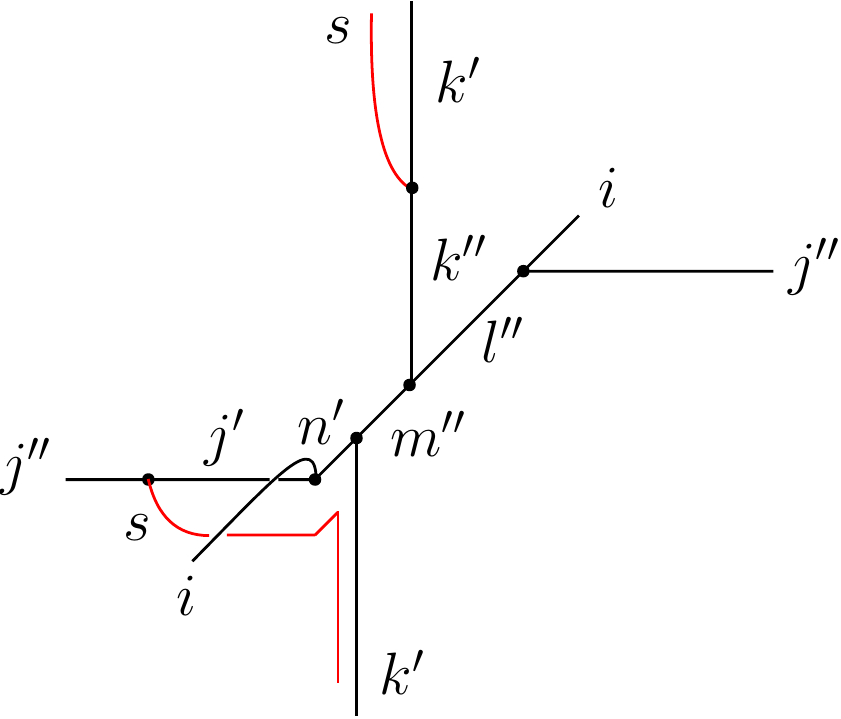}}}\\
&\xrightarrow[\text{Commute $F$ and $R$ moves}]{}\ \vcenter{\hbox{\includegraphics[width=0.45\linewidth]{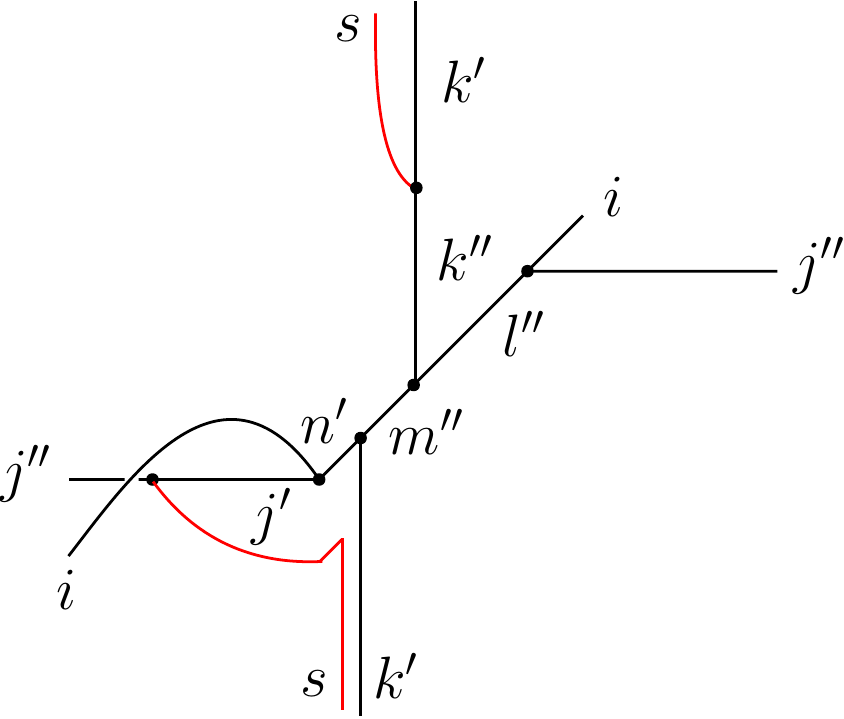}}}\\
&\xrightarrow[F \ \rm move]{\sum_{n^{\prime\prime}}F^{j^{\prime\prime}sj^{\prime}}_{n^{\prime}in^{\prime\prime}}}\ \vcenter{\hbox{\includegraphics[width=0.45\linewidth]{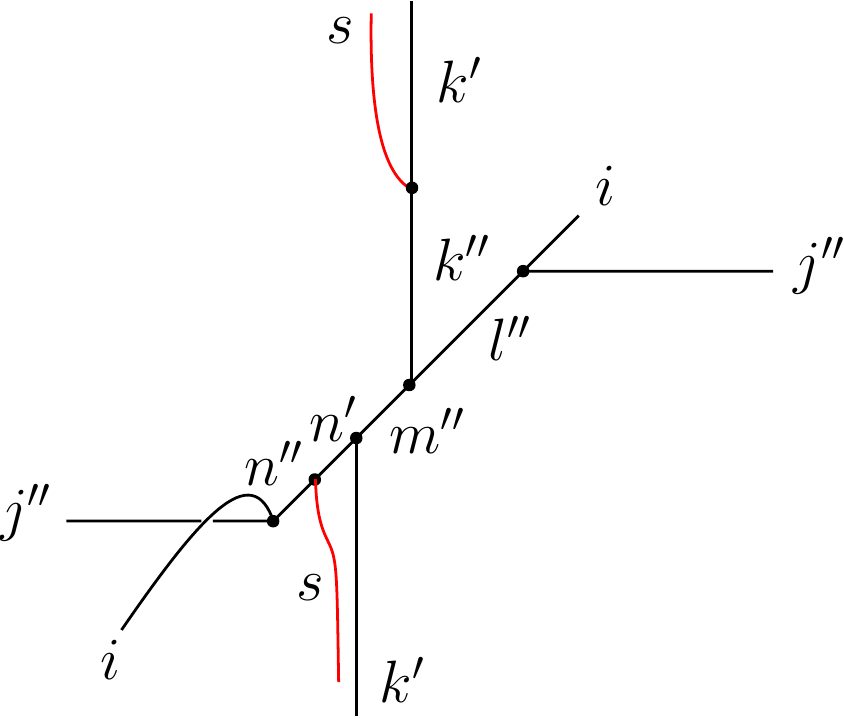}}}\\
&\xrightarrow[R \ \rm move]{(R^{ij^{\prime\prime}}_{n^{\prime\prime}})^*}\ \vcenter{\hbox{\includegraphics[width=0.45\linewidth]{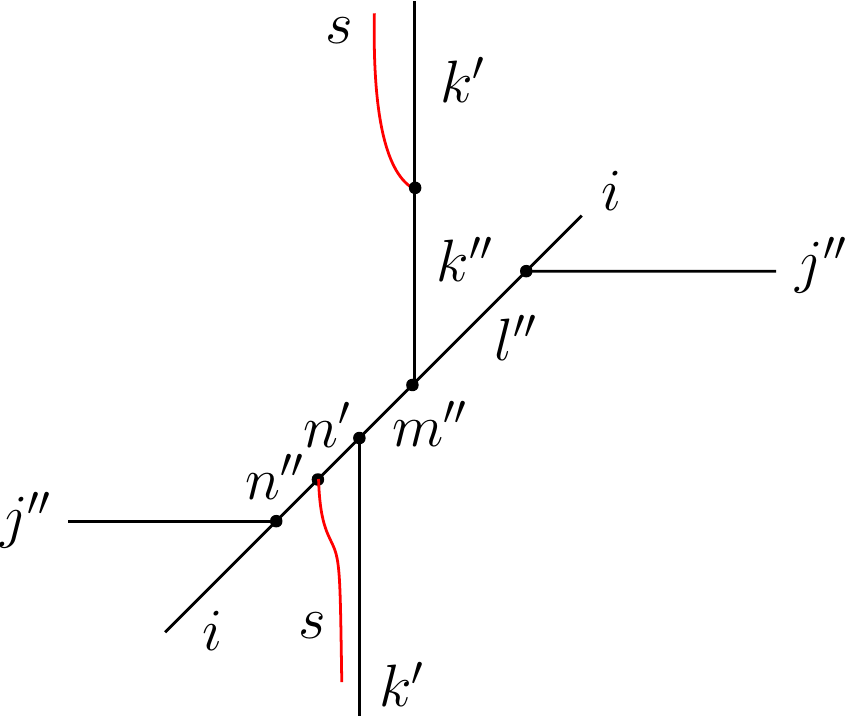}}}\\
&\xrightarrow[F \ \rm move]{\sum_{a}F^{n^{\prime\prime}sn^{\prime}}_{k^{\prime}m^{\prime\prime}a}}\ \vcenter{\hbox{\includegraphics[width=0.45\linewidth]{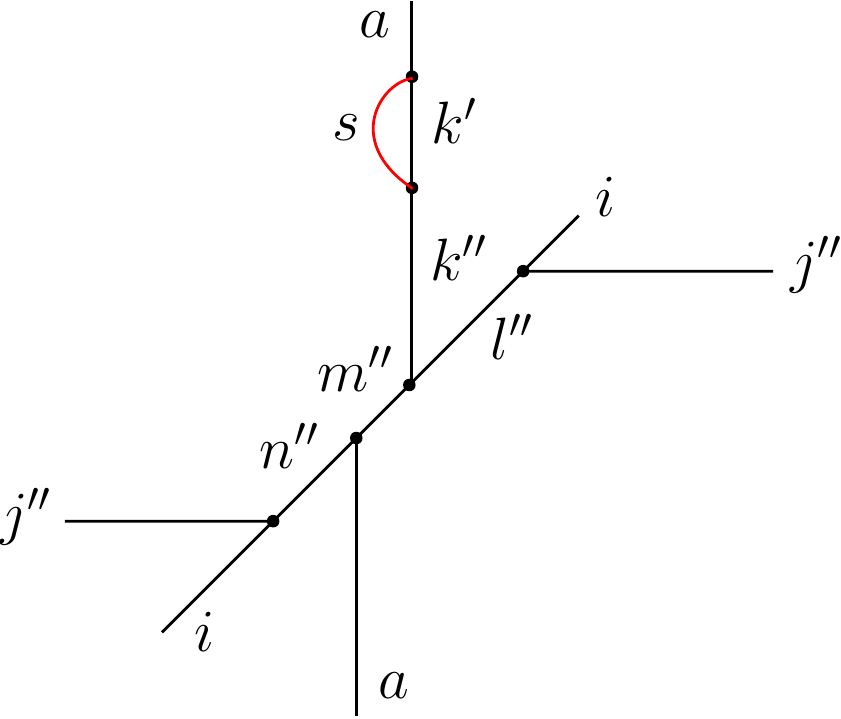}}}\\
&\xrightarrow[\text{squeezing bubbles}]{F^{sak^{\prime}}_{k^{\prime\prime}s0}d_s\delta_{ak^{\prime\prime}}}\ \vcenter{\hbox{\includegraphics[width=0.45\linewidth]{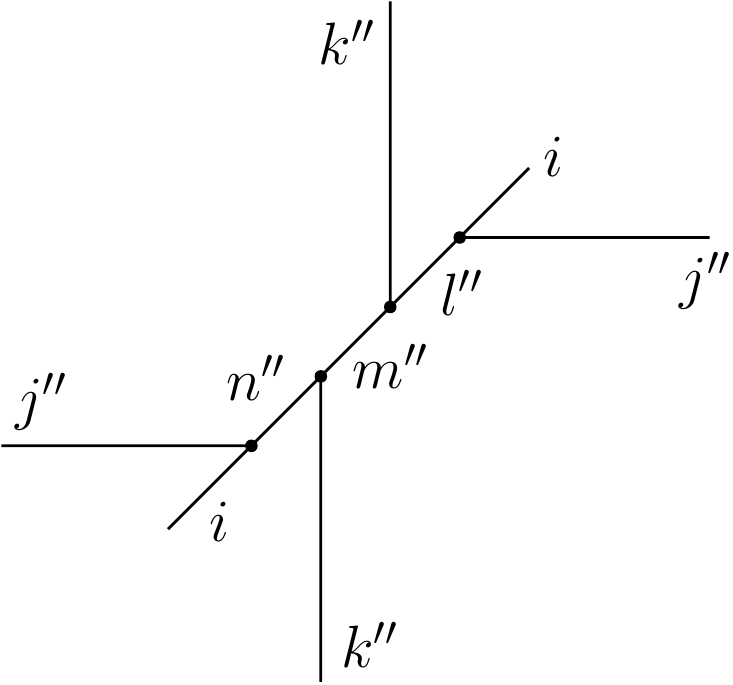}}}
\end{align*}

Collecting the coefficients from each step, we obtain 
\begin{align}
&(B^s_{yz})^{(i,j^{\prime\prime},k^{\prime\prime},l^{\prime\prime},m^{\prime\prime},n^{\prime\prime})}_{(i,j,k,l,m,n)}=
\sum_{k^{\prime},l^{\prime},j^{\prime},n^{\prime},m^{\prime}}(R^{ji}_{l})^*F^{kk0}_{ssk^{\prime}}\times \nonumber \\
&F^{mlk}_{sk^{\prime}l^{\prime}}
F^{sl^{\prime}l}_{ijj^{\prime}}
R^{j^{\prime}i}_{l^{\prime}}
F^{sj^{\prime}j}_{inn^{\prime}}
R^{ij^{\prime}}_{n^{\prime}}
F^{sn^{\prime}n}_{kmm^{\prime}}
F^{sm^{\prime}m}_{l^{\prime}k^{\prime}k^{\prime\prime}}
F^{k^{\prime}sk}_{m^{\prime}n^{\prime}m^{\prime\prime}}
F^{m^{\prime\prime}sm^{\prime}}_{l^{\prime}k^{\prime\prime}l^{\prime\prime}}\times \nonumber \\
&F^{l^{\prime\prime}sl^{\prime}}_{j^{\prime}ij^{\prime\prime}}
F^{j^{\prime\prime}sj^{\prime}}_{n^{\prime}in^{\prime\prime}}
(R^{ij^{\prime\prime}}_{n^{\prime\prime}})^*
F^{n^{\prime\prime}sn^{\prime}}_{k^{\prime}m^{\prime\prime}k^{\prime\prime}}
F^{sk^{\prime\prime}k^{\prime}}_{k^{\prime\prime}s0}d_s.
\end{align}

\section{String operators for the Walker-Wang models on the minimal lattice}
\label{app_c}

In this appendix, we calculate the string operators of a Walker-Wang model defined on the minimal lattice. The string is labeled by $s$, where $s$ is a generic anyon label in the input anyon theory $\mathcal{A}$.

\subsection{String operator along the $x$-direction}
\begin{align*}
&\vcenter{\hbox{\includegraphics[width=0.45\linewidth]{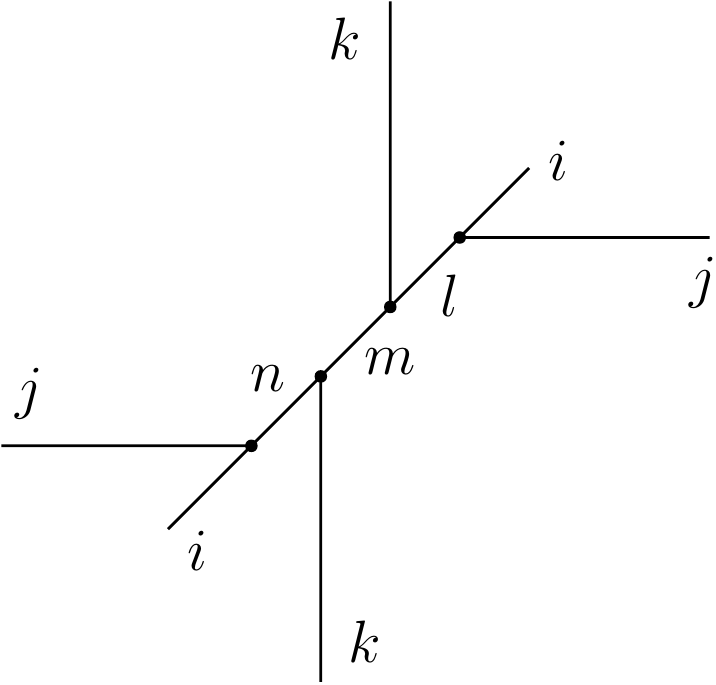}}}\\
&\xrightarrow[R \ \rm moves]{(R^{km}_{l})^*R^{ij}_{n}}\ \vcenter{\hbox{\includegraphics[width=0.45\linewidth]{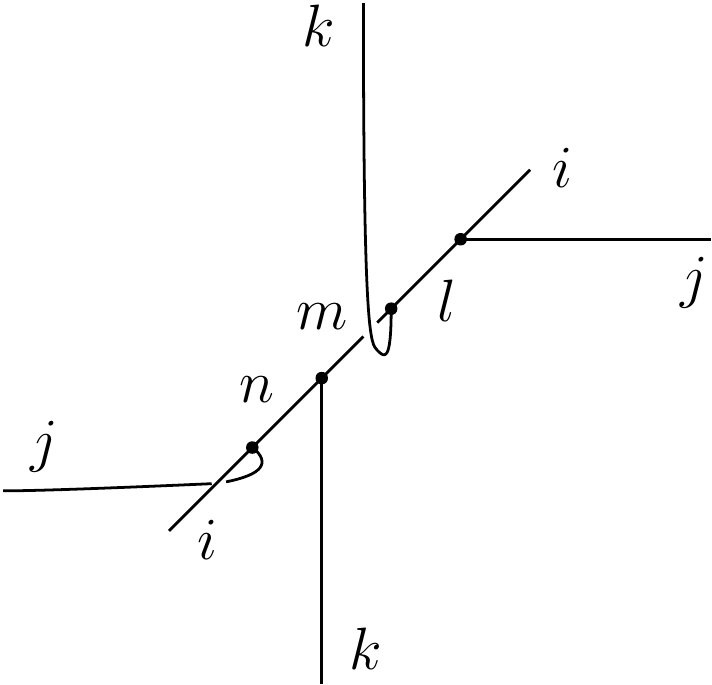}}} \\
&\xrightarrow[\text{add an $s$-loop along the $x$-axis}]{}\ \vcenter{\hbox{\includegraphics[width=0.45\linewidth]{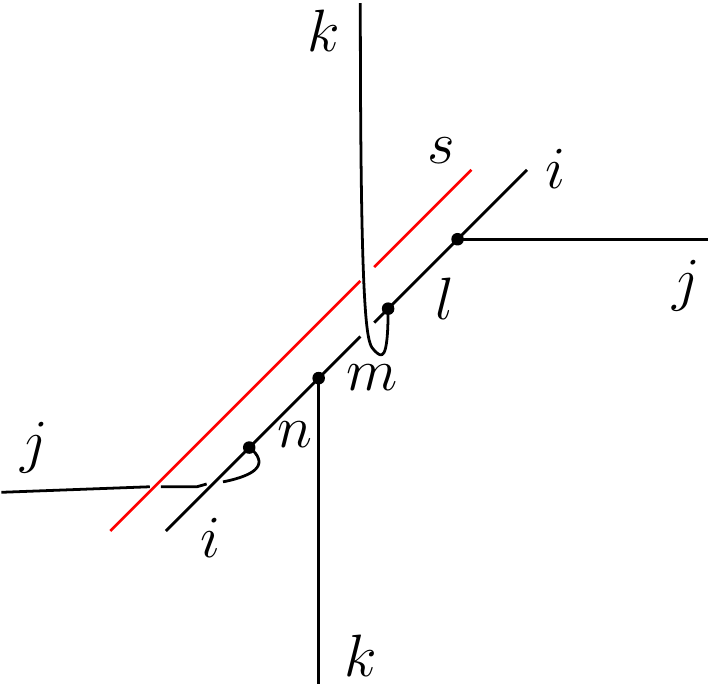}}} \\
&\xrightarrow[F \ \rm move]{\sum_{a}F^{ss0}_{iia}}\ \vcenter{\hbox{\includegraphics[width=0.45\linewidth]{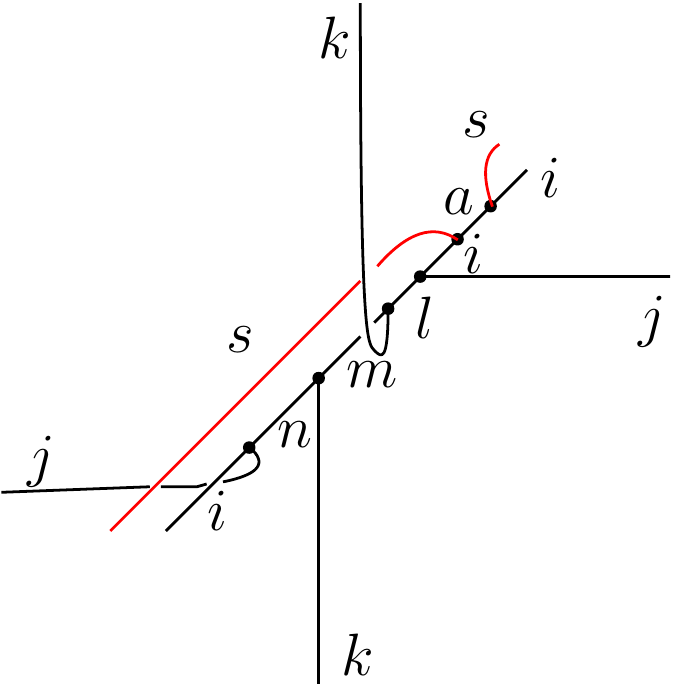}}} \\
&\xrightarrow[F \ \rm move]{\sum_{\tilde{l}}F^{asi}_{lj\tilde{l}}}\ \vcenter{\hbox{\includegraphics[width=0.45\linewidth]{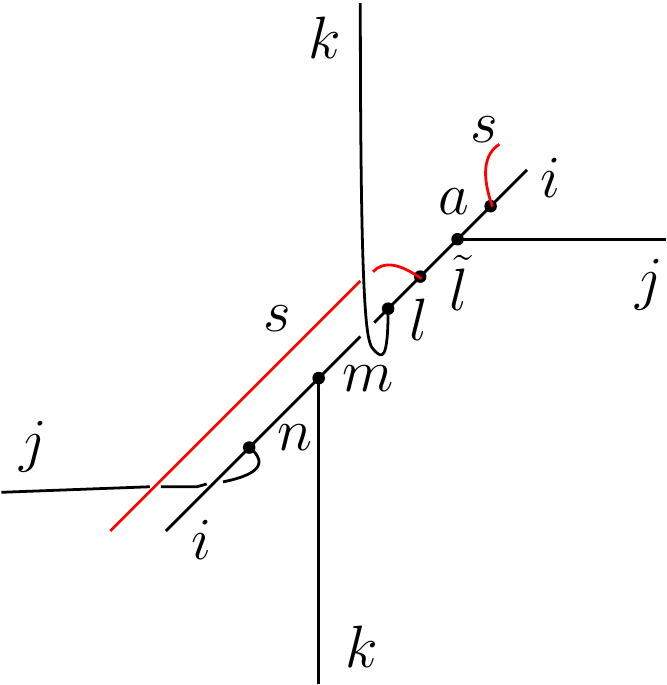}}}\\
&\xrightarrow[F \ \rm move]{\sum_{\tilde{m}}F^{\tilde{l}sl}_{mk\tilde{m}}}\ \vcenter{\hbox{\includegraphics[width=0.45\linewidth]{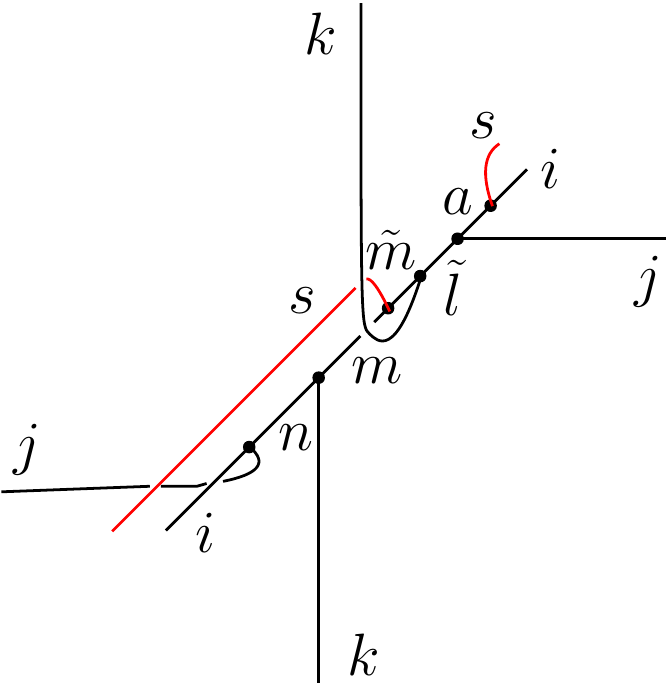}}}\\
&\xrightarrow[\text{Commute $F$ and $R$ moves}]{}\ \vcenter{\hbox{\includegraphics[width=0.45\linewidth]{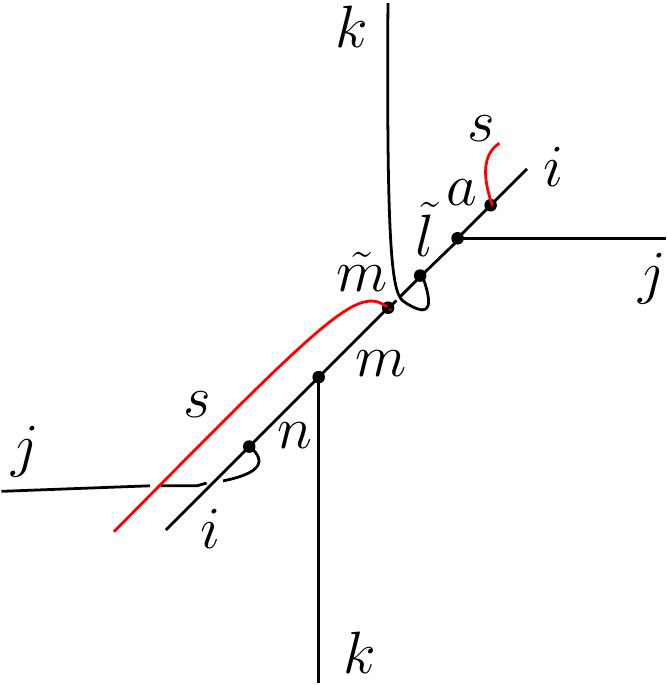}}}\\
&\xrightarrow[F \ \rm move]{\sum_{\tilde{n}}F^{\tilde{m}sm}_{nk\tilde{n}}}\ \vcenter{\hbox{\includegraphics[width=0.45\linewidth]{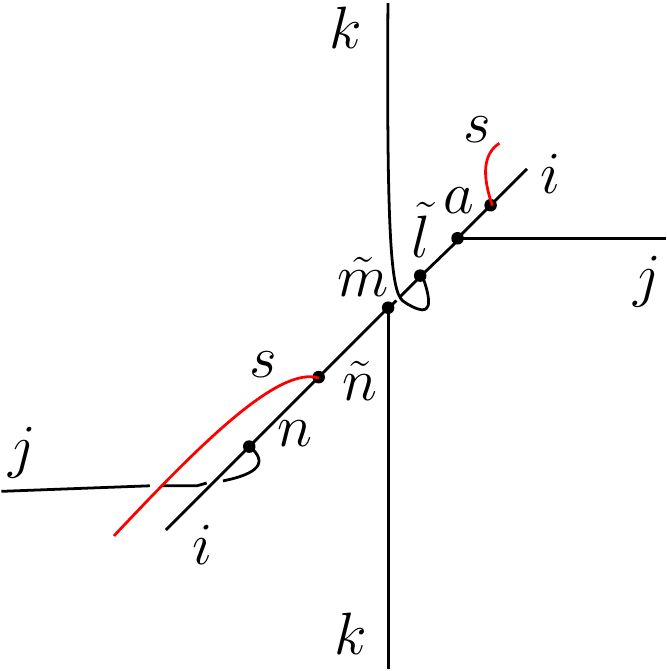}}}\\
&\xrightarrow[F \ \rm move]{\sum_{\tilde{i}}F^{\tilde{n}sn}_{ij\tilde{i}}}\ \vcenter{\hbox{\includegraphics[width=0.45\linewidth]{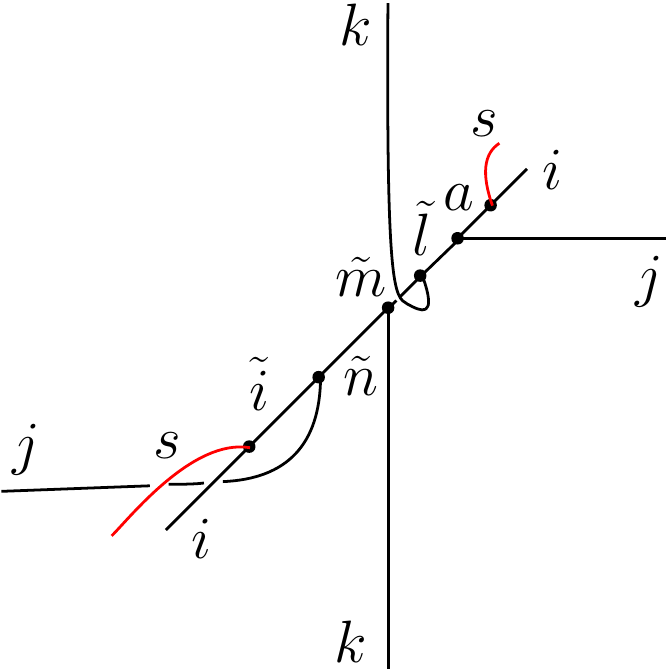}}}\\
&\xrightarrow[\text{Commute $F$ and $R$ moves}]{}\ \vcenter{\hbox{\includegraphics[width=0.45\linewidth]{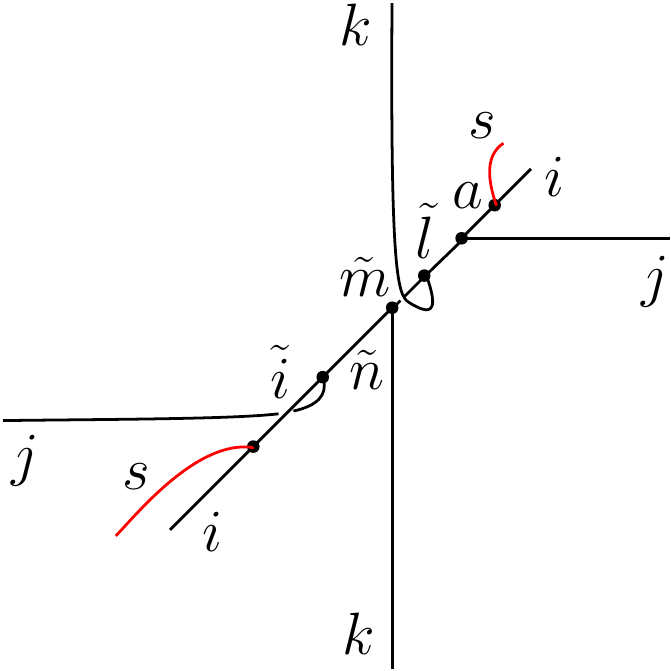}}}\\
&\xrightarrow[\text{squeezing bubbles}]{d_sF^{s\tilde{i}i}_{as0}\delta_{a\tilde{i}}}\ \vcenter{\hbox{\includegraphics[width=0.45\linewidth]{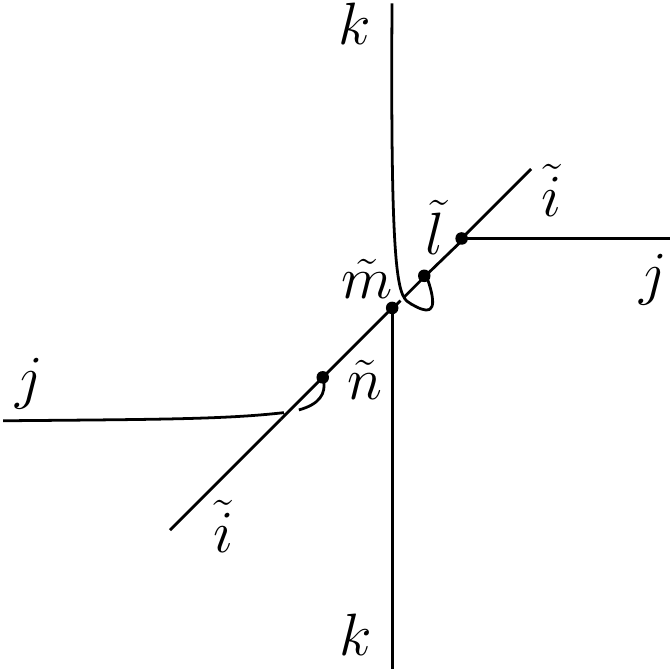}}}\\
&\xrightarrow[R \ \rm moves]{(R^{\tilde{i}j}_{\tilde{n}})^*R^{k\tilde{m}}_{\tilde{l}}}\ \vcenter{\hbox{\includegraphics[width=0.45\linewidth]{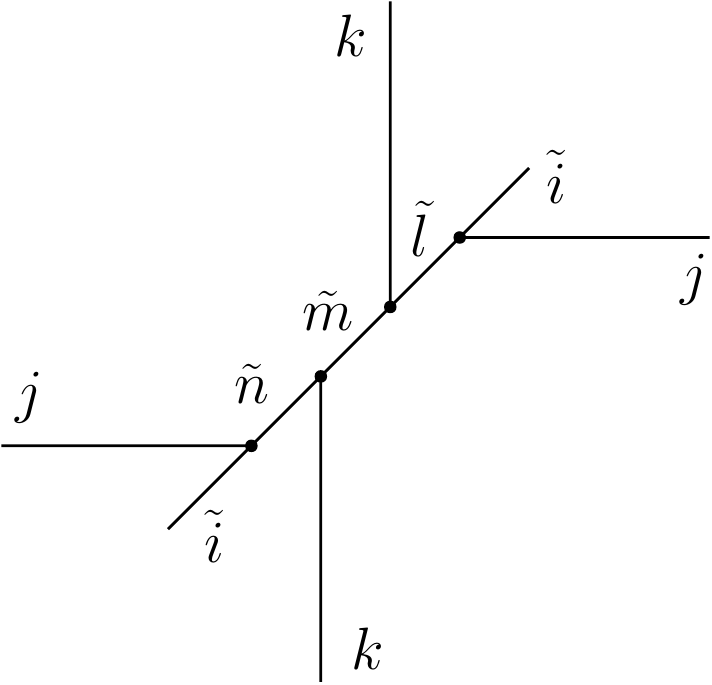}}}
\end{align*}

Collecting the coefficients from each step, we obtain 
\begin{align}
&(W^s_{x})^{(\tilde{i},j,k,\tilde{l},\tilde{m},\tilde{n})}_{(i,j,k,l,m,n)}=(R^{km}_{l})^*R^{ij}_{n}F^{ss0}_{ii\tilde{i}}F^{\tilde{i}si}_{lj\tilde{l}}F^{\tilde{l}sl}_{mk\tilde{m}}F^{\tilde{m}sm}_{nk\tilde{n}}\times  \nonumber \\
&F^{\tilde{n}sn}_{ij\tilde{i}}d_sF^{s\tilde{i}i}_{\tilde{i}s0}(R^{\tilde{i}j}_{\tilde{n}})^*R^{k\tilde{m}}_{\tilde{l}}.
\end{align}

\subsection{String operator along the $y$-direction}
\begin{align*}
&\vcenter{\hbox{\includegraphics[width=0.45\linewidth]{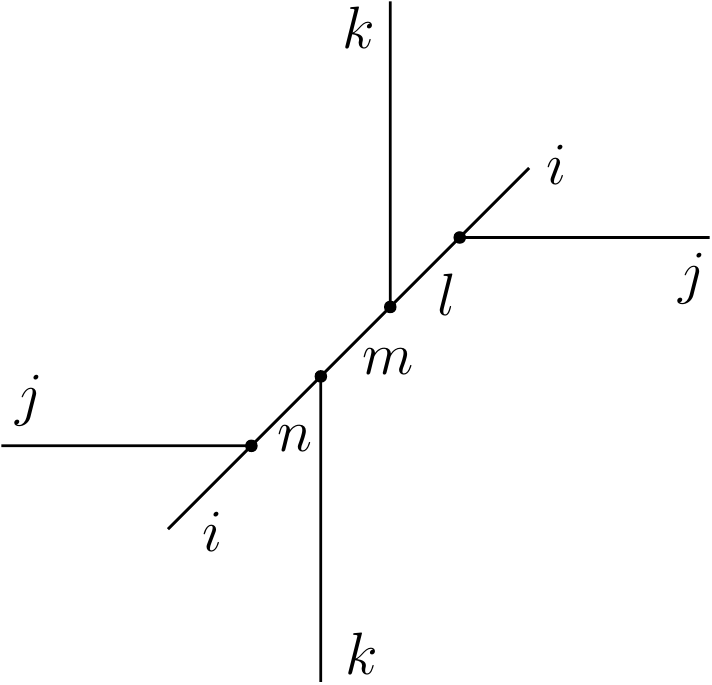}}}\\
&\xrightarrow[\text{add an $s$-loop along the $y$-axis}]{}\ \vcenter{\hbox{\includegraphics[width=0.45\linewidth]{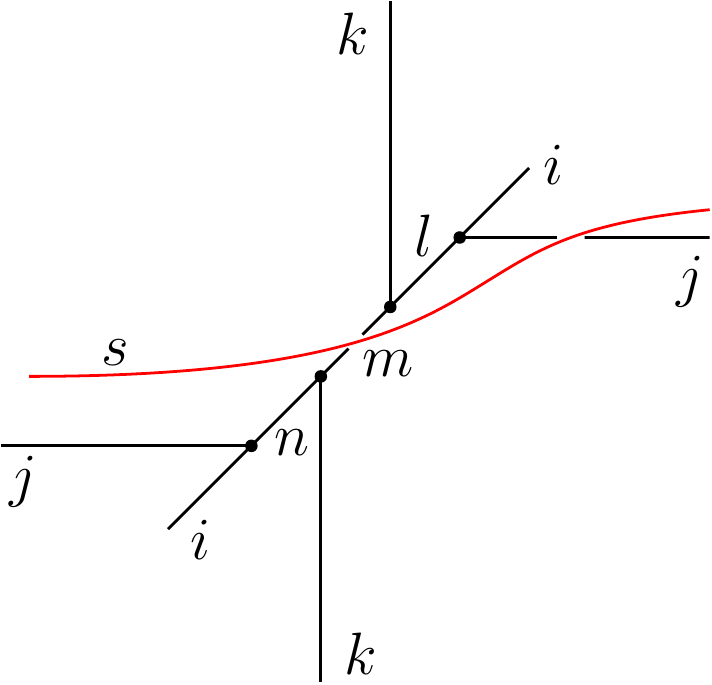}}}\\
&\xrightarrow[F \ \rm move]{\sum_{\tilde{j}}F^{jj0}_{ss\tilde{j}}}\ \vcenter{\hbox{\includegraphics[width=0.45\linewidth]{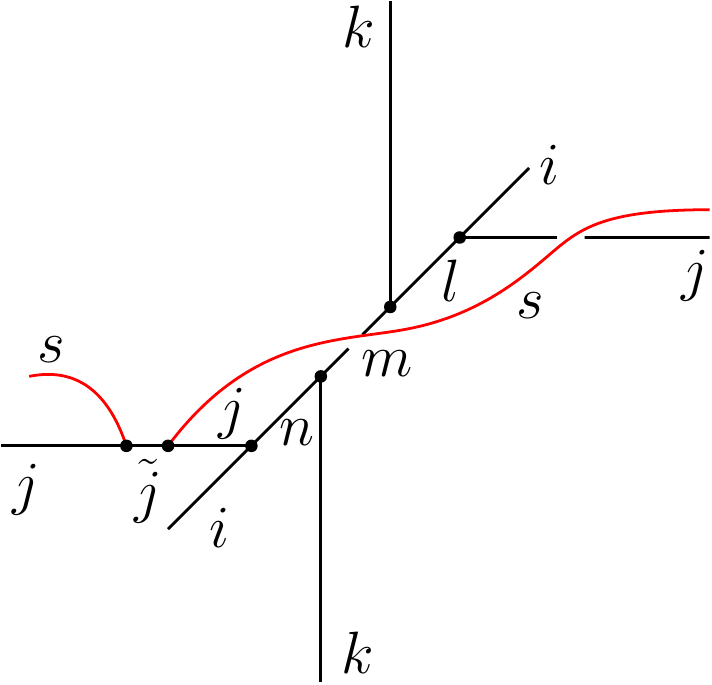}}}\\
&\xrightarrow[F \ \rm move]{\sum_{\tilde{n}}F^{s\tilde{j}j}_{in\tilde{n}}}\ \vcenter{\hbox{\includegraphics[width=0.45\linewidth]{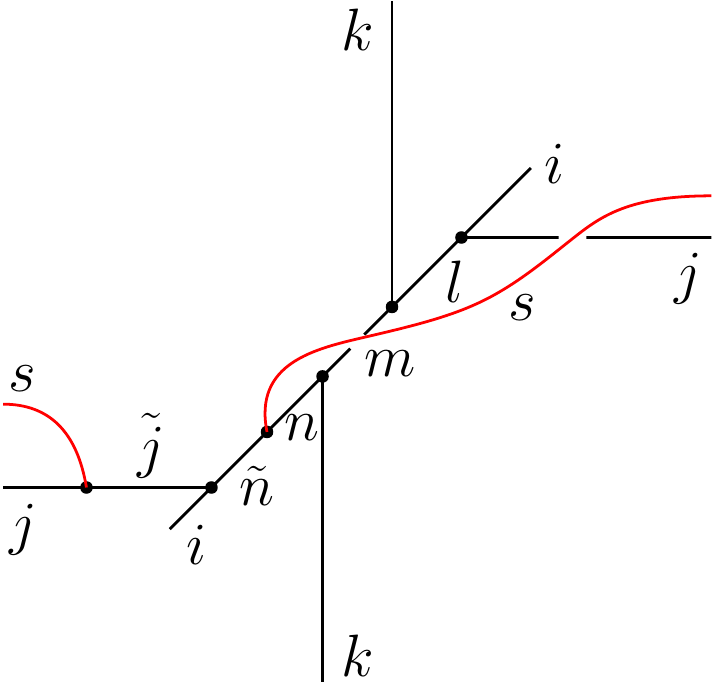}}}\\
&\xrightarrow[F \ \rm move]{\sum_{\tilde{m}}F^{s\tilde{n}n}_{km\tilde{m}}}\ \vcenter{\hbox{\includegraphics[width=0.45\linewidth]{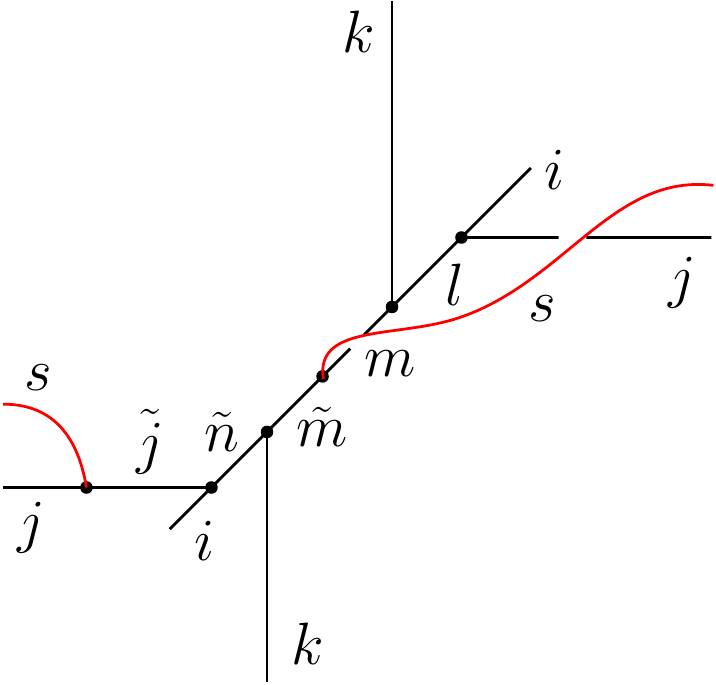}}}\\
&\xrightarrow[R \ \rm move]{R^{sm}_{\tilde{m}}}\ \vcenter{\hbox{\includegraphics[width=0.45\linewidth]{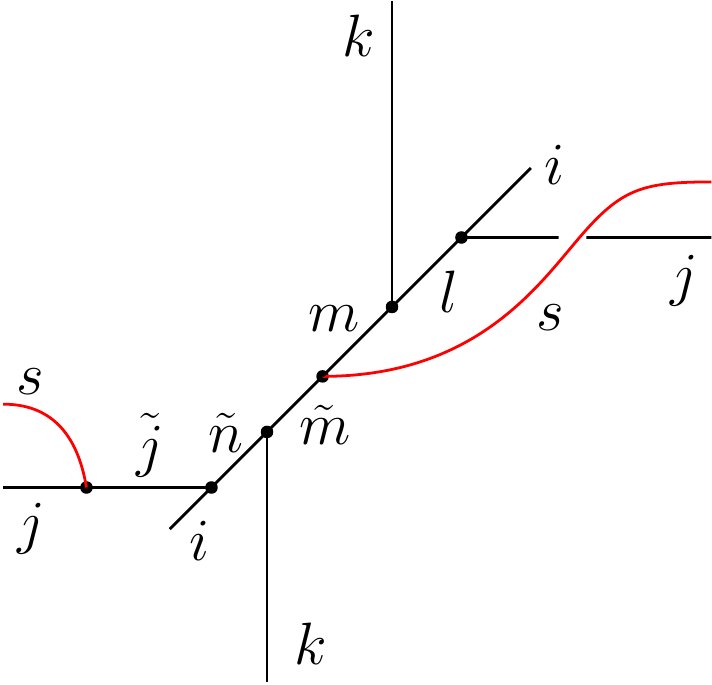}}}\\
&\xrightarrow[F \ \rm move]{\sum_{\tilde{l}}F^{\tilde{m}sm}_{lk\tilde{l}}}\ \vcenter{\hbox{\includegraphics[width=0.45\linewidth]{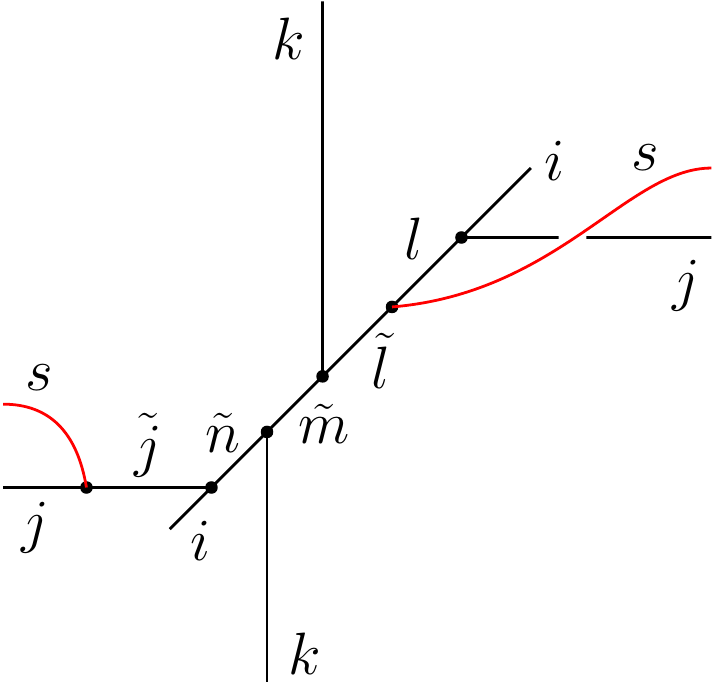}}}\\
&\xrightarrow[F \ \rm move]{\sum_{a}F^{\tilde{l}sl}_{jia}}\ \vcenter{\hbox{\includegraphics[width=0.45\linewidth]{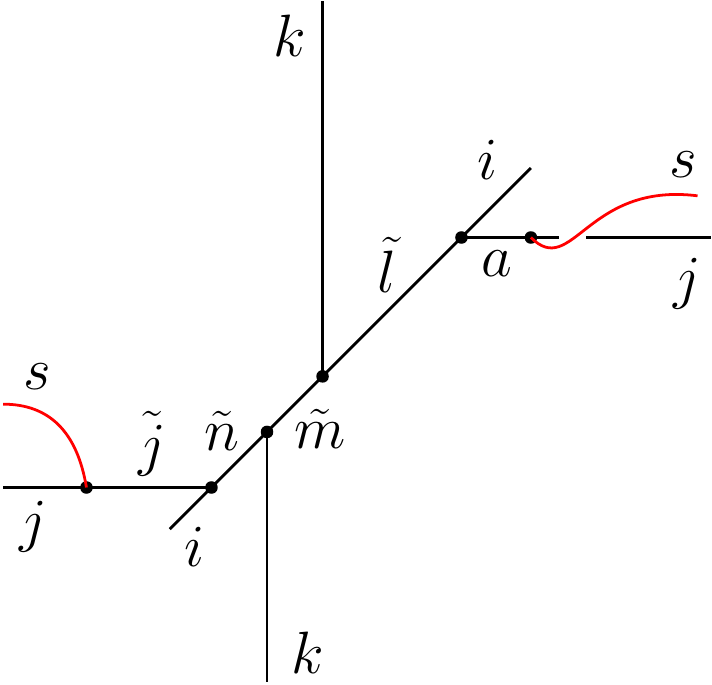}}}\\
&\xrightarrow[R \ \rm move]{(R^{sj}_a)^*}\ \vcenter{\hbox{\includegraphics[width=0.45\linewidth]{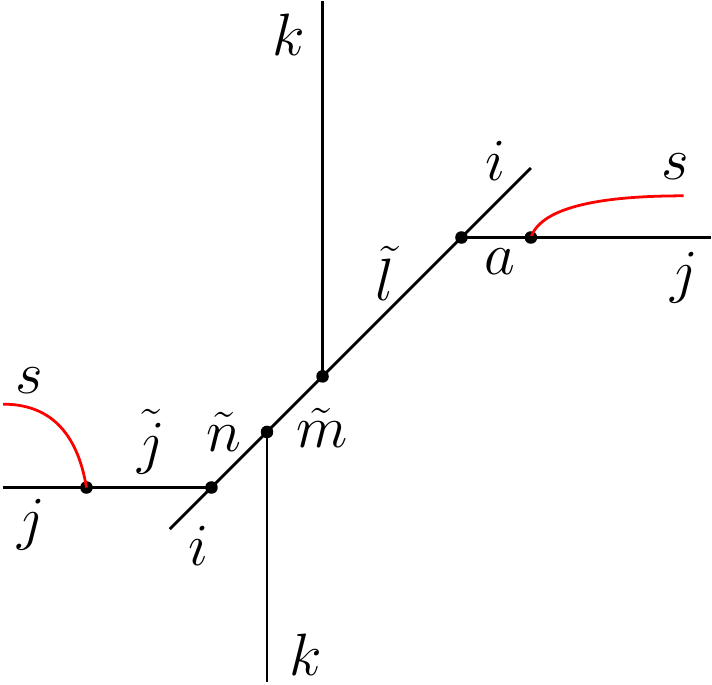}}}\\
&\xrightarrow[\text{squeezing bubbles}]{d_sF^{s\tilde{j}j}_{as0}\delta_{a\tilde{j}}}\ \vcenter{\hbox{\includegraphics[width=0.45\linewidth]{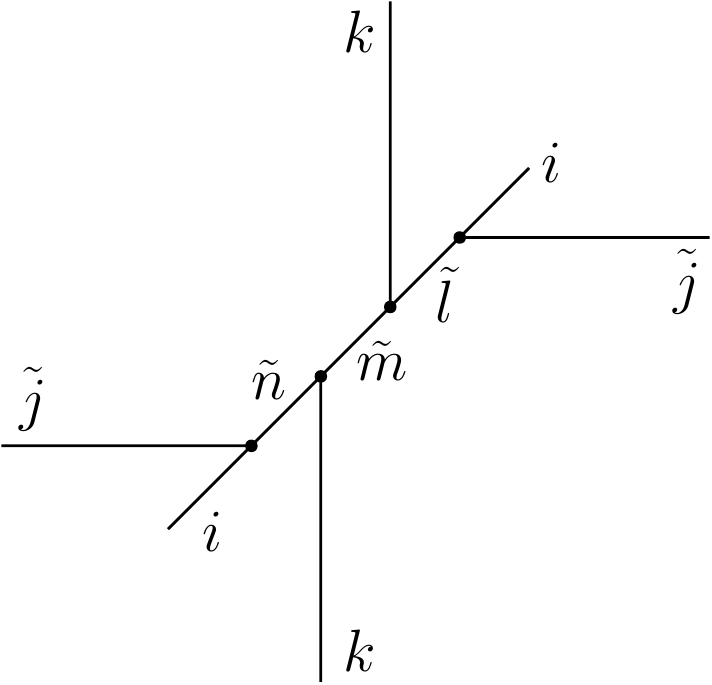}}}
\end{align*}

Collecting the coefficients from each step, we obtain 
\begin{align}
&(W^s_{y})^{(i,\tilde{j},k,\tilde{l},\tilde{m},\tilde{n})}_{(i,j,k,l,m,n)}=
F^{jj0}_{ss\tilde{j}}
F^{s\tilde{j}j}_{in\tilde{n}}
F^{s\tilde{n}n}_{km\tilde{m}}
R^{sm}_{\tilde{m}}
F^{\tilde{m}sm}_{lk\tilde{l}} \times \nonumber \\
&F^{\tilde{l}sl}_{ji\tilde{j}}
(R^{sj}_{\tilde{j}})^*
d_sF^{s\tilde{j}j}_{\tilde{j}s0}.
\end{align}

\subsection{String operator along the $z$-direction}
\begin{align*}
&\vcenter{\hbox{\includegraphics[width=0.45\linewidth]{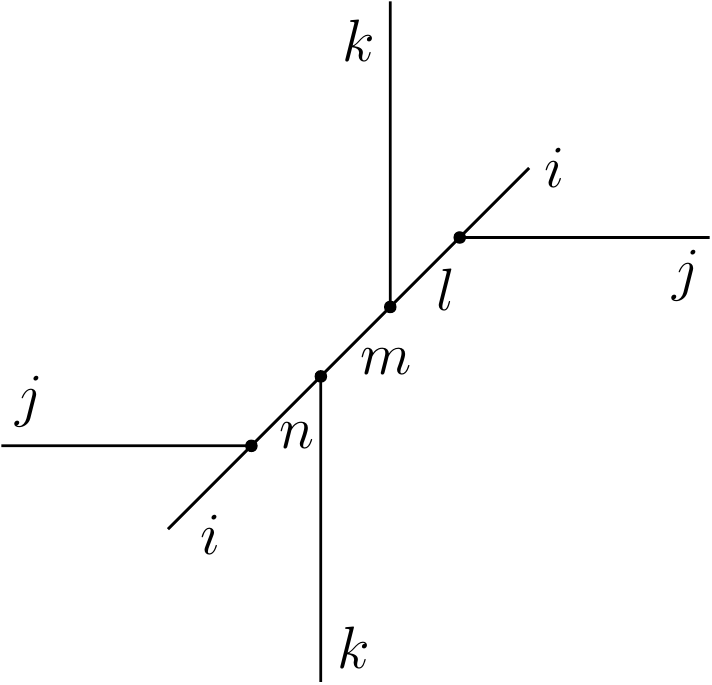}}}\\
&\xrightarrow[\text{add an $s$-loop along the $z$-axis}]{}\ \vcenter{\hbox{\includegraphics[width=0.45\linewidth]{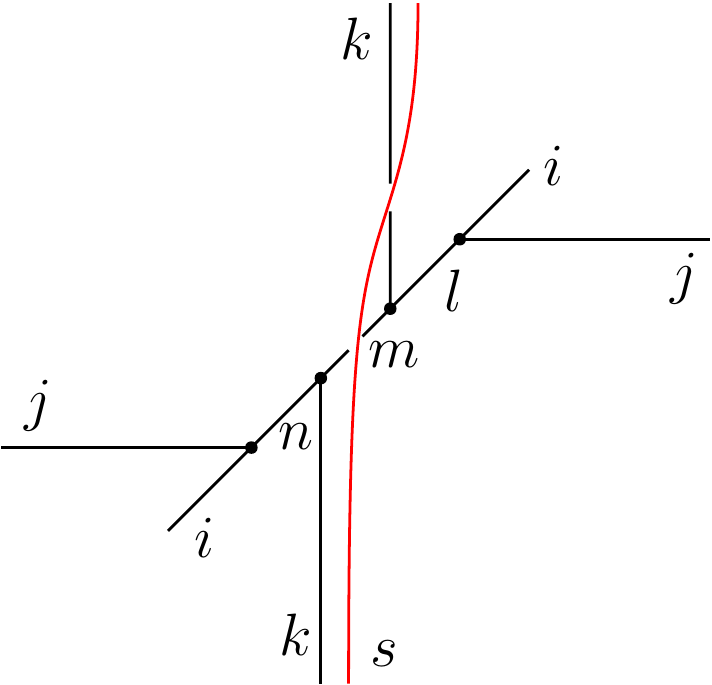}}}\\
&\xrightarrow[F \ \rm move]{\sum_{\tilde{k}}F^{kk0}_{ss\tilde{k}}}\ \vcenter{\hbox{\includegraphics[width=0.45\linewidth]{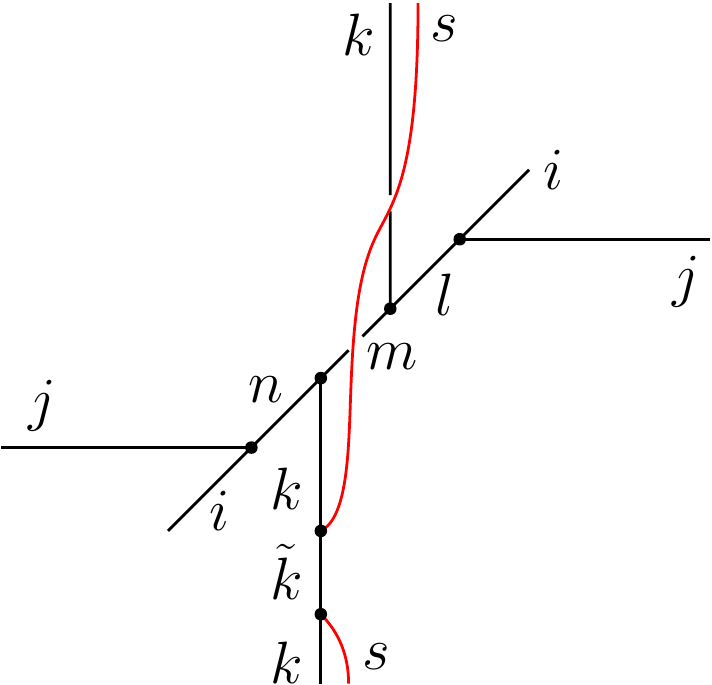}}}\\
&\xrightarrow[F \ \rm move]{\sum_{\tilde{m}}F^{\tilde{k}sk}_{mn\tilde{m}}}\ \vcenter{\hbox{\includegraphics[width=0.45\linewidth]{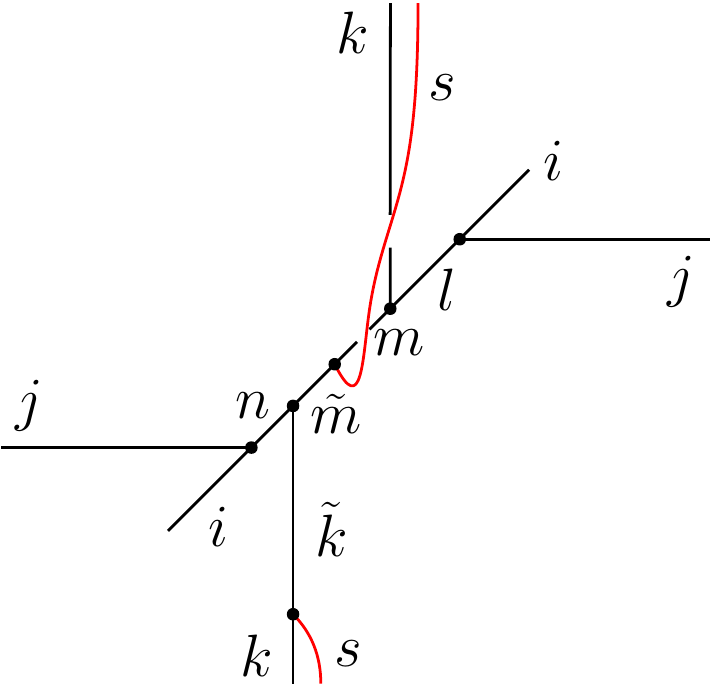}}}\\
&\xrightarrow[R \ \rm move]{(R^{sm}_{\tilde{m}})^*}\ \vcenter{\hbox{\includegraphics[width=0.45\linewidth]{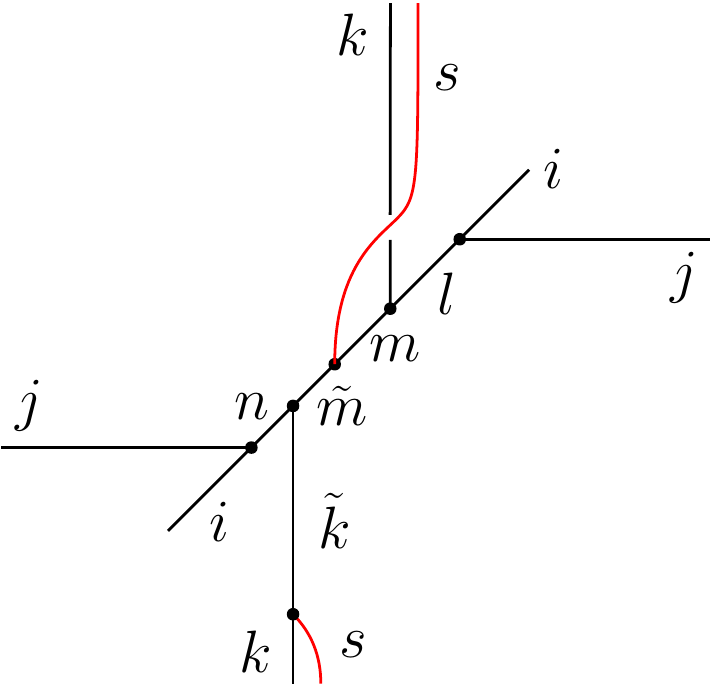}}}\\
&\xrightarrow[F \ \rm move]{\sum_{a}F^{s\tilde{m}m}_{lka}}\ \vcenter{\hbox{\includegraphics[width=0.45\linewidth]{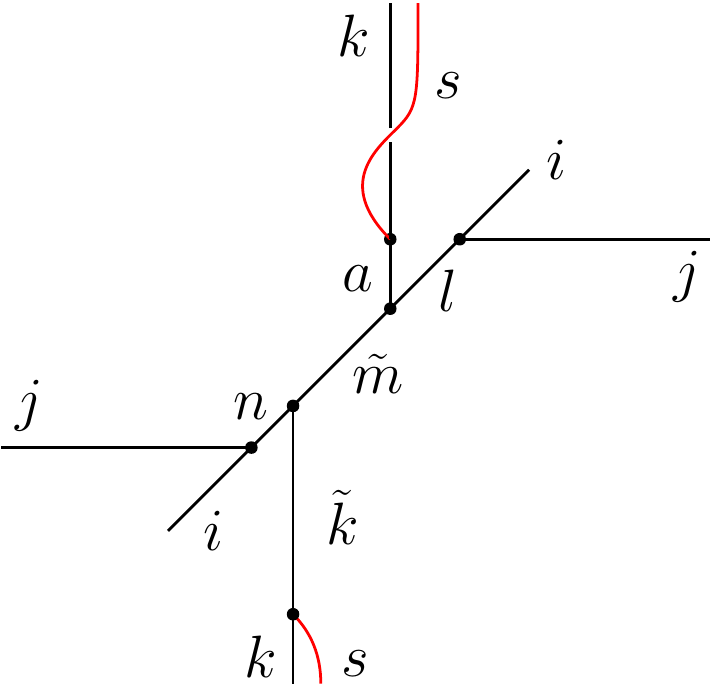}}}\\
&\xrightarrow[R \ \rm move]{R^{sk}_a}\ \vcenter{\hbox{\includegraphics[width=0.45\linewidth]{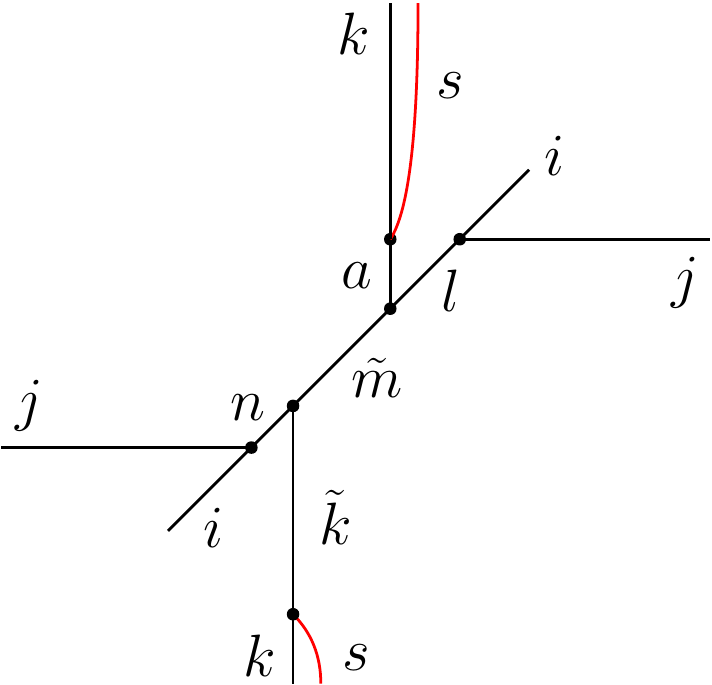}}}\\
&\xrightarrow[\text{squeezing bubbles}]{d_sF^{s\tilde{k}k}_{as0}\delta_{a\tilde{k}}}\ \vcenter{\hbox{\includegraphics[width=0.45\linewidth]{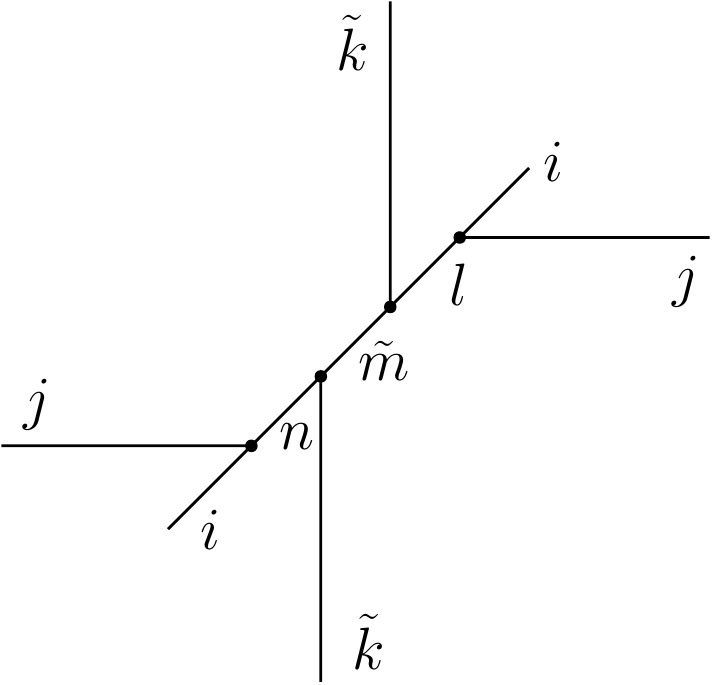}}}
\end{align*}

Collecting the coefficients from each step, we obtain 
\begin{align}
&(W^s_{z})^{(i,j,\tilde{k},l,\tilde{m},n)}_{(i,j,k,l,m,n)}=
F^{kk0}_{ss\tilde{k}}
F^{\tilde{k}sk}_{mn\tilde{m}}
(R^{sm}_{\tilde{m}})^*
F^{s\tilde{m}m}_{lk\tilde{k}}
R^{sk}_{\tilde{k}}
d_sF^{s\tilde{k}k}_{\tilde{k}s0}.
\end{align}

\section{MES basis and canonical form for $S$ and $T$ matrices}
\label{app_d}

In this appendix, we explain the necessary steps involved in transforming the 16 by 16 blocks $S_{a,b}$ and $T_{a,b}$ in Section \ref{sec:dimred} from the simultaneous eigenstates of $W^s_y$ and $W^s_z$ to the simultaneous eigenstates of $W^s_y$ and $V^s_y$ (the MES basis). We can focus on the 4-dimensional eigenspaces of the pair $(W^1_y,W^2_y)$ with fixed eigenvalues $(w^1_y,w^2_y)$, where $w^1_y, w^2_y = \pm 1$. Within each eigenspace, the problem is simplified to a basis transformation from the simultaneous eigenstates of $W^s_z$ to the simultaneous eigenstates of $V^s_y$. We denote the former (resp. latter) by $\{\ket{w^1_z,w^2_z}\}$ (resp. $\{\ket{v^1_y,v^2_y}\}$), where $w^1_z,w^2_z = \pm 1$, and $v^1_y,v^2_y = \pm$. Due to the Aharonov-Bohm interaction between charges and fluxes, the string operators satisfy the following commutation and anticommutation relations:
\begin{align}
\{W_z^1, V_y^1\} = 0, \ \ \ [W_z^1,V_y^2] = 0, \nonumber \\
\{W_z^2, V_y^2\} = 0, \ \ \ [W_z^2,V_y^1] = 0.
\end{align}
One can prove from these relations that the most general unitary change of basis from $\{\ket{w^1_z,w^2_z}\}$ to $\{\ket{v^1_y,v^2_y}\}$ is of the form
\begin{align}
&\ket{+,+} = \delta(\ket{1,1} + \alpha \ket{1,-1} + \beta \ket{-1,1} + \gamma \ket{-1,-1}), \nonumber \\
&\ket{+,-} = \delta(\ket{1,1} - \alpha \ket{1,-1} + \beta \ket{-1,1} - \gamma \ket{-1,-1}), \nonumber \\
&\ket{-,+} = \delta(\ket{1,1} + \alpha \ket{1,-1} - \beta \ket{-1,1} - \gamma \ket{-1,-1}), \nonumber \\
&\ket{-,-} = \delta(\ket{1,1} - \alpha \ket{1,-1} - \beta \ket{-1,1} + \gamma \ket{-1,-1}),
\end{align}
where $\alpha$, $\beta$, $\gamma$, and $\delta$ are independent $U(1)$ phases. We can then rewrite $S_{a,b}$ and $T_{a,b}$ in the transformed basis and try to match them with the $S$ and $T$ matrices of the 2D $\mathbb{Z}_2 \times \mathbb{Z}_2$ gauge theories. We find that by choosing the $U(1)$ phases appropriately, we can match each $S_{a,b}$ and $T_{a,b}$ to the $S$ and $T$ matrices of precisely one of the eight 2D $\mathbb{Z}_2 \times \mathbb{Z}_2$ gauge theories. The results are listed in Table \ref{tab2}.

For concreteness, we present below the explicit form of $S_{a,b}$ and $T_{a,b}$ ($a,b = \pm 1$) in the MES basis for the Walker-Wang models with input data $\text{Rep}_s(Q_8)$ and $\text{Rep}_s(D_4)$. Data for models that permute the charge labels of $D_4$ are omitted due to their similarity to those in the $\text{Rep}_s(D_4)$ case. The basis vectors in $\{\ket{w^1_y,w^2_y,v^1_y,v^2_y}\}$ are listed from large to small according to the number $v^1_y+2v^2_y+4w^1_y+8w^2_y$.

For the Walker-Wang model with input $\text{Rep}_s(Q_8)$, the data are the following:

{\footnotesize
\setlength{\arraycolsep}{1pt}
\begin{align}
&S_{1,1}=S_{-1,1}=S_{1,-1}=S_{-1,-1}= \nonumber \\ 
&\frac{1}{4}\begin{pmatrix}
 1 & 1 & 1 & 1 & 1 & 1 & 1 & 1 & 1 & 1 & 1 & 1 & 1 & 1 & 1 & 1 \\
 1 & 1 & 1 & 1 & -1 & -1 & -1 & -1 & 1 & 1 & 1 & 1 & -1 & -1 & -1 & -1 \\
 1 & 1 & 1 & 1 & 1 & 1 & 1 & 1 & -1 & -1 & -1 & -1 & -1 & -1 & -1 & -1 \\
 1 & 1 & 1 & 1 & -1 & -1 & -1 & -1 & -1 & -1 & -1 & -1 & 1 & 1 & 1 & 1 \\
 1 & -1 & 1 & -1 & 1 & -1 & 1 & -1 & 1 & -1 & 1 & -1 & 1 & -1 & 1 & -1 \\
 1 & -1 & 1 & -1 & -1 & 1 & -1 & 1 & 1 & -1 & 1 & -1 & -1 & 1 & -1 & 1 \\
 1 & -1 & 1 & -1 & 1 & -1 & 1 & -1 & -1 & 1 & -1 & 1 & -1 & 1 & -1 & 1 \\
 1 & -1 & 1 & -1 & -1 & 1 & -1 & 1 & -1 & 1 & -1 & 1 & 1 & -1 & 1 & -1 \\
 1 & 1 & -1 & -1 & 1 & 1 & -1 & -1 & 1 & 1 & -1 & -1 & 1 & 1 & -1 & -1 \\
 1 & 1 & -1 & -1 & -1 & -1 & 1 & 1 & 1 & 1 & -1 & -1 & -1 & -1 & 1 & 1 \\
 1 & 1 & -1 & -1 & 1 & 1 & -1 & -1 & -1 & -1 & 1 & 1 & -1 & -1 & 1 & 1 \\
 1 & 1 & -1 & -1 & -1 & -1 & 1 & 1 & -1 & -1 & 1 & 1 & 1 & 1 & -1 & -1 \\
 1 & -1 & -1 & 1 & 1 & -1 & -1 & 1 & 1 & -1 & -1 & 1 & 1 & -1 & -1 & 1 \\
 1 & -1 & -1 & 1 & -1 & 1 & 1 & -1 & 1 & -1 & -1 & 1 & -1 & 1 & 1 & -1 \\
 1 & -1 & -1 & 1 & 1 & -1 & -1 & 1 & -1 & 1 & 1 & -1 & -1 & 1 & 1 & -1 \\
 1 & -1 & -1 & 1 & -1 & 1 & 1 & -1 & -1 & 1 & 1 & -1 & 1 & -1 & -1 & 1 \\
\end{pmatrix}
\label{SQ8}
\end{align}}

{\footnotesize
\setlength{\arraycolsep}{1pt}
\begin{align}
&T_{1,1}=T_{-1,1}=T_{1,-1}=T_{-1,-1}= \nonumber \\
&\begin{pmatrix}
 1 & 0 & 0 & 0 & 0 & 0 & 0 & 0 & 0 & 0 & 0 & 0 & 0 & 0 & 0 & 0 \\
 0 & 1 & 0 & 0 & 0 & 0 & 0 & 0 & 0 & 0 & 0 & 0 & 0 & 0 & 0 & 0 \\
 0 & 0 & 1 & 0 & 0 & 0 & 0 & 0 & 0 & 0 & 0 & 0 & 0 & 0 & 0 & 0 \\
 0 & 0 & 0 & 1 & 0 & 0 & 0 & 0 & 0 & 0 & 0 & 0 & 0 & 0 & 0 & 0 \\
 0 & 0 & 0 & 0 & 1 & 0 & 0 & 0 & 0 & 0 & 0 & 0 & 0 & 0 & 0 & 0 \\
 0 & 0 & 0 & 0 & 0 & -1 & 0 & 0 & 0 & 0 & 0 & 0 & 0 & 0 & 0 & 0 \\
 0 & 0 & 0 & 0 & 0 & 0 & 1 & 0 & 0 & 0 & 0 & 0 & 0 & 0 & 0 & 0 \\
 0 & 0 & 0 & 0 & 0 & 0 & 0 & -1 & 0 & 0 & 0 & 0 & 0 & 0 & 0 & 0 \\
 0 & 0 & 0 & 0 & 0 & 0 & 0 & 0 & 1 & 0 & 0 & 0 & 0 & 0 & 0 & 0 \\
 0 & 0 & 0 & 0 & 0 & 0 & 0 & 0 & 0 & 1 & 0 & 0 & 0 & 0 & 0 & 0 \\
 0 & 0 & 0 & 0 & 0 & 0 & 0 & 0 & 0 & 0 & -1 & 0 & 0 & 0 & 0 & 0 \\
 0 & 0 & 0 & 0 & 0 & 0 & 0 & 0 & 0 & 0 & 0 & -1 & 0 & 0 & 0 & 0 \\
 0 & 0 & 0 & 0 & 0 & 0 & 0 & 0 & 0 & 0 & 0 & 0 & 1 & 0 & 0 & 0 \\
 0 & 0 & 0 & 0 & 0 & 0 & 0 & 0 & 0 & 0 & 0 & 0 & 0 & -1 & 0 & 0 \\
 0 & 0 & 0 & 0 & 0 & 0 & 0 & 0 & 0 & 0 & 0 & 0 & 0 & 0 & -1 & 0 \\
 0 & 0 & 0 & 0 & 0 & 0 & 0 & 0 & 0 & 0 & 0 & 0 & 0 & 0 & 0 & 1 \\
\end{pmatrix}
\label{TQ8}
\end{align}}

For the Walker-Wang model with input $\text{Rep}_s(D_4)$, the data are the following:

{\footnotesize
\setlength{\arraycolsep}{1pt}
\begin{align}
&S_{1,1}= \frac{1}{4}\begin{pmatrix}
 1 & 1 & 1 & 1 & 1 & 1 & 1 & 1 & 1 & 1 & 1 & 1 & 1 & 1 & 1 & 1 \\
 1 & 1 & 1 & 1 & -1 & -1 & -1 & -1 & 1 & 1 & 1 & 1 & -1 & -1 & -1 & -1 \\
 1 & 1 & 1 & 1 & 1 & 1 & 1 & 1 & -1 & -1 & -1 & -1 & -1 & -1 & -1 & -1 \\
 1 & 1 & 1 & 1 & -1 & -1 & -1 & -1 & -1 & -1 & -1 & -1 & 1 & 1 & 1 & 1 \\
 1 & -1 & 1 & -1 & 1 & -1 & 1 & -1 & 1 & -1 & 1 & -1 & 1 & -1 & 1 & -1 \\
 1 & -1 & 1 & -1 & -1 & 1 & -1 & 1 & 1 & -1 & 1 & -1 & -1 & 1 & -1 & 1 \\
 1 & -1 & 1 & -1 & 1 & -1 & 1 & -1 & -1 & 1 & -1 & 1 & -1 & 1 & -1 & 1 \\
 1 & -1 & 1 & -1 & -1 & 1 & -1 & 1 & -1 & 1 & -1 & 1 & 1 & -1 & 1 & -1 \\
 1 & 1 & -1 & -1 & 1 & 1 & -1 & -1 & 1 & 1 & -1 & -1 & 1 & 1 & -1 & -1 \\
 1 & 1 & -1 & -1 & -1 & -1 & 1 & 1 & 1 & 1 & -1 & -1 & -1 & -1 & 1 & 1 \\
 1 & 1 & -1 & -1 & 1 & 1 & -1 & -1 & -1 & -1 & 1 & 1 & -1 & -1 & 1 & 1 \\
 1 & 1 & -1 & -1 & -1 & -1 & 1 & 1 & -1 & -1 & 1 & 1 & 1 & 1 & -1 & -1 \\
 1 & -1 & -1 & 1 & 1 & -1 & -1 & 1 & 1 & -1 & -1 & 1 & 1 & -1 & -1 & 1 \\
 1 & -1 & -1 & 1 & -1 & 1 & 1 & -1 & 1 & -1 & -1 & 1 & -1 & 1 & 1 & -1 \\
 1 & -1 & -1 & 1 & 1 & -1 & -1 & 1 & -1 & 1 & 1 & -1 & -1 & 1 & 1 & -1 \\
 1 & -1 & -1 & 1 & -1 & 1 & 1 & -1 & -1 & 1 & 1 & -1 & 1 & -1 & -1 & 1 \\
\end{pmatrix}
\label{SD41stblock}
\end{align}}

{\footnotesize
\setlength{\arraycolsep}{1pt}
\begin{align}
&T_{1,1}= \begin{pmatrix}
1 & 0 & 0 & 0 & 0 & 0 & 0 & 0 & 0 & 0 & 0 & 0 & 0 & 0 & 0 & 0 \\
 0 & 1 & 0 & 0 & 0 & 0 & 0 & 0 & 0 & 0 & 0 & 0 & 0 & 0 & 0 & 0 \\
 0 & 0 & 1 & 0 & 0 & 0 & 0 & 0 & 0 & 0 & 0 & 0 & 0 & 0 & 0 & 0 \\
 0 & 0 & 0 & 1 & 0 & 0 & 0 & 0 & 0 & 0 & 0 & 0 & 0 & 0 & 0 & 0 \\
 0 & 0 & 0 & 0 & 1 & 0 & 0 & 0 & 0 & 0 & 0 & 0 & 0 & 0 & 0 & 0 \\
 0 & 0 & 0 & 0 & 0 & -1 & 0 & 0 & 0 & 0 & 0 & 0 & 0 & 0 & 0 & 0 \\
 0 & 0 & 0 & 0 & 0 & 0 & 1 & 0 & 0 & 0 & 0 & 0 & 0 & 0 & 0 & 0 \\
 0 & 0 & 0 & 0 & 0 & 0 & 0 & -1 & 0 & 0 & 0 & 0 & 0 & 0 & 0 & 0 \\
 0 & 0 & 0 & 0 & 0 & 0 & 0 & 0 & 1 & 0 & 0 & 0 & 0 & 0 & 0 & 0 \\
 0 & 0 & 0 & 0 & 0 & 0 & 0 & 0 & 0 & 1 & 0 & 0 & 0 & 0 & 0 & 0 \\
 0 & 0 & 0 & 0 & 0 & 0 & 0 & 0 & 0 & 0 & -1 & 0 & 0 & 0 & 0 & 0 \\
 0 & 0 & 0 & 0 & 0 & 0 & 0 & 0 & 0 & 0 & 0 & -1 & 0 & 0 & 0 & 0 \\
 0 & 0 & 0 & 0 & 0 & 0 & 0 & 0 & 0 & 0 & 0 & 0 & 1 & 0 & 0 & 0 \\
 0 & 0 & 0 & 0 & 0 & 0 & 0 & 0 & 0 & 0 & 0 & 0 & 0 & -1 & 0 & 0 \\
 0 & 0 & 0 & 0 & 0 & 0 & 0 & 0 & 0 & 0 & 0 & 0 & 0 & 0 & -1 & 0 \\
 0 & 0 & 0 & 0 & 0 & 0 & 0 & 0 & 0 & 0 & 0 & 0 & 0 & 0 & 0 & 1 \\
\end{pmatrix}
\label{TD41stblock}
\end{align}}

{\footnotesize
\setlength{\arraycolsep}{1pt}
\begin{align}
&S_{-1,1}= \frac{1}{4}\begin{pmatrix}
 1 & 1 & 1 & 1 & 1 & 1 & 1 & 1 & 1 & 1 & 1 & 1 & 1 & 1 & 1 & 1 \\
 1 & 1 & 1 & 1 & -1 & -1 & -1 & -1 & 1 & 1 & 1 & 1 & -1 & -1 & -1 & -1 \\
 1 & 1 & 1 & 1 & 1 & 1 & 1 & 1 & -1 & -1 & -1 & -1 & -1 & -1 & -1 & -1 \\
 1 & 1 & 1 & 1 & -1 & -1 & -1 & -1 & -1 & -1 & -1 & -1 & 1 & 1 & 1 & 1 \\
 1 & -1 & 1 & -1 & 1 & -1 & 1 & -1 & 1 & -1 & 1 & -1 & 1 & -1 & 1 & -1 \\
 1 & -1 & 1 & -1 & -1 & 1 & -1 & 1 & 1 & -1 & 1 & -1 & -1 & 1 & -1 & 1 \\
 1 & -1 & 1 & -1 & 1 & -1 & 1 & -1 & -1 & 1 & -1 & 1 & -1 & 1 & -1 & 1 \\
 1 & -1 & 1 & -1 & -1 & 1 & -1 & 1 & -1 & 1 & -1 & 1 & 1 & -1 & 1 & -1 \\
 1 & 1 & -1 & -1 & 1 & 1 & -1 & -1 & -1 & -1 & 1 & 1 & -1 & -1 & 1 & 1 \\
 1 & 1 & -1 & -1 & -1 & -1 & 1 & 1 & -1 & -1 & 1 & 1 & 1 & 1 & -1 & -1 \\
 1 & 1 & -1 & -1 & 1 & 1 & -1 & -1 & 1 & 1 & -1 & -1 & 1 & 1 & -1 & -1 \\
 1 & 1 & -1 & -1 & -1 & -1 & 1 & 1 & 1 & 1 & -1 & -1 & -1 & -1 & 1 & 1 \\
 1 & -1 & -1 & 1 & 1 & -1 & -1 & 1 & -1 & 1 & 1 & -1 & -1 & 1 & 1 & -1 \\
 1 & -1 & -1 & 1 & -1 & 1 & 1 & -1 & -1 & 1 & 1 & -1 & 1 & -1 & -1 & 1 \\
 1 & -1 & -1 & 1 & 1 & -1 & -1 & 1 & 1 & -1 & -1 & 1 & 1 & -1 & -1 & 1 \\
 1 & -1 & -1 & 1 & -1 & 1 & 1 & -1 & 1 & -1 & -1 & 1 & -1 & 1 & 1 & -1 \\
\end{pmatrix}
\label{SD42ndblock}
\end{align}}

{\footnotesize
\setlength{\arraycolsep}{1pt}
\begin{align}
&T_{-1,1}= \begin{pmatrix}
 1 & 0 & 0 & 0 & 0 & 0 & 0 & 0 & 0 & 0 & 0 & 0 & 0 & 0 & 0 & 0 \\
 0 & 1 & 0 & 0 & 0 & 0 & 0 & 0 & 0 & 0 & 0 & 0 & 0 & 0 & 0 & 0 \\
 0 & 0 & 1 & 0 & 0 & 0 & 0 & 0 & 0 & 0 & 0 & 0 & 0 & 0 & 0 & 0 \\
 0 & 0 & 0 & 1 & 0 & 0 & 0 & 0 & 0 & 0 & 0 & 0 & 0 & 0 & 0 & 0 \\
 0 & 0 & 0 & 0 & 1 & 0 & 0 & 0 & 0 & 0 & 0 & 0 & 0 & 0 & 0 & 0 \\
 0 & 0 & 0 & 0 & 0 & -1 & 0 & 0 & 0 & 0 & 0 & 0 & 0 & 0 & 0 & 0 \\
 0 & 0 & 0 & 0 & 0 & 0 & 1 & 0 & 0 & 0 & 0 & 0 & 0 & 0 & 0 & 0 \\
 0 & 0 & 0 & 0 & 0 & 0 & 0 & -1 & 0 & 0 & 0 & 0 & 0 & 0 & 0 & 0 \\
 0 & 0 & 0 & 0 & 0 & 0 & 0 & 0 & i & 0 & 0 & 0 & 0 & 0 & 0 & 0 \\
 0 & 0 & 0 & 0 & 0 & 0 & 0 & 0 & 0 & i & 0 & 0 & 0 & 0 & 0 & 0 \\
 0 & 0 & 0 & 0 & 0 & 0 & 0 & 0 & 0 & 0 & -i & 0 & 0 & 0 & 0 & 0 \\
 0 & 0 & 0 & 0 & 0 & 0 & 0 & 0 & 0 & 0 & 0 & -i & 0 & 0 & 0 & 0 \\
 0 & 0 & 0 & 0 & 0 & 0 & 0 & 0 & 0 & 0 & 0 & 0 & i & 0 & 0 & 0 \\
 0 & 0 & 0 & 0 & 0 & 0 & 0 & 0 & 0 & 0 & 0 & 0 & 0 & -i & 0 & 0 \\
 0 & 0 & 0 & 0 & 0 & 0 & 0 & 0 & 0 & 0 & 0 & 0 & 0 & 0 & -i & 0 \\
 0 & 0 & 0 & 0 & 0 & 0 & 0 & 0 & 0 & 0 & 0 & 0 & 0 & 0 & 0 & i \\
\end{pmatrix}
\label{TD42ndblock}
\end{align}}

{\footnotesize
\setlength{\arraycolsep}{1pt}
\begin{align}
&S_{1,-1}= \frac{1}{4}\begin{pmatrix}
 1 & 1 & 1 & 1 & 1 & 1 & 1 & 1 & 1 & 1 & 1 & 1 & 1 & 1 & 1 & 1 \\
 1 & 1 & 1 & 1 & -1 & -1 & -1 & -1 & 1 & 1 & 1 & 1 & -1 & -1 & -1 & -1 \\
 1 & 1 & 1 & 1 & 1 & 1 & 1 & 1 & -1 & -1 & -1 & -1 & -1 & -1 & -1 & -1 \\
 1 & 1 & 1 & 1 & -1 & -1 & -1 & -1 & -1 & -1 & -1 & -1 & 1 & 1 & 1 & 1 \\
 1 & -1 & 1 & -1 & 1 & -1 & 1 & -1 & -i & i & -i & i & -i & i & -i & i \\
 1 & -1 & 1 & -1 & -1 & 1 & -1 & 1 & -i & i & -i & i & i & -i & i & -i \\
 1 & -1 & 1 & -1 & 1 & -1 & 1 & -1 & i & -i & i & -i & i & -i & i & -i \\
 1 & -1 & 1 & -1 & -1 & 1 & -1 & 1 & i & -i & i & -i & -i & i & -i & i \\
 1 & 1 & -1 & -1 & -i & -i & i & i & 1 & 1 & -1 & -1 & -i & -i & i & i \\
 1 & 1 & -1 & -1 & i & i & -i & -i & 1 & 1 & -1 & -1 & i & i & -i & -i \\
 1 & 1 & -1 & -1 & -i & -i & i & i & -1 & -1 & 1 & 1 & i & i & -i & -i \\
 1 & 1 & -1 & -1 & i & i & -i & -i & -1 & -1 & 1 & 1 & -i & -i & i & i \\
 1 & -1 & -1 & 1 & -i & i & i & -i & -i & i & i & -i & -1 & 1 & 1 & -1 \\
 1 & -1 & -1 & 1 & i & -i & -i & i & -i & i & i & -i & 1 & -1 & -1 & 1 \\
 1 & -1 & -1 & 1 & -i & i & i & -i & i & -i & -i & i & 1 & -1 & -1 & 1 \\
 1 & -1 & -1 & 1 & i & -i & -i & i & i & -i & -i & i & -1 & 1 & 1 & -1 \\
\end{pmatrix}
\label{SD43rdblock}
\end{align}}

{\footnotesize
\setlength{\arraycolsep}{1pt}
\begin{align}
&T_{1,-1}= \begin{pmatrix}
 1 & 0 & 0 & 0 & 0 & 0 & 0 & 0 & 0 & 0 & 0 & 0 & 0 & 0 & 0 & 0 \\
 0 & 1 & 0 & 0 & 0 & 0 & 0 & 0 & 0 & 0 & 0 & 0 & 0 & 0 & 0 & 0 \\
 0 & 0 & 1 & 0 & 0 & 0 & 0 & 0 & 0 & 0 & 0 & 0 & 0 & 0 & 0 & 0 \\
 0 & 0 & 0 & 1 & 0 & 0 & 0 & 0 & 0 & 0 & 0 & 0 & 0 & 0 & 0 & 0 \\
 0 & 0 & 0 & 0 & 1 & 0 & 0 & 0 & 0 & 0 & 0 & 0 & 0 & 0 & 0 & 0 \\
 0 & 0 & 0 & 0 & 0 & -1 & 0 & 0 & 0 & 0 & 0 & 0 & 0 & 0 & 0 & 0 \\
 0 & 0 & 0 & 0 & 0 & 0 & 1 & 0 & 0 & 0 & 0 & 0 & 0 & 0 & 0 & 0 \\
 0 & 0 & 0 & 0 & 0 & 0 & 0 & -1 & 0 & 0 & 0 & 0 & 0 & 0 & 0 & 0 \\
 0 & 0 & 0 & 0 & 0 & 0 & 0 & 0 & 1 & 0 & 0 & 0 & 0 & 0 & 0 & 0 \\
 0 & 0 & 0 & 0 & 0 & 0 & 0 & 0 & 0 & 1 & 0 & 0 & 0 & 0 & 0 & 0 \\
 0 & 0 & 0 & 0 & 0 & 0 & 0 & 0 & 0 & 0 & -1 & 0 & 0 & 0 & 0 & 0 \\
 0 & 0 & 0 & 0 & 0 & 0 & 0 & 0 & 0 & 0 & 0 & -1 & 0 & 0 & 0 & 0 \\
 0 & 0 & 0 & 0 & 0 & 0 & 0 & 0 & 0 & 0 & 0 & 0 & i & 0 & 0 & 0 \\
 0 & 0 & 0 & 0 & 0 & 0 & 0 & 0 & 0 & 0 & 0 & 0 & 0 & -i & 0 & 0 \\
 0 & 0 & 0 & 0 & 0 & 0 & 0 & 0 & 0 & 0 & 0 & 0 & 0 & 0 & -i & 0 \\
 0 & 0 & 0 & 0 & 0 & 0 & 0 & 0 & 0 & 0 & 0 & 0 & 0 & 0 & 0 & i \\
\end{pmatrix}
\label{TD43rdblock}
\end{align}}

{\footnotesize
\setlength{\arraycolsep}{1pt}
\begin{align}
&S_{-1,-1}= \frac{1}{4}\begin{pmatrix}
 1 & 1 & 1 & 1 & 1 & 1 & 1 & 1 & 1 & 1 & 1 & 1 & 1 & 1 & 1 & 1 \\
 1 & 1 & 1 & 1 & -1 & -1 & -1 & -1 & 1 & 1 & 1 & 1 & -1 & -1 & -1 & -1 \\
 1 & 1 & 1 & 1 & 1 & 1 & 1 & 1 & -1 & -1 & -1 & -1 & -1 & -1 & -1 & -1 \\
 1 & 1 & 1 & 1 & -1 & -1 & -1 & -1 & -1 & -1 & -1 & -1 & 1 & 1 & 1 & 1 \\
 1 & -1 & 1 & -1 & 1 & -1 & 1 & -1 & -i & i & -i & i & -i & i & -i & i \\
 1 & -1 & 1 & -1 & -1 & 1 & -1 & 1 & -i & i & -i & i & i & -i & i & -i \\
 1 & -1 & 1 & -1 & 1 & -1 & 1 & -1 & i & -i & i & -i & i & -i & i & -i \\
 1 & -1 & 1 & -1 & -1 & 1 & -1 & 1 & i & -i & i & -i & -i & i & -i & i \\
 1 & 1 & -1 & -1 & -i & -i & i & i & -1 & -1 & 1 & 1 & i & i & -i & -i \\
 1 & 1 & -1 & -1 & i & i & -i & -i & -1 & -1 & 1 & 1 & -i & -i & i & i \\
 1 & 1 & -1 & -1 & -i & -i & i & i & 1 & 1 & -1 & -1 & -i & -i & i & i \\
 1 & 1 & -1 & -1 & i & i & -i & -i & 1 & 1 & -1 & -1 & i & i & -i & -i \\
 1 & -1 & -1 & 1 & -i & i & i & -i & i & -i & -i & i & 1 & -1 & -1 & 1 \\
 1 & -1 & -1 & 1 & i & -i & -i & i & i & -i & -i & i & -1 & 1 & 1 & -1 \\
 1 & -1 & -1 & 1 & -i & i & i & -i & -i & i & i & -i & -1 & 1 & 1 & -1 \\
 1 & -1 & -1 & 1 & i & -i & -i & i & -i & i & i & -i & 1 & -1 & -1 & 1 \\
\end{pmatrix}
\label{SD44thblock}
\end{align}}

{\footnotesize
\setlength{\arraycolsep}{1pt}
\begin{align}
&T_{-1,-1}= \begin{pmatrix}
 1 & 0 & 0 & 0 & 0 & 0 & 0 & 0 & 0 & 0 & 0 & 0 & 0 & 0 & 0 & 0 \\
 0 & 1 & 0 & 0 & 0 & 0 & 0 & 0 & 0 & 0 & 0 & 0 & 0 & 0 & 0 & 0 \\
 0 & 0 & 1 & 0 & 0 & 0 & 0 & 0 & 0 & 0 & 0 & 0 & 0 & 0 & 0 & 0 \\
 0 & 0 & 0 & 1 & 0 & 0 & 0 & 0 & 0 & 0 & 0 & 0 & 0 & 0 & 0 & 0 \\
 0 & 0 & 0 & 0 & 1 & 0 & 0 & 0 & 0 & 0 & 0 & 0 & 0 & 0 & 0 & 0 \\
 0 & 0 & 0 & 0 & 0 & -1 & 0 & 0 & 0 & 0 & 0 & 0 & 0 & 0 & 0 & 0 \\
 0 & 0 & 0 & 0 & 0 & 0 & 1 & 0 & 0 & 0 & 0 & 0 & 0 & 0 & 0 & 0 \\
 0 & 0 & 0 & 0 & 0 & 0 & 0 & -1 & 0 & 0 & 0 & 0 & 0 & 0 & 0 & 0 \\
 0 & 0 & 0 & 0 & 0 & 0 & 0 & 0 & i & 0 & 0 & 0 & 0 & 0 & 0 & 0 \\
 0 & 0 & 0 & 0 & 0 & 0 & 0 & 0 & 0 & i & 0 & 0 & 0 & 0 & 0 & 0 \\
 0 & 0 & 0 & 0 & 0 & 0 & 0 & 0 & 0 & 0 & -i & 0 & 0 & 0 & 0 & 0 \\
 0 & 0 & 0 & 0 & 0 & 0 & 0 & 0 & 0 & 0 & 0 & -i & 0 & 0 & 0 & 0 \\
 0 & 0 & 0 & 0 & 0 & 0 & 0 & 0 & 0 & 0 & 0 & 0 & -1 & 0 & 0 & 0 \\
 0 & 0 & 0 & 0 & 0 & 0 & 0 & 0 & 0 & 0 & 0 & 0 & 0 & 1 & 0 & 0 \\
 0 & 0 & 0 & 0 & 0 & 0 & 0 & 0 & 0 & 0 & 0 & 0 & 0 & 0 & 1 & 0 \\
 0 & 0 & 0 & 0 & 0 & 0 & 0 & 0 & 0 & 0 & 0 & 0 & 0 & 0 & 0 & -1 \\
\end{pmatrix}
\label{TD44thblock}
\end{align}}

\end{document}